\documentclass[12pt,reqno]{amsart}


\newcommand{\ca}[1]{\mathcal{#1}}

\newcommand{\field}[1]{\mathbb{#1}}
\newcommand{\rz}{\field{R}}
\newcommand{\cz}{\field{C}}
\newcommand{\zz}{\field{Z}}
\newcommand{\nz}{\field{N}}
\newcommand{\pz}{\field{P}}
\newcommand{\fr}{\frac}
\newcommand{\la}{\lambda}
\newcommand{\ep}{\varepsilon}
\newcommand{\om}{\omega}
\newcommand{\al}{\alpha}
\newcommand{\tta}{\theta}
\newcommand{\ga}{\gamma}
\newcommand{\si}{\sigma}

\newcommand{\Om}{\Omega}

\newcommand{\ti}{\tilde}
\newcommand{\pa}{{\partial}}
\newcommand{\lb}{\label}
\newcommand{\no}{\notag}

\newcommand{\ul}{\underline}

\newcommand{\py}{P_\infty}
\newcommand{\ztz}{\underset{\zeta\to 0}{=}}
\newcommand{\vez}[2]{\tta (\ul{z}({#1},{#2}))}
\newcommand{\dsk}[1][2]{\om_{\py ,#1}^{(2)}}
\newcommand{\dtk}{\om_{\py ,\hat{\nu}_0(x)}^{(3)}}
\newcommand{\abel}[1]{\ul{\al}_{P_0}({#1})}
\newcommand{\muv}[1]{\ul{\hat{\mu}}({#1})}
\newcommand{\nuv}[1]{\ul{\hat{\nu}}({#1})}

\newcommand{\hatt}{\widehat}
\newcommand{\dott}{\,\cdot\,}
\newcommand{\barr}[1]{\overline{#1}}
\newcommand{\mm}{m=3n+\ep,\,\ep\in\{1,2\},\,n\in\nz_0}
\newcommand{\rr}{r=3s+\ep',\, \ep'\in\{1,2\},\,s\in\nz_0}

\newtheorem{theorem}{Theorem}[section]
\newtheorem{defn}[theorem]{Definition}
\newtheorem{exa}[theorem]{Example}
\newtheorem{lem}[theorem]{Lemma}

\newtheorem{rem}[theorem]{Remark}

\numberwithin{equation}{section}

\DeclareMathOperator{\Bsq}{Bsq}
\newcommand{\Div}{\operatorname{Div}}
\newcommand{\DivP}{\operatorname{Div_P}}
\newcommand{\resN}{\operatorname{res}}

\renewcommand{\Im}{\text{\rm Im}}

\allowdisplaybreaks

\begin{document}

\title{Algebro-Geometric Solutions of the \\Boussinesq Hierarchy}

\author{R. Dickson}
\address{Department of Mathematics,
University of Missouri, Columbia, MO 65211, USA }
\email{dickson@picard.math.missouri.edu}
\author{F. Gesztesy}
\address{Department of Mathematics,
University of Missouri, Columbia, MO 65211, USA }
\email{fritz@math.missouri.edu}
\urladdr{http://www.math.missouri.edu/people/faculty/fgesztesypt.html}
\author{ K. Unterkofler}
\address{ Institute for Theoretical Physics,
Technical University of Graz,
A--8010 Graz, Austria}
\email{karl@itp.tu-graz.ac.at}

\begin{abstract}
We continue a recently developed systematic approach to the
Bousinesq (Bsq) hierarchy and its algebro-geometric
solutions. Our formalism includes a recursive construction
of Lax pairs and establishes associated Burchnall-Chaundy
curves, Baker-Akhiezer functions and Dubrovin-type equations
for analogs of Dirichlet and Neumann divisors. The principal
aim of this paper is a detailed theta function representation
of all algebro-geometric quasi-periodic solutions and related
quantities of the Bsq hierarchy.

\end{abstract}

\date{\today}
\maketitle
%%%%%%%%%%%%%%%%%%%%%%%%%%%%%%%%%%%%%%%%%%%%%%%%%%%%%%%%%%%%
\section{Introduction} \lb{Intro}
%%%%%%%%%%%%%%%%%%%%%%%%%%%%%%%%%%%%%%%%%%%%%%%%%%%%%%%%%%%%

The Boussinesq (Bsq) equation,
\begin{equation}
u_{tt} = u_{xx} + 3(u^2)_{xx} - u_{xxxx}, \lb{uu}
\end{equation}
 was originally introduced in 1871 as a
model for one-dimensional weakly nonlinear dispersive
water waves propagating in both directions (cf.~the recent
discussion in \cite{Pe98}).
It is customary to cast the equation in yet another form and
instead write it as the system of equations
\begin{equation}
q_{0,t} + \fr{1}{6}q_{1,xxx} + \fr{2}{3}q_1q_{1,x} = 0,
\qquad q_{1,t} - 2q_{0,x} = 0. \lb{bsq}
\end{equation}
Introducing
\begin{equation}
q_1(x,t) = -(6u(x,3^{-1/2}t) + 1)/4,
\end{equation}
equation \eqref{uu} results upon eliminating $q_0$ (cf.~also
\cite{gwr}).

The principal subject of this paper concerns
algebro-geometric quasi-periodic solutions of the completely
integrable hierarchy of Boussinesq equations, of which
\eqref{bsq} is just the first of infinitely many members.
In order
to be able to give a more precise description of the concepts
involved, we briefly recall some basic notation in connection
with the Boussinesq hierarchy.

The Boussinesq hierarchy is defined in terms of Lax pairs
$(L_3,P_m)$ of differential expressions, where $L_3$ is a
fixed one-dimensional third-order linear differential
expression,
\begin{equation}
L_3 = \fr{d^3}{dx^3} + q_1\fr{d}{dx}
+ \fr{1}{2}q_{1,x} + q_0,
\lb{L3}
\end{equation}
and $P_m$ is a differential expression of order
$m\neq 0(\text{mod }3)$, such that the commutator of
$L_3$ and
$P_m$ becomes a differential expression of order one.
For the Boussinesq equation \eqref{bsq} itself, we have
$m=2$,
that is,
\begin{equation}
P_2 = \fr{d^2}{dx^2} + \fr{2}{3}q_1, \lb{P3}
\end{equation}
and the resulting Lax commutator representation of the
Boussinesq equation then reads
\begin{equation}
\Bsq_2(q_0,q_1) = \fr{d}{dt}L_3 - [P_2,L_3] = 0,
 \ \text{ that is, } \
\begin{cases}
q_{0,t} + \fr{1}{6}q_{1,xxx} + \fr{2}{3}q_1q_{1,x} = 0, \\
q_{1,t} - 2q_{0,x} = 0.
\end{cases}
\end{equation}
A systematic, in fact, recursive approach to all
differential
expressions $P_m$ will be reviewed in Section~\ref{S:2.1}.

However, before turning to the contents of each section,
it seems appropriate to review the existing literature
on the subject and its relation to our approach. Despite
a fair
number of papers on the Boussinesq system, the current
status of
research has not yet reached  the high level of the KdV
hierarchy, or more generally, that of the AKNS hierarchy.
From
the perspective of completely integrable systems, the
reasons for
this discrepancy are easily traced back to the enormously
increased complexity when making the step from the
second-order
operator $L_2$ associated with the KdV hierarchy to the
third-order operator $L_3$ in connection with the Bsq
hierarchy.
On an algebro-geometrical level this difference amounts
to
hyperelliptic curves in the KdV (and AKNS) context as
opposed to
non-hyperelliptic ones that arise in the Bsq case.

The classical paper on the Bsq equation, or perhaps more
appropriately, the nonlinear string equation, is due to
Zakharov
\cite{zak}. In particular, he introduced the basic Lax
pair
$(L_3,P_2)$ and discussed the infinite set of polynomial
integrals
of motion. In many ways closest in spirit to our
approach is the
seminal paper by McKean \cite{mk81} (see also \cite{mk79})
describing spatially periodic solutions
of the Bsq equation. In contrast to \cite{mk81} though,
 we
 concentrate here on the algebro-geometric (i.e.,
finite-genus)
 case and make no assumptions of periodicity in order
to describe all algebro-geometric quasi-periodic solutions.
The application of inverse scattering techniques for the
third-order differential expression $L_3$ to the
initial value
problem of the Bsq equation is discussed in great detail
by Deift, Tomei, and Trubowitz \cite{dtt} and Beals,
Deift, and
Tomei \cite{bdt}. General existence theorems (local and
global in time) for
solutions of the Bsq equation can also be found,
for instance, in Craig
\cite{cra}, Bona and Sachs \cite{bosa}, and Fang and
Grillakis \cite{FG96}, and the references therein.
In particular, \cite{bdt}, \cite{bosa}, \cite{cra},
\cite{dtt}, \cite{Li97}, \cite{mk81}, and \cite{Mc84}
further
discuss and contrast the blow-up mechanism for solutions
of the
nonlinear string equation obtained by Kalantarov and
Ladyzhenskaya
\cite{kala}. Other special classes of solutions have been
considered by a variety of authors. For instance,
certain classes
of rational Bsq solutions are treated by Airault
\cite{air86},
Airault, McKean, and Moser \cite{amkm77}, Chudnovsky
\cite{chu79}, and Latham and Previato \cite{lapr95}. In
addition,
the classical dressing method of Zhakarov and Shabat to
construct
particular classes of solutions for very general
systems of integrable equations, as described, for
instance, in \cite{zm85}, \cite{zm85a}, \cite{zs74},
and \cite{zs79}, should be mentioned in this
context. Moreover, certain algebro-geometric
Bsq solutions,
obtained as special solutions of the
Kadomtsev-Petviashvili (KP)
equation or by the reduction theory of Riemann theta
functions, are briefly discussed by Dubrovin \cite{du81},
Matveev
and Smirnov \cite{masm87}, \cite{masm87a}, \cite{masm93},
Previato \cite{pre90}, \cite{pre94}, Previato and
Verdier \cite{PV93},
and Smirnov \cite{smi87}, \cite{Sm96}. The latter solutions
appear as
special cases of a general scheme of constructing
algebro-geometric solutions of completely integrable
systems developed by Krichever \cite{kr76}, \cite{kr77},
\cite{kr78} and Dubrovin \cite{du77}, \cite{du83} (see also
\cite{BBEIM94}, \cite{GH99}, \cite{NMPZ84}, \cite{SW85}).

Our principal contribution to this subject is a unified
framework
that yields all algebro-geometric quasi-periodic solutions
of the
entire Boussines hierarchy at once. In Section~\ref{S:2.1} we
briefly recall a recursive construction  of the
stationary Bsq
hierarchy following the approach first  outlined in our paper
\cite{dgu1}.  The stationary Boussinesq hierarchy is then
obtained by imposing the
$t$-independent Lax commutator relations
\begin{equation}
[P_m,L_3]=0, \quad m\neq 0\, (\text{mod }3), \lb{In1}
\end{equation}
assuming $q_0$ and $q_1$ to be $t$-independent.
{}From the differential expression $P_m$ we construct two
polynomials $S_m(z)$ and $T_m(z)$ in $z$,
which are both $x$-independent.
This leads immediately to the classical Burchnall-Chaundy
polynomial
(cf.\ \cite{bc1}, \cite{bc2}),
and hence to a (generally, non-hyperelliptic) curve
$\ca{K}_{m-1}$
of arithmetic genus
$m-1$, the central object in the analysis to follow.

In Section~\ref{S:2.2}, the stationary formalism, and in
particular, the curve $\ca{K}_{m-1}$ are briefly reviewed.
Rather than studying the Baker-Akhiezer function
$\psi$ (i.e., the
common eigenfunction $\psi$ of the commuting operators
$L_3$ and
$P_m$)
directly, our main object is a meromorphic function
$\phi$ equal
to the logarithmic $x$-derivative of $\psi$, such that
$\phi$
satisfies a nonlinear second-order differential equation.
Moreover, we describe Dubrovin-type equations for the analogs
of Dirichlet  and Neumann eigenvalues when compared to
the KdV
hierarchy.

Section~\ref{S:2.3} then presents our first set of new
results,
the explicit theta function representations of the
Baker-Akhiezer
function, the meromorphic function $\phi$, and in
particular,
that of the potentials $q_1$ and $q_0$ for the entire
Boussinesq
hierarchy (the latter being the analog of the celebrated
Its-Matveev formula \cite{itma} in the KdV context).

Sections~\ref{S:3.1} and \ref{S:3.2} then extend the
analyses of
Sections~\ref{S:2.2} and \ref{S:2.3}, respectively,
to the
time-dependent case.  Each equation in the hierarchy is
permitted to evolve in terms of an independent
deformation (time)
parameter
$t_r$. As initial data we use a stationary solution of
the $m$th
equation of the Boussinesq hierarchy and then construct a
time-dependent solution of the $r$th equation of the
Boussinesq
hierarchy.  The Baker-Akhiezer function, the
meromorphic function $\phi$, the analogs of the Dubrovin
equations, and the theta function representations of
Section~\ref{S:2.3} are all extended to the
time-dependent case.

In Appendix A we provide an introduction to the theory of
Riemann surfaces and their theta functions.  Appendix B is
a collection of results on trigonal Riemann surfaces
associated
with Bsq-type curves.

It should perhaps be noted at this point
that our elementary algebraic
approach to the Bsq hierarchy and its algebro-geometric
solutions
is in fact universally applicable to
$1+1$-dimensional hierarchies of soliton equations such as
the KdV hierarchy \cite{grt},
the AKNS hierarchy \cite{GR96},
the combined sine-Gordon and mKdV hierarchy \cite{GH97},
and the Toda and Kac-van Moerbeke
hierarchies \cite{bght} (see also \cite{GH99}).

%%%%%%%%%%%%%%%%%%%%%%%%%%%%%%%%%%%%%%%%%%%%%%%%%%%%%%%%%%%%
\section{The Recursive Approach to the Boussinesq
Hierarchy}\lb{S:2.1}
%%%%%%%%%%%%%%%%%%%%%%%%%%%%%%%%%%%%%%%%%%%%%%%%%%%%%%%%%%%%

In this section we briefly recall the necessary material
from our
previous paper \cite{dgu1} without proofs.

Suppose $q_0, q_1$ are
meromorphic on $\cz$ and introduce the third-order
differential expression
\begin{equation}
 L_3= \fr{d^3}{dx^3}+ q_1 \fr{d }{dx} +
\fr{1}{2}\, q_{1,x}+ q_0 , \quad
x \in \cz. \lb{51}
\end{equation}

For each fixed $m\in\nz_0\ (=\nz\cup\{0\})$
with $m\neq 0(\text{mod }3)$ we write
\begin{equation}
m=3n+\ep,\quad \ep\in\{1,2\},
\end{equation}
and then construct two distinct differential expressions
of order
$3n+1$ and $3n+2$, respectively, denoted by $P_m,$ where
$m=3n+1$ or $m=3n+2$.  In order for these differential
expressions $P_m$ to commute with $L_3$, one proceeds as
follows
(cf.~\cite{dgu1} for more details).

Pick   $n \in \nz_0$,   $\ep\in\{1,2\}$, and define the
sequences
$\{ f_{\ell}^{(\ep)}(x)\}_{\ell = 0,\dots,n+1}$ and
$\{ g_{\ell}^{(\ep)}(x)\}_{\ell = 0,\dots,n+1}$
recursively by
\begin{align}
(f_0^{(\ep)}, g_0^{(\ep)}) & =
(c_0^{(\varepsilon)},d_0^{(\varepsilon)})
= \begin{cases} (0,1) &
\text{for }
\ep=1,
\\ (1,d_0^{(2)}) & \text{for } \ep=2, \end{cases} \qquad
d_0^{(2)}\in\cz , \no \\[3mm]
3 f_{\ell,x}^{(\ep)} &= 2 g_{\ell-1, xxx}^{(\ep)} +
2q_1 g_{\ell-1,x}^{(\ep)} + q_{1,x} g_{\ell-1}^{(\ep)} +
3q_0 f_{\ell-1,x}^{(\ep)} + 2q_{0,x} f_{\ell-1}^{(\ep)}  ,
\lb{52} \\
3 g_{\ell,x}^{(\ep)} &=  3 q_0 g_{\ell-1,x}^{(\ep)} +
q_{0,x}g_{\ell-1}^{(\ep)} - \fr{1}{6}\,f_{\ell-1,xxxxx}^{(\ep)}
-  \fr{5}{6}\,q_1 f_{\ell-1,xxx}^{(\ep)}
- \fr{5}{4}\,q_{1,x} f_{\ell-1,xx}^{(\ep)} \no \\ &
- \big(\fr{3}{4}\,q_{1,xx} + \fr{2}{3}\, q_1^2 \big)
f_{\ell-1,x}^{(\ep)} -
 \big(\fr{1}{6}\, q_{1,xxx} + \fr{2}{3}\, q_1 q_{1,x}
\big) f_{\ell-1}^{(\ep)} ,
\quad  \ell = 1,\dots,n+1 . \no
\end{align}
However, as most of the ensuing discussion can be made for
both cases simultaneously, we write
\begin{equation}
f_\ell = f_\ell^{(\ep)}, \qquad g_\ell = g_\ell^{(\ep)},
\end{equation}
and only make the distinction explicit when necessary.

Explicitly, one computes

\noindent (i) Let $m=1\,(\text{mod }3)$ (i.e., $\ep=1$):
\begin{align}
f_0^{(1)} &= 0, \qquad g_0^{(1)} = 1, \no \\
3f_1^{(1)} &= q_1 + 3 c_1^{(1)}, \qquad 3g_1^{(1)} =
q_0 + 3d_1^{(1)}, \no \\
3f_2^{(1)} &= \fr{2}{3}\,q_{0,xx} +
\fr{4}{3}\,q_0 q_{1}
+ c_1^{(1)}\,2q_0 + d_1^{(1)}\,q_1 + 3c_2^{(1)}, \no \\
3g_2^{(1)} &= - \fr{1}{18}\,q_{1,xxxx}
- \fr{1}{6}\,q_{1,x}^2 - \fr{4}{27}q_1^3
- \fr{1}{3}\, q_1 q_{1,xx} + \fr{2}{3}\,q_0^2 \no \\
& + c_1^{(1)} \big(-\fr{1}{6} q_{1,xx}- \fr{1}{3}\,q_1^2 \big)
+ d_1^{(1)} q_0 + 3d_2^{(1)}, \lb{53} \\
&  \text{ etc.}   \no
\end{align}
(ii) Let $m=2\,(\text{mod }3)$ (i.e., $\ep=2$):
\begin{align}
f_0^{(2)} &= 1, \qquad g_0^{(2)}= d_0^{(2)} , \no \\
3f_1^{(2)} &= 2q_0 + d_0^{(2)} q_1 + 3c_1^{(2)}, \qquad
3g_1^{(2)} = - \fr{1}{6}\,q_{1,xx} - \fr{1}{3}\,
q_1^2 + d_0^{(2)} q_0 + 3d_1^{(2)}, \no \\
3f_2^{(2)} &= \big( -\fr{1}{9}\,q_{1,xxxx} -
\fr{5}{9}\,q_1 q_{1,xx}
- \fr{5}{27}\,q_1^3 - \fr{5}{12}\,q_{1,x}^2 +
\fr{5}{3}\,q_0^2 \big) \no \\ &
+ d_0^{(2)} \big( \fr{2}{3}\,q_{0,xx} + \fr{4}{3}\,
q_0 q_1 \big)
+ c_1^{(2)}\,2q_0 + d_1^{(2)}\,q_1 + 3c_2^{(2)}, \no \\
3g_2^{(2)} &= \big( -\fr{1}{9}\,q_{0,xxxx}
- \fr{5}{9}\,q_1^2 q_0 - \fr{5}{18}
\,q_0 q_{1,xx}
- \fr{5}{9}\,q_1 q_{0,xx} - \fr{5}{18}\,q_{0,x}q_{1,x}
\big) \no \\ &
+ d_0^{(2)}\,\big( - \fr{1}{18}\,q_{1,xxxx}
- \fr{1}{6}\,q_{1,x}^2 - \fr{4}{27}\,q_1^3
- \fr{1}{3}\,q_1 q_{1,xx} + \fr{2}{3}\,q_0^2
\big) \no \\ &
+ c_1^{(2)}\,\big(-\fr{1}{6}\,
q_{1,xx} - \fr{1}{3}\,q_1^2\big) + d_1^{(2)}\,q_0
+ 3d_2^{(2)}, \lb{54} \\
&  \text{ etc.,}  \no
\end{align}
where $ \{c_{\ell}^{(\ep)}\}_{\ell =1,\dots,n},
\{d_{\ell}^{(\ep)}\}_{\ell = 0,\dots,n}$
are integration constants, which arise when solving (\ref{52}).
It is convenient to introduce the homogeneous case where
all free
integration constants vanish.  We introduce
\begin{equation}
\hat{f}_\ell ^{(\ep)} =f_\ell^{(\ep)}
\mid_{c_p^{(\ep)}=d_p^{(\ep)}=0,\,p=1,\dots,\ell},
\qquad
\hat{g}_\ell^{(\ep)} =g_\ell^{(\ep)}
\mid_{c_p^{(\ep)}=d_p^{(\ep)}=0,\,p=1,\dots,\ell} \lb{2.7}
\end{equation}
and use (cf.~\eqref{52})
\begin{equation}
c_0^{(1)}=0, \quad c_0^{(2)}=1, \quad d_0^{(1)}=1, \quad
d_0^{(2)}=0. \lb{2.8}
\end{equation}
We do not list these functions
explicitly, however, this notation allows us to write
\begin{equation}
f_\ell^{(\ep)}=\sum_{p=0}^\ell
( d_p^{(\ep)}\hat{f}_{\ell-p}^{(1)}
+ c_p^{(\ep)}\hat{f}_{\ell-p}^{(2)}), \qquad
g_\ell^{(\ep)}=\sum_{p=0}^\ell
(d_p^{(\ep)}\hat{g}_{\ell-p}^{(1)} +
c_p^{(\ep)}\hat{g}_{\ell-p}^{(2)}).
\end{equation}
Given (\ref{52}) one defines the differential
expression $P_m$ of order $m$ by
\begin{eqnarray}
  P_m & =& \sum_{\ell=0}^{n}
 \Big( f_{n-\ell}^{(\ep)}\,
\fr{d^2}{dx^2}+\big(g_{n-\ell}^{(\ep)}-\fr{1}{2}\,
f_{n-\ell,x}^{(\ep)}\big)
\fr{d }{dx } \no\\ && + \big(\fr{1}{6}\,
f_{n-\ell,xx }^{(\ep)}
- g_{n-\ell,x}^{(\ep)}+\fr{2}{3}\, q_1 f_{n-\ell}^{(\ep)}
 \big) \Big)
 L_3^{ \ell}  + \sum_{\ell=0}^{n} k_{m,\ell} L_3^{\ell} ,
\lb{55}\\
&& \quad k_{m,\ell} \in \cz, \quad \ell=0,\dots, n,
\quad \mm, \no
\end{eqnarray}
and verifies that
\begin{align}
[ P _m, L_3] &= 3\, {f}_{n+1, x}^{(\ep)}\, \fr{d}{dx} +
\fr{3}{2}\,
 f_{n+1, xx}^{(\ep)} + 3\, g_{n+1, x}^{(\ep)} ,\no \\
&\hspace{20mm} \mm \lb{56}
\end{align}
(where $[\dott,\dott]$ denotes the commutator symbol).
The pair $(L_3,P_m)$ represents the Lax pair for the
Bsq hierarchy. Varying $n\in\nz_0$ and $\ep\in\{1,2\}$,
 the stationary Bsq hierarchy is then defined by the
vanishing of
the commutator of $P_m$ and $L_3$ in (\ref{56}), that
is, by
 \begin{eqnarray}
[P_m,L_3]=0 ,
\qquad \mm,  \lb{57}
\end{eqnarray}
or equivalently, by
 \begin{eqnarray}
{f}_{n+1,x}^{(\ep)}=0, \quad g_{n+1,x}^{(\ep)}=0, \qquad
\ep\in\{1,2\},\,n \in\nz_0 .
\lb{58}
\end{eqnarray}
Explicitly, one obtains for the first few equations of
the stationary Boussinesq hierarchy,
\begin{align}
m=1 & \text{ (i.e., $n=0$ and $\ep=1$)}: \no \\
& q_{0,x}=0, \quad q_{1,x}=0. \no \\
m=2 & \text{ (i.e., $n=0$ and $\ep=2$)}: \no \\
& - \fr{1}{6}\, q_{1,xxx} -\fr{2}{3}\, q_{1} q_{1,x} +
d_0^{(2)} q_{0,x}=0,
\quad 2\, q_{0,x}  + d_0^{(2)}  q_{1,x} =0 .\no\\
m=4 & \text{ (i.e., $n=1$ and $\ep=1$)}: \no \\
&  -\fr{1}{18}\, q_{1,xxxxx}
   -\fr{1}{3} \, q_1 q_{1,xxx} -\fr{2}{3}\,
q_{1,x}q_{1,xx}- \fr{4}{9}\,
   q_{1}^2 q_{1,x}  +\fr{4}{3}\,  q_{0 }q_{0,x} \no\\
& \quad  + c_1^{(1)}
\big( -\fr{1}{6}\, q_{1,xxx} -\fr{2}{3}\,
q_{1} q_{1,x}\big)+d_1^{(1)} q_{0,x} =0,
\no\\
&  \fr{2}{3}\, q_{0,xxx}
   + \fr{4}{3}\,   q_1 q_{0,x} + \fr{4}{3}\,
q_{1,x} q_{0}
   +  c_1^{(1)} 2 q_{0,x} + d_1^{(1)} q_{1,x} =0 ,
\lb{59}\\
& \quad \text{etc.} \no
\end{align}

By definition, solutions $(q_0,q_1)$ of any of the
stationary Bsq equations (\ref{59}) are called
\textbf{stationary
algebro-geometric Bsq solutions} or simply
\textbf{algebro-geometric Bsq potentials}.

Next, we introduce two polynomials $F_m$ and $G_m$,
both of degree
at most $n$ with respect to the variable $z\in\cz$,
\begin{align}
F_m(z,x) &= \sum_{\ell=0}^n f_{n-\ell}^{(\ep)}(x) z^{\ell},
\lb{510} \\
G_m(z,x) &= \sum_{\ell=0}^n g_{n-\ell}^{(\ep)}(x) z^{\ell} ,
\lb{511} \quad \mm.
\end{align}
In terms of homogeneous quantities we define
(cf.~\eqref{2.7} and
\eqref{2.8})
\begin{equation}
\widehat{F}_\ell=F_\ell\mid_{c_p^{(\ep)}
=d_p^{(\ep)}=0,\,p=1,\dots,n},
\qquad
\widehat{G}_\ell=G_\ell\mid_{c_p^{(\ep)}
=d_p^{(\ep)}=0,\,p=1,\dots,n}.
\end{equation}
We may then write
\begin{equation}
F_m=\sum_{j=0}^n ( c_{n-j}^{(\ep)}\widehat{F}_{3j+2} +
d_{n-j}^{(\ep)}\widehat{F}_{3j+1}), \qquad
G_m=\sum_{j=0}^n (c_{n-j}^{(\ep)}\widehat{G}_{3j+2} +
d_{n-j}^{(\ep)}\widehat{G}_{3j+1}). \lb{step}
\end{equation}
 Explicitly, the first few polynomials
$F_m,G_m$ read
\begin{align}
  F_1 &= 0, \quad G_1 = 1,
    \no \\
  F_2 &= 1 , \quad G_2 = d_0^{(2)},
    \no \\
  F_4 &= \fr{1}{3}\,q_1 + c_1^{(1)} , \quad
G_4 = z +\fr{1}{3} \, q_0 + d_1^{(1)}, \\
  F_5 &= z + \fr{2}{3}\,q_0 + d_0^{(2)} \,
    \fr{1}{3}\,q_1 + c_1^{(2)}, \quad
G_5 = d_0^{(2)}z - \fr{1}{18}\,q_{1,xx} -
\fr{1}{9}\,q_1^2
     + d_0^{(2)}\,\fr{1}{3}\,q_0 + d_1^{(2)},
    \no \\
& \text{ etc.}  \no
\end{align}
Given (\ref{510}) and (\ref{511}), (\ref{57})
(or equivalently, (\ref{58})) becomes
 \begin{eqnarray}
2 \,{G}_{m, xxx} +2\, q_1 {G}_{m,x}  +q_{1,x} {G}_m
-3\,(z-q_0 ) {F}_{m,x} +  2\, q_{0,x} {F}_m&=&0,
 \lb{512} \\
 \fr{1}{6}\,{F}_{m,xxxxx} + \fr{5}{6}\, q_1 {F}_{m,xxx}
+ \fr{5}{4}\,
q_{1,x}  {F}_{m,xx}   +\big( \fr{3}{4}\, q_{1,xx}+
\fr{2}{3}\, q_1^2  \big)
{F}_{m,x} \no\\+ \big(  \fr{1}{6} \, q_{1,xxx}+
\fr{2}{3}\, q_1
q_{1,x}\big) {F}_m
 + 3(z-q_{0})G_{m,x} - q_{0,x} G_m &=&0.
\lb{513}
 \end{eqnarray}
Both equations can be integrated (cf.~\cite{dgu1}) to get
\begin{align}
S_m(z) &=  -\fr{1}{6}\, F_{m,xxxx}F_m
+\fr{1}{6}\,F_{m,xxx} F_{m,x} -\fr{1}{12}\, F_{m,xx}^2
- \fr{5}{6} \,q_1 F_{m,xx} F_{m}
\no \\
&- \fr{5}{12}\,q_{1,x} F_{m,x } F_m
+ \fr{5}{12}\,q_1 F_{m,x }^2 - \fr{1}{3}\,
\big(\fr{1}{2}\, q_{1,xx} + q_1^2
\big) F_m^2 + 2 \, G_{m,xx} G_m
\no \\ &- G_{m,x}^2 + q_1 G_m^2-
3(z-q_0) F_m G_m, \lb{514}
\end{align}
where the integration constant $S_m(z)$ is a polynomial in $z$
of degree at most $2n-1+\ep,\,\mm$,
\begin{eqnarray}
S_m(z) = \sum_{p=0}^{2\, n-1+\ep} s_{m,p} z^p,\quad
\mm,
\lb{515}
\end{eqnarray}
and
\begin{align}
T_m(z) &\ =
\fr{1}{18} \, F_{m,xxxx} F_{m,xx} F_m
-\fr{1}{24} \, F_{m,xxxx} F_{m,x}^2 \no \\ &+
\fr{1}{36} \, F_{m,xxx} F_{m,xx} F_{m,x}-
\fr{1}{108} \, F_{m,xx}^3 - \fr{1}{36} F _m F _{m,xxx}^2+
\fr{1}{18}\, q_1  F_{m,xxxx} F_m^2 \no\\ &
- \fr{1}{18}\, q_{1,x} F_{m,xxx} F_m^2
- \fr{1}{9}\, q_1 F_{m,xxx} F_{m,x} F_m +
\fr{1}{18}\, q_{1,xx} F_{m,xx} F_m^2
\no\\ & +
\fr{2}{9} \, q_{1,x} F_{m,xx} F_{m,x} F_m -
\fr{7}{72} \, q_1 F_{m,xx} F_{m,x}^2 +
\fr{7}{36} \, q_1 F_{m,xx}^2 F_m
\no\\ &+
\fr{5}{18} \, q_1^2 F_{m,xx} F_m^2
- \fr{1}{24}\, q_{1,xx} F_{m,x}^2 F_m -
\fr{7}{48} \, q_{1,x} F_{m,x}^3 +
\fr{1}{12} \, q_{1,x} q_1 F_{m,x} F_m^2
\no \\ &-
\fr{1}{6} \, q_1^2 F_{m,x}^2 F_m
+\big( \fr{2}{27} \, q_1^3 - \fr{1}{36}
\, q_{1,x}^2 + \fr{1}{18} \, q_{1,xx} q_1 +(z-q_0 )^2
 \big) F_m^3
\no\\ &+
(z-q_0) G_m^3 + \fr{1}{6} \, F_{m,xxxx}
G_m^2 - \fr{1}{3} \, F_{m,xxx} G_{m,x} G_m  +
F_m G_{m,xx}^2
\no\\ &+
\fr{1}{3} \, F_{m,xx} \big( G_{m,x}^2 +
G_{m,xx} G _m \big) - F_{m,x} G_{m,xx} G_{m,x} -
q_1 (z-q_0) F_m^2 G_m
\no \\ &+
\fr{2}{3}\, q_1^2 F_m G_m^2 + \fr{5}{6} \, q_1 F_{m,xx}
  G_m^2 - \fr{4}{3} \, q_1 F_{m,x} G_{m,x} G_m +
\fr{7}{12}\, q_{1,x} F_{m,x} G _m^2
\no\\ &+
\fr{1}{3} \,q_1 F_m G_{m,x}^2  +
\fr{4}{3}\, q_1 F_m G_{m,xx} G_m +
\fr{1}{6} \, q_{1,xx} F_m G_m^2 -
\fr{1}{3}\, q_{1,x} F_m G_{m,x} G_m
\no \\ & +
(z-q_0) F_{m,x} F_m G_{m,x} -
\fr{1}{4} \,(z-q_0) F_{m,x}^2 G_m -
2(z-q_0) F_m^2 G_{m,xx }, \lb{516}
\end{align}
where the integration constant $T_m(z)$ is a monic
polynomial of degree $m$,
\begin{eqnarray}
T_m(z) = z^{m}+ \sum_{q=0}^{m-1} t_{m,q} z^q, \quad \mm.
\lb{517}
\end{eqnarray}
Next, we consider the algebraic kernel of $(L_3-z), \ z\in\cz$
(i.e., the formal nullspace in a purely algebraic sense),
\begin{equation}
\ker (L_3 - z) = \{\psi:\cz\to\cz\cup\{\infty\}
\text{ meromorphic }\mid (L_3-z)\psi = 0
\},
\quad z\in\cz.
\end{equation}
Taking into account
(\ref{57}), that is, $[P_m,L_3]=0$, computing the
restriction of $P_m$ to
$\ker(L_3\mbox{-}z)$, and using
\begin{eqnarray}
\psi_{xxx}= -q_1 \psi_{x} + \big( z- 2^{-1} q_{1,x}
- q_0\big) \psi,
\quad \text{etc.}, \lb{518}
\end{eqnarray}
to eliminate higher-order derivatives of $\psi$, one
obtains from
(\ref{52}), (\ref{55}), (\ref{58}), (\ref{510}),
(\ref{511}), (\ref{512}), and
(\ref{513})
\begin{eqnarray}
 P_m \Big{\vert}_{\ker(L_3-z)} = \Big(  F_m
\fr{d^2}{dx^2} + \big( G_m - \fr{1}{2}  F_{m,x}
\big) \fr{d}{dx} + H_m \Big)\Big{\vert}_{\ker(L_3-z)}.
\lb{519}
\end{eqnarray}
Here
\begin{eqnarray}
H_m(z,x) = \fr{1}{6}\, F_{m,xx }(z,x) +
\fr{2}{3}\,q_1(x) F_m(z,x) - G_{m,x}(z,x)
+ k_m(z) \lb{520}
\end{eqnarray}
and (cf.\ (\ref{55}))
\begin{eqnarray}
k_m(z)= \sum_{\ell=0}^{n} k_{m,\ell} z^{\ell}
\lb{521}
\end{eqnarray}
is an integration constant.  The presence of this constant
$k_m(z)$
in (\ref{520}), and hence in (\ref{519}), corresponds to adding
an arbitrary polynomial in $L_3$
to the non-trivial part of the differential expression $P_m$
(cf.\ (\ref{55})).  This polynomial in $L_3$ obviously
commutes
with
$L_3$, and without loss of generality we henceforth choose to
suppress its presence by setting
$k_m(z)=0$.

Still assuming $ f_{n+1,x}^{(\ep)}=g_{n+1,x}^{(\ep)}=0$ as
in  (\ref{58}),
$[P_m,L_3]=0$ in (\ref{55}) yields an algebraic
relationship between $P_m$ and $L_3$ by appealing to a
 result of Burchnall and Chaundy \cite{bc1}, \cite{bc2}
(see also \cite{GG91}, \cite{GP91}, \cite{Pr96}, \cite{wi85}).
In fact, one can prove
\begin{theorem} [\cite{dgu1}] \lb{t51}
Assume $f_{n+1,x}^{(\ep)}=g_{n+1,x}^{(\ep)}=0$, that is,
$[P_m,L_3]=0$,
$\mm$. Then the Burchnall-Chaundy polynomial
$\ca{F}_{m-1}(L_3,P_m)$ of the pair $(L_3,P_m)$
explicitly reads
\emph{(}cf.\ \emph{(\ref{515})} and \emph{(\ref{517}))}
\begin{align}
\ca{F}_{m-1}(L_3,P_m) &=
P_m^3 + P_m\,S_m(L_3) - T_m (L_3) = 0, \no \\
S_m(z) &=  \sum_{p=0}^{2\, n-1+\ep} s_{m,p} z^p, \quad
T_m(z) = z^{m}+ \sum_{q=0}^{m-1} t_{m,q} z^q, \lb{522} \\
 &  \hspace{10mm}   \mm . \no
\end{align}
\end{theorem}
\begin{rem} \lb{r52}
$\ca{F}_{m-1}(L_3,P_m)=0$ naturally leads to the plane
algebraic curve
$\ca{K}_{m-1}$,
\begin{equation}
\ca{K}_{m-1} : \
\ca{F}_{m-1}(z,y) =  y^3 + y \, S_m(z) - T_m(z) = 0
\lb{529}
\end{equation}
of (arithmetic) genus $m-1$. For $m\geq 4$ these
curves are non-hyperelliptic.
\end{rem}
Finally, introducing a deformation parameter $t_m\in\cz$ into
the pair $(q_0,q_1)$ (i.e.,
$q_{\ell}(x) \to q_{\ell}(x,t_m)$, $\ell =0,1$), the
time-dependent Bsq hierarchy is defined as a
 collection of evolution equations (varying $\mm$)
\begin{align}
\fr{d}{d\,t_m}L_3 (t_m) &- [P_m(t_m),L_3(t_m)] = 0, \no \\
& (x,t_m) \in\cz ^2, \, \mm, \lb{530}
\end{align}
or equivalently, by
\begin{align}
\Bsq_m(q_0,q_1) &=
\left\{ \begin{array}{l}
q_{0,t_m}-3\,g_{n+1,x}^{(\ep)}=0,\\
\\
q_{1,t_m}-3\,f_{n+1,x}^{(\ep)}=0,
\end{array} \right.
 \no \\ \no \\
& \qquad \qquad (x,t_m)\in\cz^2, \, \mm ,
\lb{531} \\
\intertext{that is, by}
 \Bsq_m(q_0,q_1)&=
\left\{ \begin{array}{l}
q_{0,t_m}
+ \fr{1}{6}\, {F}_{m,xxxxx}
+ \fr{5}{6}\,q_1 F_{m,xxx} + \fr{5}{4}\, q_{1,x}
F_{m,xx}
+ ( \fr{3}{4} \, q_{1,xx} +\fr{2}{3}\, q_1^2 ) F_{m,x}\\ \\
\quad  + ( \fr{1}{6} \, q_{1,xxx}+ \fr{2}{3}\,
q_1 q_{1,x}) F_m
+ 3(z-q_0) G_{m,x} - q_{0,x} G_m =0, \\  \\
q_{1,t_m} - 2 G_{m,xxx} - 2 q_1 G_{m,x}
- q_{1,x} G_m
+3(z-q_0) F_{m,x} -  2q_{0,x} F_m =0, \\
\end{array} \right.   \no \\
& \hspace*{4cm} (x,t_m) \in \cz^2 , \, \mm  . \lb{532}
\end{align}
Explicitly, one obtains for the first few
equations in (\ref{531}),
 \begin{eqnarray}
 && \Bsq_1(q_0,q_1)=
\left\{  \begin{array}{l}
q_{0,t_1} - q_{0,x} = 0, \\ \no \\
q_{1,t_1} - q_{1,x} = 0,
\end{array} \right.
\no \\ \no \\
&&\Bsq_2(q_0,q_1)=
\left\{ \begin{array}{l}
q_{0,t_2} + \fr{1}{6}\, q_{1,xxx} + \fr{2}{3} \,
q_{1} q_{1,x} - d_0^{(2)} q_{0,x} = 0,
\\ \no \\
q_{1,t_2} - 2\, q_{0,x} - d_0^{(2)} q_{1,x} = 0,
\end{array} \right. \no \\ \lb{533} \\
&&\Bsq_4(q_0,q_1) =
\left\{ \begin{array}{l}
q_{0,t_4} + \fr{1}{18} \, q_{1,xxxxx}
+ \fr{1}{3} \, q_1 q_{1,xxx} + \fr{2}{3} \,
q_{1,x} q_{1,xx} + \fr{4}{9} \,
q_{1}^2 q_{1,x} \\ \no\\
\qquad - \fr{4}{3} \, q_{0 } q_{0,x} + c_1^{(1)}
\big( \fr{1}{6} \, q_{1,xxx} + \fr{2}{3} \,
q_{1} q_{1,x} \big) - d_1^{(1)} q_{0,x} = 0,
\\ \no \\
q_{1,t_4} - \fr{2}{3} \, q_{0,xxx} - \fr{4}{3}\,
q_1 q_{0,x} - \fr{4}{3} \, q_{1,x} q_{0} -
c_1^{(1)} 2 q_{0,x} - d_1^{(1)} q_{1,x} = 0,
\end{array} \right.
\no\\ \no \\
 && \hspace*{2cm} \text{etc.} \no
\end{eqnarray}
%%%%%%%%%%%%%%%%%%%%%%%%%%%%%%%%%%%%%%%%%%%%%%%%%%%%%%%%%%%%%%
\section{The Stationary Boussinesq Formalism} \lb{S:2.2}
%%%%%%%%%%%%%%%%%%%%%%%%%%%%%%%%%%%%%%%%%%%%%%%%%%%%%%%%%%%%%

In this section we continue our review of the Bsq
hierarchy as discussed in \cite{dgu1} and focus our
attention on
the stationary case.
Following \cite{grt} we outline the connections
between the polynomial approach described in Section
\ref{S:2.1} and a fundamental meromorphic function
$\phi(P,x)$ defined on the Boussinesq curve $\ca{K}_{m-1}$
in (\ref{529}). Moreover, we discuss in some detail the
 associated stationary Baker-Akhiezer function
$\psi(P,x,x_0)$, the common eigenfunction of $L_3$ and
$P_m$, and associated positive divisors of degree $m-1$
on $\ca{K}_{m-1}$.  The latter topic was originally
developed by Jacobi \cite{jac} in the case of hyperelliptic
curves and applied to the KdV case by Mumford \cite{mum},
Section
III.a.1 and McKean \cite{mk85}.

Before we enter any further details we should perhaps
stress one important point. In spite of the considerable
 complexity of the formulas displayed at
various places in Sections \ref{S:2.1}--\ref{S:2.2}, the
 basic underlying formalism is a recursive one as described
 in depth in \cite{dgu1}. Consequently, the majority of our
formalism can be generated  using symbolic calculation
programs (such as Mathematica or Maple).

We recall the Bsq curve $\ca{K}_{m-1}$ in (\ref{529})
\begin{align}
\ca{K}_{m-1} :\
 \ca{F}_{m-1}(z,y) &= y^3 + y \, S_m(z) - T_m(z)
= 0,\lb{61} \no \\
 S_m(z) &= \sum_{p=0}^{2\, n-1+\ep} s_{m,p} z^p,
\quad T_m(z) = z^{m}+ \sum_{q=0}^{m-1} t_{m,q} z^q,  \\
   & \hspace{10mm} \mm, \no
\end{align}
(where $\mm$ will be fixed
throughout this section) and denote its compactification
(adding the branch
point $\py$) by the same symbol $\ca{K}_{m-1}$.
(In the following
$\ca{K}_{m-1}$ will always denote the compactified curve.)
Thus
$\ca{K}_{m-1}$ becomes a (possibly singular) three-sheeted
Riemann surface
of arithmetic genus $m-1$ in a standard manner. We will
 need a bit more
notation in this context. Points $P$ on $\ca{K}_{m-1}$ are
 represented as
pairs $P=(z,y)$ satisfying (\ref{61}) together with
$\py$, the point at
infinity. The complex structure on $\ca{K}_{m-1}$ is
defined in the usual
way by introducing local coordinates $\zeta_{P_0}:
P \to (z-z_0)$
 near points $P_0 \in \ca{K}_{m-1}$ which are neither
branch nor singular
points of $\ca{K}_{m-1}$,\ $ \zeta_{\py}: P
\to  z^{-1/3} $ near the
branch point $\py\in\ca{K}_{m-1}$ (with an
appropriate
determination of the branch of $z^{1/3} $) and similarly
at branch and/or
singular points of $\ca{K}_{m-1}$. The holomorphic map
${*}$, changing
sheets, is defined by
\begin{eqnarray}
* : \left\{ \begin{array}{l}
\ca{K}_{m-1}\to\ca{K}_{m-1},\\
P=(z,y_{j}(z)) \to P^{*} =(z, y_{j+1
(\text{mod }3)})(z) ), \quad j=1,2,3,
\end{array} \right. \quad
 P^{**}:= (P^{*})^{*},
\text{ etc.}, \lb{62}
\end{eqnarray}
where $y_{j}(z), \ j=1,2,3$ denote the three branches of
$y(P)$ satisfying $\ca{F}_{m-1}(z,y)=0$. Finally, positive
divisors on $\ca{K}_{m-1}$ of degree $m-1$ are denoted by
\begin{eqnarray}
\ca{D}_{P_1,\ldots,P_{m-1}} :
\left\{ \begin{array}{l}
\ca{K}_{m-1 } \to \nz_0,\\
P  \to \ca{D}_{P_1,\ldots,P_{m-1}}(P) =
\left\{
\begin{array}{l}
k \text{ if  $P$   occurs }  k \\
{} \hspace{ 11mm} \text{ times in } \{P_1, \ldots,P_{m-1}\} ,
\\ 0 \text{ if }  P   \not \in \{P_1, \ldots,P_{m-1}\} .
\end{array} \right.
\end{array} \right. \lb{63}
\end{eqnarray}
Specific details on curves of Bsq-type
(i.e., trigonal curves with a triple point at $\py$)
as defined in \eqref{61}
can be found in Appendix \ref{app-b}.

Given these preliminaries, let $\psi(P,x,x_0)$ denote
the common normalized
eigenfunction of $L_3$ and $P_m$, whose existence is
guaranteed by the
commutativity of $L_3$ and $P_m$ (cf., e.g.,
\cite{bc1}, \cite{bc2}), that is, by
\begin{eqnarray}
{} [P_m, L_3 ]=0, \quad m=3n+\ep
\lb{64}
\end{eqnarray}
for a given $\ep\in\{1,2\},$ and $n \in \nz_0$, or
equivalently, by the
requirement
\begin{eqnarray}
f_{n+1,x}^{(\ep)}=0,  \qquad g_{n+1,x}^{(\ep)}=0. \lb{65}
\end{eqnarray}
Explicitly, this yields
\begin{eqnarray}
L_3 \psi(P,x,x_0) = z(P)\,\psi(P,x,x_0),\quad P_m
\psi(P,x,x_0)=
y(P) \, \psi(P,x,x_0), \lb{66} \\
 P=(z,y)  \in \ca{K}_{m-1} \backslash \{\py \},
\ \ x\in\cz. \no
\end{eqnarray}
Assuming the normalization,
\begin{eqnarray}
\psi(P,x_0,x_0)=1, \qquad  P \in \ca{K}_{m-1}
\backslash \{ \py \}
\lb{67}
\end{eqnarray}
for some fixed $x_0\in\cz,$ $\psi(P,x,x_0)$ is called the
stationary  Baker-Akhiezer function for the Bsq hierarchy.
Closely
related to $\psi(P,x,x_0)$ is the following
meromorphic function
$\phi(P,x)$ on $\ca{K}_{m-1}$ defined by
\begin{eqnarray}
\phi(P,x)=\fr{\psi_{x}(P,x,x_0)}{\psi(P,x,x_0)},
\quad P\in \ca{K}_{m-1} ,
 \ x\in\cz, \lb{68}
\end{eqnarray}
such that
\begin{eqnarray}
\psi(P,x,x_0)= \exp\bigg( \int_{x_0}^{x} d \, x'
\phi(P,x')\bigg),
\qquad  P \in \ca{K}_{m-1}
\backslash \{ \py \}  . \lb{69}
\end{eqnarray}
Since $\phi(P,x)$ is a fundamental object for the
stationary Bsq hierarchy,
we next intend to establish its
 connection with the recursion formalism of Section~\ref{S:2.1}.
In pursuit of this connection, it is necessary to define a
variety of further polynomials $A_m,$ $B_m,$ $C_m,$ $D_{m-1},$
$E_m,$ $J_m,$ and  $N_m$ with respect to $z\in\cz$,
\begin{align}
&A_m(z,x) =  - G_m(z,x)^2 -
   \fr{1}{3} \,q_1(x)\, F_m(z,x)^2 +
   \fr{1}{4} \, F_{m,x}(z,x)^2
   - \fr{1}{3} \, F_m(z,x) \, F_{m,xx}(z,x), \lb{610} \\
&B_m(z,x) =  \,
   (z-q_0(x)) \, \big( -2\, F_m(z,x)^2 \,
   G_m(z,x) + \fr{1}{2}\, F_m(z,x)^2 \, F_{m,x}(z,x)
\big) \no\\
& \hspace*{5mm} - G_m(z,x)^2 \, G_{m,x}(z,x) +
   \fr{1}{4} \, F_{m,x}(z,x) ^2 \, G_{m,x}(z,x)   \no\\
& \hspace*{5mm} -  \fr{1}{6} \, q_{1,x}(x) \, F_m(z,x)^2 \,
G_m(z,x)
   - \fr{1}{2} \, q_{1,x}(x) \, F_m(z,x)^2 \, F_{m,x}(z,x)
   \no\\
& \hspace*{5mm} + \fr{1}{6}\, G_m(z,x)^2 \, F_{m,xx}(z,x) -
   \fr{11}{18} \, q_{1}(x)\, F_m(z,x)^2 \, F_{m,xx}(z,x)
   \no \\
& \hspace*{5mm} - \fr{1}{24}\, F_{m,x}(z,x)^2 \, F_{m,x x}(z,x)
        + \fr{1}{36} \, F_m(z,x)\, F_{m,xx}(z,x)^2 \no \\
& \hspace*{5mm} +\fr{2}{3} \, q_1(x) \,  F_m(z,x) \,
G_m(z,x)^2 -
   \fr{2}{9} \, q_1(x)^2 \, F_m(z,x)^3 \no \\
& \hspace*{5mm} -\fr{2}{3} \,q_{1}(x) F_m(z,x) \, G_m(z,x)\,
F_{m,x}(z,x) + \fr{1}{6} \,q_{1}(x) F_m(z,x) \, F_{m,x}(z,x)^2
   \no \\
& \hspace*{5mm} + F_m(z,x) \, G_m(z,x) \, G_{xx}(z,x) -
   \fr{1}{2} \, F_m(z,x) \, F_{m,x}(z,x) \, G_{m,x x}(z,x)
   \no \\
& \hspace*{5mm} - \fr{1}{6} \, q_{1,xx}(x) \, F_m(z,x)^3 -
   \fr{1}{6} \, F_m(z,x) \, G_m(z,x) \, F_{m,xxx}(z,x)
   \no \\
& \hspace*{5mm} + \fr{1}{12} \, F_m(z,x) \, F_{m,x}(z,x) \,
      F_{m,xxx}(z,x) - \fr{1}{6} \, F_m(z,x)^2
\,F_{m,xxxx}(z,x)
\no \\
& \hspace*{5mm} - F_m(z,x) \, G_{m,x}(z,x)^2 ,\lb{611} \\
&C_m(z,x) =  \, F_m(z,x)\,J_m(z,x) - (G_m(z,x) +
\fr{1}{2}F_{m,x}(z,x))H_m(z,x), \lb{612} \\
&D_{m-1}(z,x) =
  (F_m(z,x)\,B_m(z,x) -A_m ^2(z,x) - S_m(z)\, F_m^2(z,x))
\no \\
& \hspace*{5mm} \times \ep(m)\,(G_m(z,x)+\fr{1}{2}\,
F_{m,x}(z,x))^{-1}, \lb{612a} \\
&E_m(z,x)=-(A_m(z,x)\,C_m(z,x) -B_m(z,x)(G_m(z,x)+\fr{1}{2}\,
F_{m,x}(z,x))  \no \\
& \hspace*{5mm}
+S_m(z)\,F_m(z,x)\,(G_m(z,x)+\fr{1}{2}\,
F_{m,x}(z,x)))F_m(z,x)^{-1}, \lb{612b} \\
&J_m(z,x) =  \, H_{m,x}(z,x) +
\big(z-q_0(x)-\fr{1}{2} \,
  q_{1,x}(x)\big)\,F_m(z,x) ,  \lb{654} \\
&N_m(z,x) =
  (C_m^2(z,x) + E_m(z,x)\, (G_m(z,x)+\fr{1}{2}\, F_{m,x}(z,x))
\no \\
 & \hspace*{5mm} + S_m(z)(G_m(z,x)+\fr{1}{2}\, F_{m,x}(z,x))^2)
\ep(m)\, F_m(z,x)^{-1},
\lb{612c}
\end{align}
where
\begin{equation}
\ep(m) =
\begin{cases}
\phantom{-}1 & \text{ for } m=2 \, (\text{mod }3), \\
-1 & \text{ for } m=1 \, (\text{mod }3).
\end{cases} \lb{617}
\end{equation}
Explicit (though rather lengthy) formulas for $C_m,$, $D_{m-1},$
$E_m,$ and $N_m,$ directly in terms of $F_m$ and $G_m$ and their
$x$-derivatives, which prove their polynomial character with
respect to
$z$, can be found in
\cite{dgu1}. Moreover we recall the relations (cf.~\cite{dgu1}),
\begin{align}
&B_m\,C_m +A_m\, E_m + S_m\,\big( A_m\,(G_m+
\fr{1}{2}\, F_{m,x}) - F_m\,C_m\big)
 - T_m\,F_m\,(G_m+\fr{1}{2}\, F_{m,x})=0, \lb{619} \\
& B_m= \fr{2}{3}\, S_m\, F_m +  \fr{1}{3}\,
\ep(m)\,D_{m-1,x}, \lb{620} \\
& \ep(m)\, C_m  \, D_{m-1} = T_m\,F_m^2 - A_m B_m, \lb{622}\\
&  D_{m-1} N_m =
B_m\,E_m -T_m \big( A_m\,(G_m+\fr{1}{2}\, F_{m,x})
- F_m\, C_m \big) , \lb{623} \\
& \ep(m)\, A_m \, N_m =
  T_m\,(G_m+\fr{1}{2}\, F_{m,x})^2 -C_m E_m , \lb{624}\\
& \ N_{m,x}   \big(  G_m +\fr{1}{2}\,F_{m,x} \big)  =
 N_m \big( q_{1}\, F_m +  F_{m,xx} \big)
-\ep(m) \, J_m\,  \big( 2\,\big( G_m  +
\fr{1}{2}\,  F_{m,x} \big)\, S_m + 3\, E_m \big).
\lb{735a}
\end{align}
Next we recall explicit expressions for $\phi(P,x)$.
\begin{lem} [\cite{dgu1}] \lb{l62}
Let $ P=(z,y) \in \ca{K}_{m-1}$ and $(z,x)
\in\cz^2$. Then
\begin{align}
\phi(P,x)  = &\ \fr{(G_m(z,x)+2^{-1}F_{m,x}(z,x))
y(P) + C_m(z,x) }{F_m(z,x)y(P) - A_m(z,x) }
\lb{629} \\[3mm]
  =&\ \fr{ F_m(z,x)
y(P)^2+A_m(z,x)y(P)+B_m(z,x)}{\ep(m)D_{m-1}(z,x)}
\lb{630} \\[3mm]
  =&  \ \fr{-\ep(m)N_m(z,x)}
{ (G_m(z,x)+2^{-1}F_{m,x}(z,x))
y(P)^2-C_m(z,x)y(P)-E_m(z,x)} .
\lb{631}
\end{align}
\end{lem}
By inspection of (\ref{510}) and (\ref{511}) one infers
that $D_{m-1}$ and $N_m$
are monic polynomials with respect to $z$ of degree $m-1$
and $m$, respectively.
Hence we may write
\begin{equation}
D_{m-1}(z,x)= \prod_{j=1}^{m-1} \left( z- \mu_{j}(x)\right),
\quad
N_m(z,x)= \prod_{\ell=0}^{m-1} \left( z- \nu_{\ell}(x)\right).
\lb{633}
\end{equation}
 Defining
\begin{align}
\hat \mu_{j}(x)& =\big(\mu_{j}(x),
\fr{A_m(\mu_{j}(x),x)}{F_m(\mu_{j}(x),x) } \big) \in
\ca{K}_{ m-1 }, \quad j=1,\dots, m-1, \ \  x \in \cz,
\lb{634}\\
\hat \nu_{\ell}(x) &= \big(\nu_{\ell}(x),-
\fr{C_m( \nu_{\ell}(x),x)}{G_m( \nu_{\ell}(x),x) +
\fr{1}{2}\,F_{m,x}( \nu_{\ell}(x),x) }
\big) \in
\ca{K}_{ m-1 }, \no\\
&  \hspace{66mm} \ell=0,\dots, m-1, \ \  x \in \cz,
\lb{635}
\end{align}
one infers from (\ref{629})  that the divisor $\big(
\phi(P,x)\big)$ of
$  \phi(P,x) $ is given by (cf. (\ref{63}))
\begin{eqnarray}
  \big( \phi(P,x)\big) = \ca{D}_{\hat \nu_0(x),\ldots,
\hat \nu_{m-1}(x)}(P)-
 \ca{D}_{\py,\hat \mu_1(x),\ldots, \hat
\mu_{m-1}(x)}(P). \lb{636}
\end{eqnarray}
That is,  $\hat \nu_0(x),\ldots, \hat \nu_{m-1}(x)$ are
the $m$
zeros of $\phi(P,x)$ and $P_{\infty},\hat \mu_1(x),
\ldots, \hat \mu_{m-1}(x)$
its $m$ poles.

Further properties of $\phi(P,x)$ and $\psi(P,x,x_0)$ are
summarized in
\begin{theorem} [\cite{dgu1}] \lb{l63}
Assume \emph{(\ref{64})}--\emph{(\ref{68})}, $ P=(z,y)
\in \ca{K}_{ m-1 }
\backslash \{\py\},$ and let $(z,x,x_0) \in
\cz^3.$
Then
\begin{align}
(i)\ &  \phi(P,x)  \text{ satisfies the second-order
equation}
\no \\
&  \phi_{x x}(P,x)+3\, \phi_{x }(P,x)\phi (P,x)
+\phi(P,x)^3 + q_{1}(x)\,\phi(P,x)
  = z-q_0(x)-\fr{1}{2}\,
q_{1,x}(x).
\lb{637}
\\ (ii)\ & \phi(P,x)\, \phi(P^{*},x)\,\phi(P^{**},x)=
\fr{N_m(z,x)}{D_{m-1}(z,x)} .
\lb{638}\\
(iii)\ & \phi(P,x)+\phi(P^{*},x)+\phi(P^{**},x)=
\fr{D_{m-1,x}(z,x)}{D_{m-1}(z,x)}.
\lb{639}\\
(iv)\ & y(P)\, \phi(P,x)+y(P^{*})\,
\phi(P^{*},x)+ y(P^{**})\, \phi(P^{**},x) \no \\ &
\hspace{55mm} =
\fr{3\, T_m(z)\, F_m(z,x)- 2\, S_m(z) \, A_m(z,x)}
{ \ep(m) D_{m-1}(z,x)}.
 \lb{640} \\
(v)\ & \psi(P,x,x_0)\,\psi(P^{*},x,x_0)\,
\psi(P^{**},x,x_0)  =
\fr{D_{m-1}(z,x)}{D_{m-1}(z,x_0)}.
\lb{641}\\
(vi)\ &\psi_x(P,x,x_0)\, \psi_x(P^{*},x,x_0)\,
\psi_x(P^{**},x,x_0)
=  \fr{N_m(z,x)}{D_{m-1}(z,x_0)}.
 \lb{642}\\
(vii) \ &\psi(P,x,x_0)= \left(
\fr{D_{m-1}(z,x)}{D_{m-1}(z,x_0)}\right)^{1/3}\exp\bigg(
\int_{x_0}^{x}d\, x' \ep(m) D_{m-1}(z,x')^{-1}
\no\\ &
\hspace{20mm}  \times
\big( F_m(z,x')\,y(P)^2+
A_m(z,x')\,y(P)+ \fr{2}{3}\,
F_m(z,x')\, S_m(z) \big)
\bigg). \lb{643}
\end{align}
\end{theorem}
Thus, up to normalizations, $D_{m-1}$ represents the
product of the three
branches of $\psi$ and $N_m$ the product of the three
branches of $\psi_x$, their zeros represent the analogs
of Dirichlet and Neumann eigenvalues of $L_3$ with the
corresponding boundary conditions imposed at the point
$x\in\cz$ when compared to the KdV Lax expression $L_2$.

Returning to $D_{m-1}(z,x)$ and $N_m(z,x)$ for a
moment, we note
that
(\ref{52}), (\ref{510}), (\ref{511}), (\ref{612a}), and
(\ref{612c}) yield
\begin{align}
D_{0}&= 1 , \no\\
D_{1}&= z -q_0(x) -6^{-1}\, q_{1,x}(x) - d_0^{(2)}\,
q_{1}(x)  -(d_0^{(2)})^3  , \lb{650} \\
 & \text{etc.},  \no \\
\intertext{and}
N_{1}&= z-q_{0}(x),\no\\
N_{2}&= \big(z-q_0(x) +6^{-1}\,  q_{1,x}(x)\big)^2-
d_0^{(2)} \big( (z-q_0(x))q_1(x)-6^{-1}\,
q_1(x) q_{1,x}(x) \big)\no\\&
-6^{-1}\, (d_0^{(2)})^2    q_{1,xx}(x)
-(d_0^{(2)})^3 (z-q_0(x)),
 \lb{651} \\
 & \text{etc.} \no\end{align}
 Concerning the dynamics of the zeros $\mu_j(x)$ and
$\nu_{\ell}(x)$
of $D_{m-1}(z,x)$ and $N_m(z,x)$ one obtains
the following Dubrovin-type equations.
\begin{lem} [\cite{dgu1}] \lb{l64}
Suppose the curve $\ca{K}_{m-1}$ is nonsingular and
assume \emph{(\ref{65})} to hold.  \\
(i)\ Suppose the zeros
$\{\mu_j(x)\}_{j=1,\dots,m-1}$ of $D_{m-1}(\dott,x)$
remain distinct in $\Om_\mu$, where $\Om_\mu\subseteq\cz$ is
open and connected. Then
$\{\mu_j(x)\}_{j=1,\dots,m-1}$ satisfy the
system of differential equations
\begin{equation}
 \mu_{j,x}(x) =
\fr{-\ep(m) \, F_m(\mu_{j}(x),x)
\big( 3 y(\hat \mu_{j}(x))^2+
  S_m(\mu_{j}(x))\big)}{
 {\displaystyle
\prod_{\substack{k=1\\k \neq  j}}^{m-1}} \big(
\mu_{j}(x)-\mu_{k}(x)\big)}, \ \ j=1,\dots,m-1, \lb{652}
\end{equation}
with initial conditions
\begin{equation}
\{\hat{\mu}_j(x_0)\}_{j=1,\dots,m-1}\subset\ca{K}_{m-1},
\lb{652a}
\end{equation}
for some fixed $x_0\in\Om_\mu$.  The initial value
problem \emph{(\ref{652})},
\emph{(\ref{652a})} has
a unique solution
$\{\hat{\mu}_j(x)\}_{j=1,\dots,m-1}\subset\ca{K}_{m-1}$
satisfying
\begin{equation}
\hat{\mu}_j\in C^\infty (\Om_\mu,\ca{K}_{m-1}), \quad
j=1,\dots,m-1.
\end{equation}
(ii)\ Suppose the zeros
$\{\nu_\ell(x)\}_{\ell=0,\dots,m-1}$ of $N_m(\dott,x)$
remain distinct in $\Om_\nu$, where $\Om_\nu\subseteq\cz$
is open and connected. Then
$\{\nu_\ell(x)\}_{\ell=0,\dots,m-1}$ satisfy the
system of differential equations
\begin{equation}
 \nu_{\ell,x}(x) =
\fr{-\ep(m) \, J_m(\nu_{\ell}(x),x)
\big( 3 y(\hat \nu_{\ell}(x))^2+
 S_m(\nu_{\ell}(x))\big)}
{{\displaystyle
\prod_{\substack{k=0\\k \neq  \ell}}^{m-1}}\big(
\nu_{\ell}(x)-\nu_{k}(x)\big) }, \quad
\ell =0,\dots,m-1, \lb{653}
\end{equation}
with initial conditions
\begin{equation}
\{\hat{\nu}_\ell(x_0)\}_{\ell=0,\dots,m-1}\subset\ca{K}_{m-1},
\lb{653a}
\end{equation}
for some fixed $x_0\in\Om_\nu$.
The initial value problem \emph{(\ref{653})},
\emph{(\ref{653a})} has a unique solution
$\{\hat{\nu}_\ell(x)\}_{\ell=0,\dots,m-1}\subset\ca{K}_{m-1}$
satisfying
\begin{equation}
\hat{\nu}_\ell\in C^\infty (\Om_\nu,\ca{K}_{m-1}), \quad
\ell=0,\dots,m-1.
\end{equation}
\end{lem}

For trace formulas expressing certain combinations of $q_0, q_1$
and their $x$-derivatives in terms of $\mu_{j}(x)$ and
$\nu_{\ell}(x)$ we refer to \cite{dgu1}.

The following example illustrates our recursion formalism
for the simplest genus $g=1$ case. Further examples can be
found in \cite{dgu1}.
\begin{exa} \lb{e1} $m=2 \ \ (\text{genus } g=1)$:
\begin{align}
&q_0(x)=0, \quad q_1(x)=-3\wp(x), \lb{ell} \\
&L_3 = \frac{d^3}{dx^3}-3\,  \wp(x)\,\frac{d }
{dx}-\frac{3}{2}\,\wp'(x),
 \qquad P_2  =  \frac{d^2}{d\,x^2}-2 \,
\wp(x) ,\label{841}
\\ &
\ca{F}_{1}(z,y)= y^3-\frac{g_2}{4}\,
y -z^2-\frac{g_3}{4} =0,
\label{842} \\
& F_{2}(z,x)= 1,\qquad G_{2}(z,x)= 0, \label{843}\\
&D_{1}(z,x) = z + \frac{1}{2}\, \wp'(x),\qquad
 N_{2}(z,x) =  \big(z-\frac{1}{2}\,  \wp'(x)\big)^2 ,
\label{844}\\
\phi_j(z,x)&=
\frac{z-\frac{1}{2}\, \wp'(x)}{y_j - \wp(x)}\label{845a} \\&=
\frac{y_j^2 + y_j\, \wp(x) + \wp(x)^2
-\frac{g_2}{4}}{z+\frac{1}{2} \, \wp'(x)}
\label{845b} \\ &=
\frac{(z-\frac{1}{2}\, \wp'(x))^2}{(z-\frac{1}{2}\,
\wp'(x))y_j -
\wp(x)(z-\frac{1}{2}\, \wp'(x))}  ,
\qquad  1 \leq j \leq 3 ,  \label{845}
\end{align}
where $y_j, \ 1\leq j \leq 3$ denote the roots
of (\ref{842}) and $\wp(x)$ denotes the elliptic Weierstrass
function (cf.,~e.g., \cite{as}, Ch.~18).
\end{exa}
\section{Stationary Algebro-Geometric
 Solutions of the Boussinesq Hierarchy}
\lb{S:2.3}

In this section we continue our study of the stationary Bsq
hierarchy, but now direct our efforts towards obtaining
explicit Riemann theta function representations for the
fundamental quantities $\phi$ and $\psi$, introduced in
Section~\ref{S:2.2}, and especially, for each of the potentials
$q_0$ and $q_1$ associated with the differential expression
$L_3$. As a result of our preparatory material in
Sections~\ref{S:2.1} and \ref{S:2.2}, we are now able to
simultaneously treat the class of algebro-geometric
quasi-periodic solutions of the entire Bsq hierarchy, one of our
principal aims in this paper.

In the following we freely employ the notation established in
Appendices \ref{app-a} and \ref{app-b} and refer to this material
whenever appropriate.

\begin{lem} \lb{l:lem1}
Let $x\in\cz$.  Near $\py\in\ca{K}_{m-1}$,
in terms of the local coordinate $\zeta=z^{-1/3}$, one has
\begin{equation}
\phi (P,x) \ztz \fr{1}{\zeta} \sum_{j=0}^\infty
\beta_j(x) \zeta^j
\text{ as } P\to\py,\lb{e:phiinf}
\end{equation}
where
\begin{align}
&\beta_0=1,\quad \beta_1=0,\quad\beta_2=-\fr{1}{3}q_1, \quad
\beta_3=-\fr{1}{3}q_0+\fr{1}{6}q_{1,x}, \no \\
&\beta_j = -\fr{1}{3}\Big( \beta_{j-2,xx} + q_1\beta_{j-2} +
\sum_{k=2}^{j-1} (3\beta_{k,x}\beta_{j-k-1}
+ \beta_k\beta_{j-k})
+\sum_{\ell =1}^{j-1}\sum_{k=0}^\ell
 \beta_k\beta_{\ell -k}\beta_{j-\ell}
\Big), \quad j\geq 4.
\end{align}
\end{lem}
\begin{proof}
In terms of the local coordinate $\zeta=z^{-1/3}$,
(\ref{637})
reads
\begin{equation}
\phi_{xx} + 3\phi\phi_x  + \phi^3 + q_1\phi =
 \zeta^{-3} - q_0 - 2^{-1}\, q_{1,x}.
\lb{e:phinon}
\end{equation}
A power series ansatz in (\ref{e:phinon})
then yields the indicated Laurent series.
\end{proof}
Let $\tta(\ul{z})$ denote the Riemann theta function (cf.
(\ref{aa49})) associated with $\ca{K}_{m-1}$ and an appropriately
fixed homology basis.  Next, choosing a convenient base point
$P_0\in\ca{K}_{m-1}\backslash\{\py\}$, the vector of Riemann
constants $\ul{\Xi}_{P_0}$ is given by \eqref{aa55},
and the Abel maps $\ul{A}_{P_0}(\dott)$
and $\ul{\al}_{P_0}(\dott)$ are defined by
(\ref{aa46}) and (\ref{aa47}), respectively.
For brevity, define the function
$\ul{z} : \ca{K}_{m-1}\times
\si^{m-1}\ca{K}_{m-1} \to \cz^{m-1}$ by
\begin{equation}
\ul{z}(P,\ul{Q}) = \ul{\Xi}_{P_0} -
\ul{A}_{P_0}(P) + \abel{\ul{Q}},\quad P\in\ca{K}_{m-1},\,
\ul{Q}=(Q_1,\dots,Q_{m-1})\in\si^{m-1}\ca{K}_{m-1}.
\end{equation}
We note that by (\ref{aa70a}) and (\ref{aa71}),
$\ul{z}(\dott,\ul{Q})$ is independent of the choice of base
point $P_0$.

The normalized differential $\dtk$ of the
third kind (\textit{dtk}) is the unique differential
holomorphic on $\ca{K}_{m-1}\backslash\{\py,\nu_0(x)\}$ with
simple poles at $\py$ and $\hat{\nu}_0(x)$ with residues
$\pm 1,$
respectively, that is,
\begin{equation}
\dtk (P) \ztz \big(\zeta^{-1} + O(1)\big)d\zeta
\text{ as } P\to\py \lb{e:dtk}.
\end{equation}
Then
\begin{equation}
\int_{P_0}^P \dtk \ztz \ln(\zeta) + e^{(3)}(P_0) + O(\zeta)
\text{ as } P\to\py , \lb{e:int3}
\end{equation}
where $e^{(3)}(P_0)$ is an appropriate constant. Furthermore,
let $\dsk$ denote the normalized differential defined by
\begin{equation}
\dsk (P) = - \sum_{j=1}^{m-1} \la_j\eta_j(P) -
\fr{1}{3y(P)^2+S_m(z)}
\begin{cases}
z^{2n}dz, & m=3n+1, \\ y(P)z^n dz, & m=3n+2,
\end{cases} \lb{e:dsk}
\end{equation}
where the constants $\{\la_j\}_{j=1,\dots,m-1}$ are
determined by the normalization condition
\begin{equation}
\int_{a_j}\dsk = 0, \quad j=1,\dots,m-1,
\end{equation}
and the differentials $\{\eta_j(P)\}_{j=1,\dots,m-1}$
(defined in
(\ref{e:basis})) form a basis for the space of holomorphic
differentials.
The $b$-periods of the differential $\dsk$ are denoted by
\begin{equation}
\ul{U}_2^{(2)} = (U^{(2)}_{2,1},\dots,U^{(2)}_{2,m-1}), \quad
U^{(2)}_{2,j} = \fr{1}{2\pi i}\int_{b_j} \dsk ,
\quad j=1,\dots,m-1.
\lb{e:1.11}
\end{equation}

A straightforward Laurent expansion of \eqref{e:dsk} near
$P_\infty$ yields the following result.
\begin{lem} \lb{l:dsk}
Assume the curve $\ca{K}_{m-1}$ is nonsingular.
Then the differential $\dsk$ defined in \eqref{e:dsk} is a
differential of
the second kind \emph{(}\textit{dsk}\emph{)}, holomorphic on
$\ca{K}_{m-1}\backslash\{\py\}$ with a pole of order 2
at $\py$.
In particular, near $\py$ in the local coordinate $\zeta$,
the differential $\dsk$ has the Laurent series
\begin{equation}
\dsk (P) \ztz \big( \zeta^{-2} + u + w\zeta + O(\zeta^2)
\big)d\zeta
\text{ as } P\to\py,
\end{equation}
where
\begin{align}
u &=
\begin{cases}
\la_{m-1} - c_1^{(1)} & \text{ for } m=1\, (\text{mod }3), \\
\la_{m-n-1} - (d_0^{(2)})^2 & \text{ for }
m=2\, (\text{mod }3), \\
\end{cases} \lb{uv1} \\
\intertext{and}
w &=
\begin{cases}
\la_{m-n-1} - 2d_1^{(1)} & \text{ for } m=1\, (\text{mod }3),\\
(d_0^{(2)})^3 - c_1^{(2)} - d_0^{(2)}\la_{m-n-1}
+ \la_{m-1} & \text{ for }
m=2\, (\text{mod }3). \\
\end{cases} \lb{uw}
\end{align}
\end{lem}
{}From Lemma \ref{l:dsk} one infers
\begin{equation}
\int_{P_0}^P\dsk\ztz - \zeta^{-1}   + e_2^{(2)}(P_0)+
u\zeta + 2^{-1}w\zeta^2 + O(\zeta^3)
\text{ as } P\to\py, \lb{e:r}
\end{equation}
where $e_2^{(2)}(P_0)$ is an appropriate constant.

The theta function representation of $\phi(P,x)$ then reads as
follows.
\begin{theorem} \lb{t:phi}
Let $P=(z,y)\in\ca{K}_{m-1}
\backslash \{ \py \}$, $(z,x)\in\cz^2$.
Suppose that $\ca{D}_{\ul{\hat{\mu}}(x)}$, or equivalently,
$\ca{D}_{\ul{\hat{\nu}}(x)}$ is nonspecial.
Then
\begin{equation}
\phi(P,x) = \fr{\vez{\py}{\muv{x}}}{\vez{\py}{\nuv{x}}}
		 \fr{\vez{P}{\nuv{x}}}{\vez{P}{\muv{x}}}
	  	 \exp\bigg(e^{(3)}(P_0)-\int_{P_0}^P \dtk \bigg).
\lb{e:phi}
\end{equation}
\end{theorem}
\begin{proof}
Let $\Phi$ be defined by the right-hand side of (\ref{e:phi})
with the
aim to prove that $\phi=\Phi$.  From (\ref{e:int3}) it
follows that
\begin{equation}
\exp\bigg(e^{(3)}(P_0)-\int_{P_0}^P \dtk \bigg) \ztz
\zeta^{-1} + O(1).
\lb{e:q}
\end{equation}
Using (\ref{636}) we
immediately see that $\phi$
has simple poles at $\muv{x}$ and $\py$, and
simple zeros at $\hat{\nu}_0(x)$ and $\nuv{x}$.  By
\eqref{e:q} and a
special case of Riemann's vanishing theorem
(Theorem \ref{taa23}),
we see that $\Phi$ has
the same properties.  Using the Riemann-Roch theorem
(Theorem \ref{taa12}), we conclude that
the holomorphic function $\Phi/\phi=c,$ a constant with
respect
to $P$. Using (\ref{e:q}) and Lemma \ref{l:lem1}, one computes
\begin{equation}
\fr{\Phi}{\phi} \ztz \fr{(1+O(\zeta ))
( \zeta^{-1}+O(1))}{ \zeta^{-1}+O(\zeta)}
\ztz 1+O(\zeta) \text{ as } P\to\py,
\end{equation}
from which one concludes $c=1.$
\end{proof}
Similarly, the theta function representation of the
Baker-Akhiezer function $\psi(P,x,x_0)$ is summarized in the
following theorem.
\begin{theorem} \lb{t:psi}
Assume that the curve $\ca{K}_{m-1}$ is nonsingular.
Let $P=(z,y)\in\ca{K}_{m-1}
\backslash \{ \py \}$ and let $x,\,x_0\in\Om_\mu$, where
$\Om_\mu\subseteq\cz$ is open and connected.
Suppose that $\ca{D}_{\ul{\hat{\mu}}(x)}$, or equivalently,
$\ca{D}_{\ul{\hat{\nu}}(x)}$ is nonspecial, for $x\in\Om_\mu$.
Then
\begin{equation}
\psi(P,x,x_0) = \fr{\vez{P}{\muv{x}}\,\vez{\py}{\muv{x_0}}}
{\vez{\py}{\muv{x}}\,\vez{P}{\muv{x_0}}}
\exp\bigg( (x-x_0)\big(
e_2^{(2)}(P_0)-\int_{P_0}^P \dsk \big)\bigg).
\lb{e:psi}
\end{equation}
\end{theorem}
\begin{proof}
Assume temporarily that
\begin{equation}
\mu_j(x)\neq \mu_{j'}(x) \text{ for } j\neq  j' \text{ and }
x\in\widetilde{\Om}_\mu\subseteq\Om_\mu, \lb{coll}
\end{equation}
where $\widetilde{\Om}_\mu$ is open and connected.
For the Baker-Akhiezer function $\psi$ we will use the
same strategy
as was used in the previous proof. However, the situation is
slightly more involved in that
$\psi$ has an essential singularity at $\py$.  Let $\Psi$ denote
the right-hand side of (\ref{e:psi}).  In order to prove that
$\psi=\Psi$, one first observes that since
\begin{equation}
\psi(P,x,x_0)=\exp\bigg(\int_{x_0}^x dx'\phi(P,x')\bigg),
\end{equation}
the zeros and poles of $\psi$ can come only from simple
poles in the
integrand (with positive and negative residues respectively).
Using (\ref{634}) and (\ref{652}), one computes
\begin{equation}
\begin{split}
\phi &= \fr{ F_m y^2 + A_m y + \fr{2}{3}F_m S_m +
\fr{1}{3} \ep(m) D_{m,x} }
          { \ep(m) D_m } \no \\
     &= \fr{1}{3}\fr{F_m}{\ep(m) D_m}\big( 3 y^2 + S_m \big) +
        \fr{1}{3}\fr{3 A_m y + F_m S_m}{\ep(m) D_m} +
        \fr{1}{3}\fr{D_{m,x}}{D_m}
\no \\&=
      \fr{2}{3}\fr{F_m}{\ep(m) D_m}\big( 3 y^2 + S_m \big) -
      \fr{1}{3}\sum_{k=1}^{m-1} \fr{\mu_{k,x}}{z-\mu_k} + O(1)
      \no \\ &= - \fr{\mu_{j,x}}{z-\mu_j} + O(1),
      \text{ as } P\to{\hat \mu}_j(x).
\end{split}
\end{equation}
More concisely,
\begin{equation}
\phi(P,x') = \fr{\pa}{\pa x'}\ln(z-\mu_j(x')) +
 O(1) \text{ for $P$ near }
\hat{\mu}_j(x').
\end{equation}
Hence
\begin{equation}
\begin{split}
&\exp\Big(\int_{x_0}^x dx'\big(\fr{\pa}{\pa x'}
\ln(z-\mu_j(x'))+O(1)\big)\Big) \\[2mm]
&= \begin{cases}
(z-\mu_j(x))O(1) &
\text{ for $P$ near } \hat{\mu}_j(x) \neq  \hat{\mu}_j(x_0), \\
 O(1) & \text{ for $P$ near } \hat{\mu}_j(x) =
 \hat{\mu}_j(x_0), \\
(z-\mu_j(x_0))^{-1}O(1) &
\text{ for $P$ near } \hat{\mu}_j(x_0) \neq  \hat{\mu}_j(x),
\end{cases}
\end{split} \lb{e:sing}
\end{equation}
where $O(1)\neq  0$ in (\ref{e:sing}).  Consequently,
all zeros of $\psi$
and $\Psi$ on $\ca{K}_{m-1}\backslash\{\py\}$ are simple
and coincide.  It
remains to identify the essential singularity of $\psi$ and
$\Psi$ at $\py$.
{}From (\ref{e:phiinf}), we infer
\begin{equation}
\int_{x_0}^x dx'\phi(P,x') \ztz (x-x_0)(\zeta^{-1} + O(\zeta))
\text{ as } P \to P_\infty.
\end{equation}
Looking at \eqref{e:r} we see that this coincides with the
singularity
in the exponent of $\Psi$ near $\py$.
The uniqueness result in Lemma~\ref{lem34} for Baker-Akhiezer
functions then completes the proof that $\Psi=\psi$ as both
functions share  the same
singularities and zeros.
The extension of this result from $x\in\widetilde{\Om}_\mu$
to $x\in\Om_\mu$
then simply follows from the continuity of $\ul{\al}_{P_0}$ and
the hypothesis of $\ca{D}_{\ul{\hat{\mu}}(x)}$ being
nonspecial for
$x\in\Om_\mu$.
\end{proof}

Next it is necessary to introduce two further polynomials
$K_m$ and $L_m$ with respect to the variable $z\in\cz$,
\begin{align}
&K_m(z,x)= (\ep(m) N_m(z,x)
- J_m(z,x)C_m(z,x))(G_m(z,x)+2^{-1} F_{m,x}(z,x))^{-1},
\lb{e:N_m} \\
& L_m(z,x) = (\ep(m) D_{m-1}(z,x) -
(G_m(z,x)-2^{-1} F_{m,x}(z,x))A_m(z,x))F_m(z,x)^{-1}.
\lb{e:D_m}
\end{align}
In analogy to our polynomials $A_m$--$N_m$ introduced in
\eqref{610}--\eqref{612c}, it is possible to derive explicit
expressions of
$K_m$ and $L_m$ directly in terms of $F_m$ and $G_m$ and their
$x$-derivatives. These expressions then prove, in
particular, the
polynomial character of $K_m$ and $L_m$ with respect to $z,$ but
we here omit the rather lengthy formulas since they can be
generated with the help of symbolic calculation programs
such as Maple or Mathematica.
\begin{lem}  \lb{l:1.10}
Let $x\in\cz$.  Then
\begin{equation}
L_m(\mu_j(x),x) = -\big(G_m(\mu_j(x),x)-
2^{-1} F_{m,x}(\mu_j(x),x)\big)y(\hat{\mu}_j(x)),
 \lb{e:c2}
\end{equation}
for $j=1,\dots,m-1$ and
\begin{equation}
K_m(\nu_\ell(x),x) = J_m(\nu_\ell(x),x)y(\hat{\nu}_\ell(x)),
\lb{e:c1}
\end{equation}
for $\ell=0,\dots,m-1$.
\end{lem}
The well-known linearization property of the Abel map for
completely integrable systems of soliton-type, is next verified
 in the context of the Bsq hierarchy.
\begin{theorem} \lb{t:lin}
Assume that the curve $\ca{K}_{m-1}$ is
nonsingular and let $x,\,x_0\in\cz$. Then
\begin{gather}
\abel{ \ca{D}_{\muv{x}}} = \abel{ \ca{D}_{\muv{x_0}}} +
 \ul{U}_2^{(2)}(x-x_0), \lb{ab1} \\
\ul{A}_{P_0}(\hat{\nu}_0(x)) + \abel{\ca{D}_{\nuv{x}}} =
\ul{A}_{P_0}(\hat{\nu}_0(x_0)) +\abel{\ca{D}_{\nuv{x_0}}} +
 \ul{U}_2^{(2)}(x-x_0) \lb{ab2}.
\end{gather}
\end{theorem}
\begin{proof}
We prove only (\ref{ab1}) as (\ref{ab2}) follows
\textit{mutatis mutandis} (or from \eqref{ab1} and Abel's
theorem, Theorem~\ref{taa14}). Assume temporarily that
\begin{equation}
\mu_j(x)\neq \mu_{j'}(x) \text{ for } j\neq  j' \text{ and }
x\in\widetilde{\Om}_\mu\subseteq\cz, \lb{coll2}
\end{equation}
where $\widetilde{\Om}_\mu$ is open and connected.
Then
using (\ref{652}), (\ref{e:basis}), and (\ref{b26}),
one computes
\begin{align}
&\fr{d}{dx} \al_{P_0,\ell} (\ca{D}_{\muv{x}}) =
\sum_{j=1}^{m-1} \mu_{j,x}(x) \om_\ell({\hat \mu}_j(x))
\no \\ & \quad =
-\ep(m) \sum_{k=1}^{m-n-1} e_{\ell}(k) \sum_{j=1}^{m-1}
\mu_j(x)^{k-1}F_m(\mu_j(x),x) {{\displaystyle
\prod_{\substack{p=1\\p \neq  j}}^{m-1}}\big(
\mu_j(x)-\mu_p(x)\big)^{-1}  }\no \\ &
\quad  \phantom{=} -\ep(m)
\sum_{k=1}^n e_{\ell}(k+m-n-1) \sum_{j=1}^{m-1}
 \mu_j(x)^{k-1}A_m(\mu_j(x),x) {{\displaystyle
\prod_{\substack{p=1\\p \neq  j}}^{m-1}}\big(
\mu_j(x)-\mu_p(x)\big)^{-1} } . \lb{e:eqn}
\end{align}
Next we consider the two cases $m=3n+1$ and $m=3n+2$
separately and substitute the polynomials $F_m(\mu_j(x),x)$ and
$A_m(\mu_j(x),x)$ in the variable $\mu_j(x)$
into (\ref{e:eqn}).~Using a standard Lagrange
interpolation argument then yields
\begin{equation}
\fr{d}{dx}\al_{P_0,\ell}(\ca{D}_{\muv{x}}) = -
\begin{cases}
e_{\ell}(m-1), & m=3n+1, \\ e_{\ell}(m-n-1), & m=3n+2.
\end{cases} \lb{ben}
\end{equation}
The result now follows for $x\in\widetilde{\Om}_\mu,$ using
(\ref{e:1.11}), (\ref{ben}), (\ref{def}),
and (\ref{ben1}).~By continuity of $\ul{\al}_{P_0},$ this result
extends from $x\in\widetilde{\Om}_\mu$ to $x\in\cz.$
\end{proof}
We conclude this section with the theta function
representations for the stationary Bsq solutions $q_0, q_1$ (the
analog of the Its-Matveev formula in the KdV context).
\begin{theorem} \lb{t:q0q1}
Assume that the curve $\ca{K}_{m-1}$ is nonsingular
and let $x\in\Om_\mu$, where
$\Om_\mu\subseteq\cz$ is open and connected.
Suppose that $\ca{D}_{\ul{\hat{\mu}}(x)}$, or equivalently,
$\ca{D}_{\ul{\hat{\nu}}(x)}$ is nonspecial for $x\in\Om_\mu$.
Then
\begin{align}
q_0(x) &= 3\,\pa_{\ul{U}_3^{(2)}}\pa_x\ln(\vez{\py}{\muv{x}}) +
(3/2)w, \lb{q0}\\
q_1(x) &= 3\,\pa_x^2\ln(\vez{\py}{\muv{x}}) + 3u, \lb{q1}
\end{align}
with $u$ and $w$ defined in (\ref{uv1}) and (\ref{uw}), that is,
\begin{align}
u &=
\begin{cases}
\la_{m-1} - c_1^{(1)} & \text{ for } m=1\, (\text{mod }3), \\
\la_{m-n-1} - (d_0^{(2)})^2 & \text{ for }
m=2\, (\text{mod }3), \\
\end{cases} \lb{itu1} \\
\intertext{and}
w &=
\begin{cases}
\la_{m-n-1} - 2d_1^{(1)} & \text{ for } m=1\, (\text{mod }3),\\
(d_0^{(2)})^3 - c_1^{(2)} - d_0^{(2)}\la_{m-n-1}
+ \la_{m-1} & \text{ for }
m=2\,
(\text{mod }3). \\
\end{cases} \lb{itw1}
\end{align}
\end{theorem}
\begin{proof}
Using
Lemma~\ref{l:dsk} and Theorem~\ref{t:psi}, one can write
$\psi$ near $\py$ in the coordinate $\zeta$, as
\begin{align}
\psi(P,x,x_0) & \ztz \big(1+\al_1(x)\zeta+
\al_2(x)\zeta^2+O(\zeta^3)\big) \no \\ &\hspace{1mm}\times
\exp\big( (x-x_0)(\zeta^{-1} - u\zeta -
2^{-1}w\zeta^2 + O(\zeta^3)) \big)
\text{ as } P\to \py, \lb{e:ch}
\end{align}
where the terms $\al_1(x)$ and $\al_2(x)$ in
(\ref{e:ch}) come from the Taylor
expansion about $\py$ of the ratios of the theta
functions in (\ref{e:psi}), and the exponential
term stems from substituting
(\ref{e:r}) into (\ref{e:psi}).
Using (\ref{e:ch}) and its $x$-derivatives one can show that
\begin{equation}
\psi_{xxx} + 3(u-\al_{1,x})\psi_x +3(2^{-1}w - \al_{1,xx} +
\al_1\al_{1,x} - \al_{2,x})\psi - \zeta^{-3}\psi = O(\zeta)\psi.
\end{equation}
Since $O(\zeta)\psi$ is another Baker-Akhiezer function with the
same essential singularity at $\py$ and the same divisor on
$\ca{K}_{m-1}\backslash\{\py\}$, the uniqueness theorem for
Baker-Akhiezer functions (cf.~ Lemma~\ref{lem34})
then yields $O(\zeta)=0$. Hence
\begin{align}
q_0(x) &= 3\big(2^{-1}w - 2^{-1}\al_{1,xx}(x) +
 \al_1(x)\al_{1,x}(x) - \al_{2,x}(x)\big) \lb{e:q0}, \\
q_1(x) &= 3(u-\al_{1,x}(x)), \lb{e:q1}
\end{align}
where
\begin{gather}
\al_{1,x}(x) = -\pa_x^2\ln\vez{\py}{\muv{x}} \lb{e:al1}, \\
-2^{-1}\al_{1,xx}(x) + \al_1(x)\al_{1,x}(x) - \al_{2,x}(x) =
\pa_{\ul{U}_3^{(2)}}\pa_x\ln\vez{\py}{\muv{x}} \lb{elp}.
\end{gather}
Here
\begin{equation}
\pa_{\ul{U}_3^{(2)}}=
\sum_{j=1}^{m-1} U_{3,j}^{(2)}\fr{\pa}{\pa z_j} \lb{dir}
\end{equation}
denotes the directional derivative in the direction of the vector
of $b$-periods $\ul{U}_3^{(2)}$, defined by
\begin{equation}
\ul{U}_3^{(2)} = (U_{3,1}^{(2)},\dots,U_{3,m-1}^{(2)}), \quad
U_{3,j}^{(2)} = \fr{1}{2\pi i}\int_{b_j} \dsk[3],
\quad j=1,\dots,m-1,
\end{equation}
with $\dsk[3]$ the \textit{dsk} holomorphic on
$\ca{K}_{m-1}\backslash\{\py\}$ with a pole of order 3
at $\py$,
\begin{equation}
\dsk[3](P) \ztz \big(\zeta^{-3} + O(1)\big)d\zeta \text{ as }
P\to\py.
\end{equation}
Combining (\ref{e:q0})--(\ref{elp}) then proves \eqref{q0} and
\eqref{q1}.
\end{proof}

For interesting spectral characterizations of third-order (in
fact, odd-order) self-adjoint differential operators with
quasi-periodic coefficients we refer to \cite{GGK97}.

%%%%%%%%%%%%%%%%%%%%%%%%%%%%%%%%%%%%%%%%%%%%%%%%%%%%%%%%%%%%%
\section{The Time-Dependent Boussinesq Formalism} \lb{S:3.1}
%%%%%%%%%%%%%%%%%%%%%%%%%%%%%%%%%%%%%%%%%%%%%%%%%%%%%%%%%%%%%

\setcounter{equation}{0}
\setcounter{theorem}{0}

In this section we return to the recursive approach outlined in
Section~\ref{S:2.1} and briefly recall our treatment of the
time-dependent Bsq hierarchy in \cite{dgu1}.

We start with a stationary algebro-geometric solution
$(q_0^{(0)}(x),q_1^{(0)}(x))$ associated  with
$ \ca{K}_{m-1}$ satisfying

\begin{equation}
\Bsq_m(q_0^{(0)},q_1^{(0)})=
\begin{cases}
-3\,f_{n+1,x}^{(\ep)}=0, \\[3mm]
-3\,g_{n+1,x}^{(\ep)} =0 ,
\end{cases}
\quad x\in\cz, \ m=3n+\ep \lb{71}
\end{equation}
for some fixed $\ep\in\{1,2\}$,
$n \in\nz_0$, and  a given set of
integration constants $\{c_{\ell}^{(\ep)}\}_{\ell=1,\dots,n},$
 $\{ d_{\ell}^{(\ep)}\}_{\ell =0,\dots,n} $. Our aim is to
construct the $r$th
Bsq flow
\begin{eqnarray}
\Bsq_r(q_0 ,q_1 ) =  0,
 \quad (q_{0}(x,t_{0,r}),q_{1}(x,t_{0,r}) )=
(q_{0}^{(0)}(x),q_{1}^{(0)}(x) ) ,
  \quad x \in \cz, \, r=3s+\varepsilon^{\prime}   \lb{72}
\end{eqnarray}
for some fixed $\varepsilon^{\prime}\in\{1,2\},$ $s \in\nz_0,$
and $t_{0,r}\in \cz$.
In terms of Lax pairs this amounts to solving
\begin{eqnarray}
\fr{d}{d\,t_r}  L_3(t_r) - [\widetilde{P}_r(t_r),
L_3(t_r)] & = & 0,
\quad t_r \in \cz, \lb{73} \\
{ } [P_m(t_{0,r}),L_3(t_{0,r})] &=& 0 .
\lb{74}
\end{eqnarray}
As a consequence one obtains
\begin{gather}
[P_m(t_r),L_3(t_r)]=0, \quad \ t_r\in\cz, \lb{75} \\
P_m(t_r)^3 + P_m(t_r)\,S_m(L_3(t_r)) - T_m(L_3(t_r)) = 0,
\quad \ t_r \in \cz, \lb{76}
\end{gather}
since the Bsq flows are isospectral deformations of
$L_3(t_{0,r})$.

We emphasize that the integration constants
$\{\ti{c}_\ell^{(\ep')}\}$ and $\{\ti{d}_\ell^{(\ep')}\}$
 in $\widetilde{P}_r$, and
$\{c_\ell^{(\ep)}\}$ and
$\{d_\ell^{(\ep)}\}$ in $P_m$, are independent of each
other (even for $r=m$).  Hence we shall employ the
notation $\widetilde{P}_r$, $\widetilde{F}_r$,
$\widetilde{G}_r$, $\widetilde{H}_r$,
etc., in order to distinguish them from $P_m$, $F_m$,
$G_m$, $H_m$, etc.  In addition we follow a more
elaborate approach inspired by Hirota's
$\tau$-function approach and indicate the individual
$r$th Bsq flow by a separate time variable $t_r\in\cz$.
(The latter notation suggests considering all Bsq flows
simultaneously by introducing $\ul{t} = (t_1,t_2,t_4,t_5,
\dots)$.)

Instead of working directly with (\ref{73}) and
(\ref{75}) we find it preferable to take
the following two equations as our point of departure
(never mind
their somewhat intimidating size),
\begin{align}
q_{0,t_r}  = & \ - \fr{1}{6}\, \widetilde{F}_{r,xxxxx}
- \fr{5}{6} \, q_1 \, \widetilde{F}_{r,xxx} - \fr{5}{4} \,
q_{1,x} \, \widetilde{F}_{r,xx}
- ( \fr{3}{4} \, q_{1,xx} + \fr{2}{3} \, q_1^2 ) \,
\widetilde{F}_{r,x} \no \\ &
- ( \fr{1}{6} \, q_{1,xxx}+ \fr{2}{3} \, q_1 q_{1,x}) \,
\widetilde{F}_r - 3(z-q_0) \, \widetilde{G}_{r,x} + q_{0,x} \,
\widetilde{G}_r ,
\lb{77} \\
q_{1,t_r} = &\ 2\, \widetilde{G}_{r,xxx} + 2\, q_1 \,
\widetilde{G}_{r,x} + q_{1,x} \, \widetilde{G}_r
-3 \, (z-q_0) \, \widetilde{F}_{r,x} +
 2 \, q_{0,x} \, \widetilde{F}_r   ,
 \quad \ (x,t_r) \in \cz^2 , \no
\end{align}
\begin{align}
       -\fr{1}{6}\, F_{m,xxxx} F_m
+\fr{1}{6}\,
F_{m,xxx} F_{m,x} - \fr{1}{12} \,
F_{m,xx}^2 - \fr{5}{6} \,q_1
F_{m,xx} F_m
& \no\\
- \fr{5}{12} \, q_{1,x} F_{m,x} F_m
+ \fr{5}{12}\,
 q_1 F_{m,x}^2 - \fr{1}{3}\,\big(\fr{1}{2}
\, q_{1,xx} + q_1^2 \big)
 F_m^2 & \lb{78} \\
 + 2 \, G_{m,xx} G_m - G_{m,x}^2 +
q_1 G_m^2-
3 \, (z- q_0 ) F_m G_m =
& \ S_m(z)   ,
\no
\quad (x,t_r) \in \cz^2 ,
\end{align}
\begin{align}
& \hspace{-3mm} \fr{1}{18} \, F_{m,xxxx}
F_{m,xx} F_m -
\fr{1}{24} \, F_{m,xxxx} F_{m,x}^2
+ \fr{1}{18}\, q_1 F_{m,xxxx} F_m^2+
\fr{1}{36} \, F_{m,xxx} F_{m,xx} F_{m,x}
\no\\ &
-\fr{1}{36} F_m F_{m,xxx}^2
- \fr{1}{18}\, q_{1,x} F_{m,xxx} F_m^2
- \fr{1}{9}\, q_1 F_{m,xxx}
F_{m,x} F_m
-\fr{1}{108} \, F_{m,xx}^3
\no\\ &+
\fr{2}{9} \, q_{1,x} F_{m,xx}
F_{m,x} F_m +
\fr{1}{18}\, q_{1,xx} F_{m,xx}
F_m^2 -
\fr{7}{72} \, q_1 F_{m,xx} F_{m,x}^2+
\fr{5}{18} \, q_1^2 F_{m,xx}
F_m^2
\no\\ &+
\fr{7}{36} \, q_1 F_{m,xx }^2 F_m
- \fr{1}{24} \, q_{1,xx} F_{m,x}^2
F_m -
\fr{7}{48} \, q_{1,x} F_{m,x}^3-
\fr{1}{6}  \, q_1^2 F_{m,x}^2 F_{m} +
\fr{1}{12} \, q_{1,x} q_1 F_{m,x}
F_m^2
\no\\ &
+\big( \fr{2}{27} \, q_1^3 - \fr{1}{36} \,
q_{1,x}^2  +
\fr{1}{18} \, q_{1,xx} q_1 +(z-q_0 )^2 \big)
F_m^3+
(z-q_0) G_m^3 +  \fr{1}{6} \,
F_{m,xxxx} G_m^2
\no\\ &-
\fr{1}{3} \, F_{m,xxx} G_{m,x}
G_m +
F_m G_{m,xx}^2+
\fr{1}{3} \, F_{m,xx} \big( G_{m,x}^2 +
G_{m,xx} G_m \big) -
F_{m,x} G_{m,xx} G_{m,x}
\no\\ &-
q_1 (z-q_0) F_m^2 G_m+
\fr{2}{3}\, q_1^2 F_m G_m^2  +
\fr{5}{6} \, q_1 F_{m,xx} G_m^2  -
\fr{4}{3} \, q_1 F_{m,x} G_{m,x}
G_m +
\fr{1}{3} \,q_1 F_m G_{m,x}^2
\no\\ & +
\fr{7}{12}\, q_{1,x} F_{m,x} G_m^2 +
\fr{4}{3}\, q_1 F_m G_{m,xx}
G_m +
\fr{1}{6} \, q_{1,xx} F_m G_m^2-
\fr{1}{3}\, q_{1,x} F_m G_{m,x} G_m
\no\\ & +
(z-q_0) F_{m,x} F_m  G_{m,x}  -
\fr{1}{4} \,(z-q_0) F_{m,x }^2
G_{m} - 2\,(z-q_0) F_m^2
G _{m,xx} =T_m(z)   ,
\lb{79}\\
& \hspace{110mm}(x,t_r) \in \cz^2 , \no
\end{align}
where (cf.\ (\ref{510}), (\ref{511}))
\begin{align}
F_{{m}} (z,x,t_r) &= \sum_{\ell=0}^{n}
f_{n-\ell}^{(\varepsilon)}(x,t_r) z^{\ell}, \quad
F_{{m}} (z,x,t_{0,r}) = \sum_{\ell=0}^{n}
f_{n-\ell}^{(\varepsilon),(0)}(x ) z^{\ell},
 \lb{710} \\
G_{{m}} (z,x,t_r) &= \sum_{\ell=0}^{n}
g_{n-\ell}^{(\varepsilon)}(x,t_r) z^{\ell} , \quad
G_{{m}} (z,x,t_{0,r}) = \sum_{\ell=0}^{n}
g_{n-\ell}^{(\varepsilon),(0)}(x ) z^{\ell}
\lb{713}
\end{align}
for fixed $t_{0,r} \in \cz,\ m =3n+\varepsilon,$
$r=3s+\varepsilon^{\prime},$ $n,s\in\nz_0,$ $\varepsilon,
\varepsilon^{\prime}\in\{1,2\}.$ Here
$f_{ \ell}^{(\varepsilon)}(x,t_r),
g_{ \ell}^{(\varepsilon)}(x,t_r)$ and
$f_{ \ell}^{(\varepsilon),(0)}(x ),
g_{ \ell}^{(\varepsilon),(0)}(x )$ are defined
as in (\ref{52}) with
$(q_{0}(x ),q_{1}(x ))$ replaced by
$(q_{0}(x,t_r),$ $q_{1}(x,t_r))$, and
$(q_{0}^{(0)}(x ),q_{1}^{(0)}(x ))$, respectively.

In analogy to (\ref{633}) one introduces
\begin{equation}
D_{m-1}(z,x,t_r)=
  \prod_{j=1}^{m-1} \left( z- \mu_{j}(x,t_r)\right),
\quad
N_{m}(z,x,t_r)=
  \prod_{\ell=0}^{m-1} \left( z- \nu_{\ell}( x,t_r)\right),
\lb{715}
\end{equation}
where  $D_{m-1}$ and $N_m$ are defined as in
(\ref{612a}) and (\ref{612c}).
This implies in
particular (cf.\ (\ref{623})),
\begin{align}
  D_{m-1}(z,x,t_r) N_{m}(z,x,t_r)  = & \
  B_{m}(z,x,t_{r})\,E_{m}(z,x,t_{r}) -T_{m}(z)
\big( A_{m}(z,x,t_{r})\, \no\\ &
\hspace{-26mm} \times \big(G_{m}(z,x,t_{r})+
2^{-1} \, F_{m,x}(z,x,t_{r}) \big)
- F_{m}(z,x,t_{r})\, C_{m}(z,x,t_{r}) \big),
\lb{716}
\end{align}
and $A_{m}$, $B_{m}$, $C_{m}$, $D_{m-1}$, $E_{m}$, $J_m$,
and $N_m$ are
defined  as in
\eqref{610}--\eqref{612c}. Hence \eqref{619}--\eqref{735a}
also hold in the present context. Moreover, we recall

\begin{lem} [\cite{dgu1}] \lb{l72}
Assume \emph{(\ref{77})}--\emph{(\ref{713})} and let
$(z,x, t_r ) \in \cz^3$. Then
\begin{align}
(i) \ D_{m-1,t_r}(z,x,t_r) &= D_{m-1,x}(z,x,t_r)
\Big(  \widetilde G_{r}  (z,x,t_r)-\fr{1}{2}\,
\widetilde  F_{r,x} (z,x,t_r)  \no\\
&  \hspace{-20mm}
 - \fr{\widetilde F_{r} (z,x,t_r)}{F_{m}(z,x,t_r)}
\big(G_{m}  (z,x,t_r)-\fr{1}{2}\,  F_{m,x} (z,x,t_r)\big)
\Big)  + D_{m-1 }(z,x,t_r)\,
 \no\\
& \hspace{-20mm}\times  3\,  \Big(
  \widetilde H_{r }(z,x,t_r)
- \fr{\widetilde F_{r} (z,x,t_r)} {F_{m}(z,x,t_r)}\,
   H_{m }(z,x,t_r)  \Big) .
\lb{732}\\
(ii) \ N_{m,t_r}(z,x,t_r) & =
 N_{m,x}(z,x,t_r)  \Big(
 \widetilde G_{r}  (z,x,t_r)+\fr{1}{2}\,
\widetilde  F_{r,x} (z,x,t_r)
-\fr{\widetilde J_{r}(z,x,t_r)}{  J_{m}(z,x,t_r)}
 \no\\
& \hspace{-20mm}
\times \big(  G_{m}  (z,x,t_r)+\fr{1}{2}\,
F_{m,x} (z,x,t_r)\big) \Big)
  - N_{m}(z,x,t_r) \Big(   q_{1}(x,t_r) \,
\widetilde  F_{r} (z,x,t_r)
\no\\
& \hspace{-20mm}  +
\widetilde F_{r,x x}  (z,x,t_r)
-\fr{\widetilde J_{r}(z,x,t_r)}{  J_{m}(z,x,t_r)}
\big( q_{1}(x,t_r) \, F_{m} (z,x,t_r) +
F_{m,x x}  (z,x,t_r) \big) \Big)  .
  \lb{733}
\end{align}
 \end{lem}

Similarly, Lemma~\ref{l62} remains valid and one obtains
\begin{align}
\phi(P,x,t_{r} ) =&\ \fr{(G_{m}(z,x,t_r)+\fr{1}{2}\,
F_{m,x}(z,x,t_r))
y(P) + C_{m}(z,x,t_r) }
{ F_{m}(z,x,t_r) y(P)- A_{m}(z,x,t_r) }
\lb{717} \\[3mm]
  & \hspace{-10mm} = \ \fr{ F_{m}(z,x,t_r)
y(P)^2 + A_{m}(z,x,t_r)
y(P) + B_{m}(z,x,t_r) }
{ \ep(m)  D_{m-1}(z,x,t_r)   }
\lb{718} \\[3mm]
     & \hspace{-10mm} =  \fr{- \ep(m) N_m(z,x,t_r)}
{(G_m(z,x,t_r)+\fr{1}{2}\, F_{m,x}(z,x,t_r))
y(P)^2 - C_m(z,x,t_r)y(P) - E_m(z,x,t_r)},
\lb{719}
\\ & \hspace{60mm} P=(z,y) \in \ca{K}_{m-1}. \no
\end{align}
 In analogy to (\ref{634}) and (\ref{635}) one
then introduces (the analogs of) Dirichlet and Neumann
data by
\begin{align}
\hat \mu_{j}(x,t_{r}) &
=\big(\mu_{j}(x,t_{r}),
\fr{A_{m}(\mu_{j}(x,t_{r}), x,t_{r} )}
{F_{m}(\mu_{j}(x,t_{r}),x,t_{r}) } \big)
\in \ca{K}_{ m-1 }, \no\\
& \hspace{45mm}  j=1,\dots ,m-1,
\ \  (x,t_r) \in \cz^2, \lb{720}\\
\hat \nu_{\ell}(x,t_{r})  &= \big(\nu_{\ell}(x,t_{r}), -
\fr{C_{m}( \nu_{\ell}(x,t_{r}),x,t_{r})}
{G_{m}( \nu_{\ell}(x,t_{r}),x,t_{r}) +
\fr{1}{2}\,F_{m,x}( \nu_{\ell}(x,t_{r}),x,t_{r}) }
\big) \in
\ca{K}_{m-1 }, \no\\
&  \hspace{45mm}  \ell=0,\dots,m-1,
\ \  (x,t_r) \in \cz^2  \lb{721}
\end{align}
and  hence infers  that the divisor
$\big( \phi(P,x,t_{r})\big)$ of
$  \phi(P,x,t_{r}) $ is given by
\begin{eqnarray}
  \big( \phi(P,x,t_{r})\big) =
\ca{D}_{\hat \nu_0(x,t_{r}),\ldots, \hat
\nu_{m-1}(x,t_{r})}(P)-
 \ca{D}_{\py,\hat \mu_1(x,t_{r}),\ldots,
\hat \mu_{m-1}(x,t_{r})}(P).
\lb{722}
\end{eqnarray}
Next we define the time-dependent BA-function
$\psi(P,x,x_0,t_r,t_{0,r})$
\begin{align}
\psi(P,x,x_0,t_r,t_{0,r})&\ = \exp\bigg(
\int_{x_0}^{x} d \, x' \phi(P,x',t_r) +
\int_{t_{0,r}}^{t_r} d \, s
\big( \widetilde{F}_{r}(z,x_0,s) \no \\ &
\hspace{-20mm} \times \big(\phi_x(P,x_0,s)
+ \phi(P,x_0,s)^2\big)  +
(\widetilde{G}_r(z,x_0,s) - \fr{1}{2}\,
\widetilde{F}_{r,x}(z,x_0,s)) \phi (P,x_0,s)
 \no \\
& \hspace{-20mm} + \big( \fr{1}{6}\,
\widetilde{F}_{r,xx}(z,x_0,s)
 +\fr{2}{3}\,q_1(x,s) \widetilde{F}_{r}(z,x_0,s)-
  \widetilde{G}_{r,x}(z,x_0,s) \big) \big)
   \bigg),  \lb{723} \\
&\hspace{35mm} P \in \ca{K}_{m-1 }\backslash
\{ \py \},
\quad (x,t_r) \in \cz^2, \no\end{align}
with fixed $(x_0,t_{0,r}) \in \cz^2$. The following
theorem recalls the basic properties
of $\phi(P,x,t_r) $ and $ \psi(P,x,x_0,t_r,t_{0,r}) $.

\begin{theorem} [\cite{dgu1}] \lb{l71}
 Assume \emph{(\ref{77})}--\emph{(\ref{713})},
$P =(z,y) \in
\ca{K}_{m-1}\backslash \{ \py \}$ and let
$  (z,x,x_0,t_r,t_{0,r}) \\ \in \cz^5$. Then
\begin{align}
(i)\ &  \phi(P,x,t_r)  \text{ satisfies  }\no\\
&\phi_{x x}(P,x,t_r)+3\, \phi_{x }(P,x,t_r)\,\phi (P,x,t_r)
+\phi(P,x,t_r)^3+ q_{1}(x,t_r)\,\phi(P,x,t_r) \no\\
& \hspace{15mm} =z-q_0(x,t_r)- 2^{-1}\, q_{1,x}(x,t_r),
\lb{724} \\&
\phi_{t_r}(P,x,t_r)= \partial_{x}
\big( \widetilde F_{r}(z,x,t_r)
(\phi(P,x,t_r)^2+\phi_x(P,x,t_r)) \no\\
& \hspace{15mm}+
 (\widetilde G_{r}(z,x,t_r)-2^{-1}\,
\widetilde F_{r,x}(z,x,t_r))
\phi(P,x,t_r) +\widetilde  H_{r}(z,x,t_r) \big) .
\lb{725} \\
 (ii)\ &  \psi(P,x,x_0,t_r,t_{0,r})  \text{ satisfies }\no\\
&\psi_{x x x}(P,x,x_0,t_r,t_{0,r}) + q_{1}(x,t_r)
 \psi_{x }(P,x,x_0,t_r,t_{0,r}) \no\\ &
 \hspace{15mm} +
(q_0(x,t_r) +2^{-1}\, q_{1,x}(x,t_r)-z)
\psi(P,x,x_0,t_r,t_{0,r})=0,
\lb{726} \\
& \psi_{t_r}(P,x,x_0,t_r,t_{0,r})=
 \big(\widetilde F_{r}(z,x,t_r)(\phi(P,x,t_r)^2
+\phi_x(P,x,t_r)) \no\\ & +
 (\widetilde G_{r}(z,x,t_r) - 2^{-1}
\widetilde F_{r,x}(z,x,t_r))
\phi(P,x,t_r) +\widetilde  H_{r}(z,x,t_r) \big)
\psi (P,x,x_0,t_r,t_{0,r}) \lb{727} \\
& \hspace{-0mm}(\text{i.e.,} \  (L_3-z) \psi =0,
\  (P_{m}-y ) \psi =0,\
\psi_{t_r} = \widetilde P_{r} \psi ) .\no\\
(iii)\ & \phi(P,x,t_r)\, \phi(P^{*},x,t_r)\,
\phi(P^{**},x,t_r)=
 \fr{N_{m}(z,x,t_r)}{D_{m-1}(z,x,t_r)} .
\lb{728}\\
(iv)\ & \phi(P,x,t_r)+\phi(P^{*},x,t_r)+
\phi(P^{**},x,t_r)=
\fr{D_{m-1,x}(z,x,t_r)}{D_{m-1}(z,x,t_r)}.
\lb{729} \\
(v).\ & y(P)\, \phi(P,x,t_r)+
y(P^{*})\,\phi(P^{*},x,t_r)\no\\ &   +
y(P^{**})\, \phi(P^{**},x,t_r) =
\fr{3\, T_{m}(z)\, F_{m}(z,x,t_r)- 2\, S_{m}(z)
\, A_{m}(z,x,t_r)}
{ \ep(m) D_{m-1}(z,x,t_r)}.
  \lb{730} \\
(vi)\ & \psi(P,x,x_0,t_r, t_{0,r})
\psi(P^{*},x,x_0,t_r, t_{0,r} )
\psi(P^{**},x,x_0,t_r, t_{0,r} )
\no\\ &  \hspace{70mm} =
\fr{D_{m-1}(z,x,t_r)}{D_{m-1}(z,x_0,t_{0,r})} .
\lb{736} \\
(vii)\ &\psi_x(P,x,x_0,t_r, t_{0,r})
\psi_x(P^{*},x,x_0,t_r, t_{0,r})
\psi_x(P^{**},x,x_0,t_r, t_{0,r})
\no\\ & \hspace{70mm}  =
\fr{N_{m}(z,x,t_r)}{D_{m-1}(z,x_0,t_{0,r})} .
 \lb{737}\\
(viii)\ &\psi(P,x,x_0,t_r, t_{0,r})=
 \left(
\fr{D_{m-1}(z,x,t_r)}
{D_{m-1}(z,x_0,t_{0,r})}\right)^{1/3} \exp \bigg(
\int_{x_0}^{x}d\, x'
\ep(m) D_{m-1}(z,x',t_r)^{-1}
\no\\ &
  \times
\big[F_{m}(z,x',t_r)\,y(P)^2+
A_{m}(z,x',t_r)\,y(P)
+  \fr{2}{3}\, F_{m}(z,x',t_r)\, S_{m}(z) \big]
\no\\ &
- \int_{t_{0,r}}^{t_r} d\, s \
\bigg(\ep(m) D_{m-1}(z,x_0,s)^{-1}
\big[ F_{m}(z,x_0,s) y(P)^2 +
A_{m}(z,x_0,s)\,y(P) \no\\ &
+  \fr{2}{3}\, F_{m}(z,x_0,s)\, S_{m}(z)
\big]
\times \big[ \widetilde G_{r}  (z,x_0,s)
-\fr{1}{2}\,\widetilde  F_{r,x} (z,x_0,s)
\no\\ & - \big(G_{m}  (z,x_0,s)-\fr{1}{2}\,
F_{m,x} (z,x_0,s) \big)\,
\fr{\widetilde F_{r} (z,x_0,s)}{F_{m}(z,x_0,s)}\big]
+ y(P) \, \fr{\widetilde F_r (z,x_0,s)}
{F_{m}(z,x_0,s)} \bigg)
\bigg) .
\lb{738}
\end{align}
\end{theorem}
The dynamics of the zeros $\mu_j(x,t_r)$ and
$\nu_{\ell}(x,t_r)$
of $D_{m-1}(z,x,t_r)$ and $N_{m}(z,x,t_r)$, in
analogy to Lemma \ref{l64}, are
then described in  terms of Dubrovin-type equations
as follows.
\begin{lem} [\cite{dgu1}] \lb{l74}
Suppose \emph{(\ref{77})}--\emph{(\ref{713})}
 and assume that the curve $\ca{K}_{m-1}$ is nonsingular. \\
(i) Suppose the zeros $\{\mu_j(x,t_r)\}_{j=1,
\dots,m-1}$ of $D_{m-1}(\dott,x,t_r)$ remain distinct
for $(x,t_r)\in\Om_\mu$, where $\Om_\mu\subseteq\cz^2$
is open and
connected.  Then
$\{\mu_j(x,t_r)\}_{j=1,\dots,m-1}$ satisfy
the system of differential
equations,
\begin{align}
\mu_{j,x}(x,t_r) &=
-\ep(m)\,F_{m}(\mu_{j}( x,t_r ),x,t_r)\,
\fr{ \big(3 y(\hat \mu_{j}(x,t_r))^2+
 S_{m}(\mu_{j}(x,t_r))\big)}{
 {\displaystyle
\prod_{\substack{k=1\\k \neq  j}}^{m-1}} \big(
\mu_{j}(x,t_r)-\mu_{k}(x,t_r)\big)}, \no \\ &
\hspace{79mm}  j=1,\dots,m-1, \lb{739} \\
\mu_{j,t_r}(x,t_r) &= -\ep(m)\Big(
F_{m}(\mu_{j}(x,t_r),x,t_r)
\big(\widetilde G_{r}  (\mu_{j}(x,t_r),x,t_r)-2^{-1}
\widetilde  F_{r,x} (\mu_{j}(x,t_r),x,t_r)\big)
\no\\ &
+  \widetilde  F_{r}(\mu_{j}(x,t_r),x,t_r)
\big(G_{m}  (\mu_{j}(x,t_r),x,t_r)-2^{-1}
F_{m,x} (\mu_{j}(x,t_r),x,t_r)\big)
\Big)\no\\ & \times
 \fr{\big(3 y(\hat \mu_{j}(x,t_r))^2+
  S_{m}(\mu_{j}(x,t_r))\big)}{
 {\displaystyle
\prod_{\substack{k=1\\k \neq  j}}^{m-1}} \big(
\mu_{j}(x,t_r)-\mu_{k}(x,t_r)\big)},
\quad j=1,\dots,m-1, \lb{740}
\end{align}
with initial conditions
\begin{equation}
\{\hat{\mu}_j(x_0,t_{0,r})\}_{j=1,\dots,m-1}
\in\ca{K}_{m-1} \lb{740a},
\end{equation}
for some fixed $(x_0,t_{0,r})\in\Om_\mu$.  The initial
value problem
\emph{(\ref{740})}, \emph{(\ref{740a})} has a unique solution
satisfying
\begin{equation}
\hat{\mu}_j\in C^\infty (\Om_\mu,\ca{K}_{m-1} ),
\quad j=1,\dots,m-1.
\end{equation}
(ii) Suppose the zeros $\{\nu_\ell(x,t_r)\}_{\ell=0,
\dots,m-1}$ of $N_{m}(\dott,x,t_r)$ remain distinct
for $(x,t_r)\in\Om_\nu$, where $\Om_\nu\subseteq\cz^2$ is
open and connected. Then
$\{\nu_\ell(x,t_r)\}_{\ell=0,\dots,m-1}$ satisfy
the system of differential
equations,
\begin{align}
\nu_{\ell,x}(x,t_r) &=
-\ep (m) \, J_{m}(\nu_{\ell}(x),x,t_r) \,
\fr{\big(3y(\hat \nu_{\ell}(x,t_r))^2+
 S_{m}(\nu_{\ell}(x,t_r))\big)}
{{\displaystyle
\prod_{\substack{k=0\\k \neq  \ell}}^{m-1}}\big(
\nu_{\ell}(x,t_r)-\nu_{k}(x,t_r)\big) },
\no \\ & \hspace{79mm} \ell=0,\dots,m-1,
\lb{741}\\
 \nu_{\ell,t_r}(x,t_r) &= -\ep (m)\Big(
 J_{m}(\nu_{\ell}(x,t_r),x,t_r)
\big(\widetilde G_{r}
(\nu_{\ell}(x,t_r),x,t_r)+2^{-1}
\widetilde  F_{r,x} (\nu_{\ell}(x,t_r),x,t_r) \big)
 \no\\ & -
 \widetilde J_{r}(\nu_{\ell}(x,t_r),x,t_r)
\big(  G_{m}  (\nu_{\ell}(x,t_r),x,t_r)+2^{-1}\,
 F_{m,x} (\nu_{\ell}(x,t_r),x,t_r) \big) \Big)
\no\\ &  \times
\fr{ \big(3 y(\hat \nu_{\ell}(x,t_r))^2+
 S_{m}(\nu_{\ell}(x,t_r))\big)}
{{\displaystyle
\prod_{\substack{k=0\\k \neq  \ell}}^{m-1}}\big(
\nu_{\ell}(x,t_r)-\nu_{k}(x,t_r)\big) },
\quad \ell=0,\dots,m-1,
\lb{742}
\end{align}
with initial conditions
\begin{equation}
\{\hat{\nu}_\ell(x_0,t_{0,r})\}_{\ell=0,\dots,m-1}
\in\ca{K}_{m-1} \lb{742a},
\end{equation}
for some fixed $(x_0,t_{0,r})\in\Om_\nu$.  The initial
value problem
\emph{(\ref{742})}, \emph{(\ref{742a})} has a unique solution
satisfying
\begin{equation}
\hat{\nu}_\ell\in C^\infty (\Om_\nu,\ca{K}_{m-1}),
\quad \ell=0,\dots,m-1.
\end{equation}
(iii) The initial condition
\begin{eqnarray}
(q_0(x, t_{0,r}),q_1(x, t_{0,r}))
 = (q_0^{(0)}(x),q_1^{(0)}(x)), \quad x \in
\cz \lb{743}
\end{eqnarray}
effects
\begin{align}
\hat \mu_{j}(x,  t_{0,r}) & = \hat \mu_{j}^{(0)}(x),
\quad  j=1,\dots,m-1,
\quad x \in \cz, \lb{744}\\
\hat \nu_{\ell}(x,  t_{0,r})&  =
\hat \nu_{\ell}^{(0)}(x),
\quad \ell=0,\dots,m-1,
\quad x \in \cz \lb{745}
\end{align}
(cf.\  (\ref{710})--(\ref{715})).
\end{lem}

%%%%%%%%%%%%%%%%%%%%%%%%%%%%%%%%%%%%%%%%%%%%%%%%%%%%%%%%%%%%
\section{Time-Dependent Algebro-Geometric Solutions \\
of the Boussinesq Hierarchy} \lb{S:3.2}
%%%%%%%%%%%%%%%%%%%%%%%%%%%%%%%%%%%%%%%%%%%%%%%%%%%%%%%%%%%%

In our final and principal section we extend the results of
Section~\ref{S:2.3} from the stationary Bsq hierarchy, to the
time-dependent case. In particular, we obtain Riemann
theta function representations for the time-dependent
Baker-Akhiezer function and the  time-dependent
meromorphic function $\phi$. We finish this section with
the corresponding theta function representation for general
time-dependent algebro-geometric quasi-periodic Bsq solutions
$q_0,q_1.$

We start with the theta function representation of our
fundamental object $\phi(P,x,t_r).$
\begin{theorem}
Let $P=(z,y)\in\ca{K}_{m-1}\backslash\{ \py \}$,
$(z,x,t_r)\in\cz^3$. Suppose that
$\ca{D}_{\hat{\ul{\mu}}(x,t_r)}$, or equivalently,
$\ca{D}_{\hat{\ul{\nu}}(x,t_r)}$ is nonspecial.  Then
\begin{equation}
\phi(P,x,t_r) =
	   \fr{\vez{\py}{\muv{x,t_r}}}{\vez{\py}{\nuv{x,t_r}}}
	   \fr{\vez{P}{\nuv{x,t_r}}}{\vez{P}{\muv{x,t_r}}}
  	   \exp\left(e^{(3)}(P_0)-\int_{P_0}^P
	   \om_{\py ,\hat{\nu}_0(x,t_r)}^{(3)}\right) . \lb{e:phit}
\end{equation}
\end{theorem}
\begin{proof}
The proof carries over \textit{ad verbatim} from the
stationary case, Theorem \ref{t:phi}.
\end{proof}
Let $\dsk[r],\,\rr$, be the normalized \textit{dsk} holomorphic
 on $\ca{K}_{m-1}\backslash\{\py\}$, with a pole of order
$r$ at $\py$,
\begin{equation}
\dsk[r](P) \ztz (\zeta^{-r} + O(1))d\zeta \text{ as }
P\to\py,
\quad \rr.
\end{equation}
Furthermore, define the normalized \textit{dsk}
\begin{align}
\widetilde\Om_{\py,r+1}^{(2)} =
\sum_{\ell =0}^s
\ti{c}_{s-\ell}^{(\varepsilon^{\prime})}\,(3\ell
+2)\,\dsk[3\ell +3] &+
\sum_{\ell =0}^s
\ti{d}_{s-\ell}^{(\varepsilon^{\prime})}\,(3\ell
+1)\,\dsk[3\ell +2] ,
\no
\\ &\rr, \lb{clap}
\end{align}
where (cf.~\eqref{52})
\begin{equation}
(\ti{c}_0^{(\varepsilon^{\prime})},
\ti{d}_0^{(\varepsilon^{\prime})})
=\begin{cases}
(0,1) & \text{ for } \varepsilon^{\prime}=1, \\
(1,\ti{d}_0^{(2)}) & \text{ for } \varepsilon^{\prime}=2,
\lb{cd} \end{cases} \qquad \ti{d}_0^{(2)}\in\cz.
\end{equation}
In addition, we define the vector of $b$-periods of the
\textit{dsk}
$\widetilde\Om_{\py,r+1}^{(2)}$
\begin{align}
\ul{\widetilde{U}}_{r+1}^{(2)} =
(\widetilde{U}^{(2)}_{r+1,1},
\dots,\widetilde{U}^{(2)}_{r+1,m-1}), \quad
\widetilde{U}^{(2)}_{r+1,j} &= \fr{1}{2\pi i}\int_{b_j}
\widetilde\Om_{\py,r+1}^{(2)},
\quad j=1,\dots,m-1 \no \\
 & \hspace{11mm} \rr.
\end{align}
Motivated by the second integrand in (\ref{723}) one defines
the function $I_r(P,x,t_r)$, meromorphic
 on $\ca{K}_{m-1}\times\cz^2$ by
\begin{align}
I_r(P,x,t_r) &= \widetilde{F}_r(z,x,t_r)
(\phi_x(P,x,t_r)+\phi (P,x,t_r)^2) \no \\ &+
(\widetilde{G}_r(z,x,t_r)-2^{-1}
\widetilde{F}_{r,x}(z,x,t_r))\phi(P,x,t_r) +
\widetilde{H}_r(z,x,t_r), \lb{e:I_rr}
\end{align}
for $\rr$.
Denote by $\widehat {I}_r(P,x,t_r)$ the associated
homogeneous quantity replacing
$\widetilde{F}_r,
\widetilde{G}_r,
\widetilde{H}_r$ by the
corresponding homogeneous polynomials
$\widehat{\widetilde{F}}_r,
\widehat{\widetilde{G}}_r,
\widehat{\widetilde{H}}_r.$
\begin{theorem}
Let $\rr,\,(x,t_r)\in\cz^2$, and $\zeta= z^{-1/3}$ be
the local coordinate near $\py$.
Then
\begin{equation}
\widehat{I}_r(P,x,t_r) \ztz  \zeta^{-r} + O(\zeta)
\text{ as } P\to\py. \lb{e:I_r}
\end{equation}
\end{theorem}
\begin{proof}
One easily verifies \eqref{e:I_r} by direct
 computation for $r=1$ and $r=2$.
Assume \eqref{e:I_r} is true with $\rr$.
 Then one may rewrite (\ref{e:I_r})
as
\begin{equation}
\widehat{I}_r(P,x,t_r) \ztz  \zeta^{-r} +
\sum_{j=1}^\infty \delta_j(x,t_r)\,\zeta^j
 \text{ as } P\to\py, \lb{Sun}
\end{equation}
for some coefficients $\{\delta_j(x,t_r)\}_{j\in\nz}$.
Compare coefficients of $\zeta$ in (\ref{e:phiinf})
and (\ref{Sun}) by means of (\ref{725}) and (\ref{e:I_rr})
to obtain
\begin{align}
\delta_{1,x}(x,t_r) &= -\fr{1}{3}q_{1,t_r}(x,t_r), \\
\delta_{2,x}(x,t_r) &= \fr{1}{6}q_{1,t_rx}(x,t_r) -
\fr{1}{3}q_{0,t_r}(x,t_r), \\
\delta_{3,x}(x,t_r) &= \fr{1}{3}q_{0,t_rx}(x,t_r) -
\fr{1}{18}q_{1,t_r xx}(x,t_r).
\end{align}
{}From \eqref{531} one infers
\begin{align}
\delta_1(x,t_r) &= \ga_1(t_r) -
\hat{f}_{s+1}^{(\varepsilon^{\prime})}(x,t_r), \\
\delta_2(x,t_r) &= \ga_2(t_r) +
2^{-1}\hat{f}_{s+1,x}^{(\varepsilon^{\prime})}(x,t_r)
- \hat{g}_{s+1}^{(\varepsilon^{\prime})}(x,t_r), \\
\delta_3(x,t_r) &= \ga_3(t_r)
- 6^{-1}\hat{f}_{s+1,xx}^{(\varepsilon^{\prime})}(x,t_r) +
\hat{g}_{s+1,x}^{(\varepsilon^{\prime})}(x,t_r),
\end{align}
where $\ga_1(t_r),\,\ga_2(t_r),$ and $\ga_3(t_r)$ are
integration
 constants. Next we note that the coefficients of the power
series for $\phi(P,x,t_r)$ in the coordinate
$\zeta$ near $\py$ (cf. Lemma \ref{l:lem1}),
and the coefficients of the homogeneous polynomials
$\widehat{\widetilde{F}}_r
(\zeta,x,t_r)$ and
$\widehat{\widetilde{G}}_r(\zeta,x,t_r)$,
(and hence those of
$\widehat{\widetilde{H}}_r(\zeta,x,t_r)$)
are differential polynomials in $q_0$
and $q_1$, with no arbitrary integration
constants in their construction.
{}From the definition of $\widehat{I}_r$
in (\ref{e:I_rr}) it follows that
it also can have
no arbitrary integration constants, and must
 consist purely of differential
polynomials in $q_0$ and $q_1$.  From these
considerations it follows that
$\ga_1(t_r) = \ga_2(t_r) = \ga_3(t_r) = 0$.
Hence one concludes
\begin{align}
\widehat{I}_r (P,x,t_r) & \ztz \, \zeta^{-r} -
\hat{f}_{s+1}^{(\varepsilon^{\prime})}\zeta +
\big(2^{-1}\hat{f}_{s+1,x}^{(\varepsilon^{\prime})}(x,t_r) -
\hat{g}_{s+1}^{(\varepsilon^{\prime})}(x,t_r) \big)\,\zeta^2
\no \\ & \hspace{2mm} +
\big( \hat{g}_{s+1,x}^{(\varepsilon^{\prime})}(x,t_r)
- 6^{-1}\hat{f}_{s+1,xx}^{(\varepsilon^{\prime})}(x,t_r)
\big)\,\zeta^3 + O(\zeta^4) \text{ as } P\to\py, \lb{newpf}
\end{align}
where the functions  $f_s^{(\varepsilon^{\prime})}(x,t_r)$
and $g_s^{(\varepsilon^{\prime})}(x,t_r)$
are defined as in (\ref{52}) with
$(q_0(x),q_1(x))$ replaced by
$(q_0(x,t_r),q_1(x,t_r))$.
We note that one may write
\begin{equation}
\widehat{\widetilde{F}}_{r+3}(\zeta,x,t_r) =
\zeta^{-3}\widehat{\widetilde{F}}_r(\zeta,x,t_r) +
\hat{f}_{s+1}^{(\varepsilon^{\prime})}(x,t_r),
\end{equation}
with analogous expressions for $\widehat{\widetilde{G}}_r$ and
$\widehat{\widetilde{H}}_r$.  It follows that
\begin{align}
\widehat{I}_{r+3}(P,x,t_r) &=
\zeta^{-3} \widehat{I}_r(P,x,t_r)
+  \hat{f}_{s+1}^{(\varepsilon^{\prime})}(x,t_r)
\big(\phi_x(P,x,t_r) +\phi(P,x,t_r)^2\big) \no \\ &+
\big(\hat{g}_{s+1}^{(\varepsilon^{\prime})}(x,t_r)
- \fr{1}{2}\,
\hat{f}_{s+1,x}^{(\varepsilon^{\prime})}(x,t_r)\big)
\phi(P,x,t_r) \no \\ &  +
 \fr{1}{6}\,
\hat{f}_{s+1,xx }^{(\varepsilon^{\prime})}(x,t_r) +
\fr{2}{3}\,
q_1(x,t_r) \hat{f}_{s+1}^{(\varepsilon^{\prime})}(x,t_r) -
\hat{g}_{s+1,x}^{(\varepsilon^{\prime})}(x,t_r)  .
\lb{uphere}
\end{align}
Using Lemma \ref{l:lem1}
and (\ref{newpf}), \eqref{uphere} yields
\begin{equation}
\widehat{I}_{r+3}(P,x,t_r) \ztz \zeta^{-r-3} + O(\zeta)
\text{ as } P\to\py,
\end{equation}
and the result follows by induction.
\end{proof}
By \eqref{step} one infers
\begin{equation}
I_r = \sum_{\ell=0}^s \ti{c}_{s-\ell}^{(\varepsilon^{\prime})}
 \, \widehat{I}_{3\ell+2} +
      \sum_{\ell=0}^s \ti{d}_{s-\ell}^{(\varepsilon^{\prime})}
\, \widehat{I}_{3\ell+1},
      \quad \rr.
\end{equation}
Thus,
\begin{equation}
\int_{t_{0,r}}^{t_r} I_r(P,x,\tau)d\tau \ztz (t_r-t_{0,r})
\sum_{\ell=0}^s
\Big( \ti{c}_{s-\ell}^{(\varepsilon^{\prime})}
\fr{1}{\zeta^{3\ell+2}} +
\ti{d}_{s-\ell}^{(\varepsilon^{\prime})}
 \fr{1}{\zeta^{3\ell+1}} \Big) + O(\zeta)
\text{ as } P\to\py.
\lb{view}
\end{equation}
Furthermore, integrating \eqref{clap} yields
\begin{equation}
\begin{split}
\int_{P_0}^P \widetilde\Om_{\py,r+1}^{(2)} & \ztz
\sum_{\ell=0}^s \ti{c}_{s-\ell}^{(\varepsilon^{\prime})}
 \, (3\ell+2)
\int_{\zeta_0}^\zeta \fr{d\xi}{\xi^{3\ell+3}} +
\sum_{\ell=0}^s \ti{d}_{s-\ell}^{(\varepsilon^{\prime})}
 \, (3\ell+1)
\int_{\zeta_0}^\zeta \fr{d\xi}{\xi^{3\ell+2}} \\
& \hspace*{1mm}= -
\sum_{\ell=0}^s \ti{c}_{s-\ell}^{(\varepsilon^{\prime})}
 \, \fr{1}{\zeta^{3\ell+2}} -
\sum_{\ell=0}^s \ti{d}_{s-\ell}^{(\varepsilon^{\prime})}
 \, \fr{1}{\zeta^{3\ell+1}}
 + e_{r+1}^{(2)}(P_0) + O(\zeta)
\text{ as } P\to\py,
\end{split}
\lb{sonic}
\end{equation}
where $e_{r+1}^{(2)}(P_0)$ is a constant that arises from
evaluating all the integrals at their lowers limits $P_0$, and
summing accordingly. Combining (\ref{view}) and (\ref{sonic})
yields
\begin{equation}
\int_{t_{0,r}}^{t_r} I_r(P,x,s)ds  \ztz
(t_r-t_{0,r}) \Big( e_{r+1}^{(2)}(P_0)-\int_{P_0}^P
\widetilde\Omega_{\py,r+1}^{(2)} \Big) + O(\zeta) \text{ as }
 P\to\py \lb{e:IrOm}.
\end{equation}
Given these preparations, the theta function representation of
$\psi(P,x,x_0,t_r,t_{0,r})$ reads as follows.
\begin{theorem} \lb{t:psit}
Assume that the curve $\ca{K}_{m-1}$ is nonsingular.
Furthermore, let $P=(z,y)\in\ca{K}_{m-1}
\backslash \{ \py \}$, and let $(x,t_r),\,(x_0,t_{0,r})
\in\Om_\mu$,
where $\Om_\mu\subseteq\cz^2$ is
open and connected.
Suppose also that $\ca{D}_{\ul{\hat{\mu}}(x,t_r)}$, or
equivalently,
$\ca{D}_{\ul{\hat{\nu}}(x,t_r)}$ is nonspecial for
$(x,t_r)\in\Om_\mu$.
Then
\begin{equation}
\begin{split}
\psi(P,&x,x_0,t_r,t_{0,r})
	 = \fr{\vez{P}{\muv{x,t_r}}}{\vez{\py}{\muv{x,t_r}}}
	   \fr{\vez{\py}{\muv{x_0,t_{0,r}}}}{\vez{P}
{\muv{x_0,t_{0,r}}}} \\
	   &\times \exp\Big( (x-x_0)\big(e_2^{(2)}(P_0)-\int_{P_0}^P
\dsk\big)
	 + (t_r - t_{r,0})\big(e_{r+1}^{(2)}(P_0)-\int_{P_0}^P
	 \widetilde\Omega_{\py,r+1}^{(2)} \big)\Big). \lb{e:psit}
\end{split}
\end{equation}
\end{theorem}
\begin{proof}
We present only a proof of the time variation here, and
refer the reader
to Theorem \ref{t:psi} for the argument concerning the
space variation.
Let $\psi(P,x,x_0,t_r,t_{0,r})$ be defined as in (\ref{723})
and denote the right-hand side of (\ref{e:psit}) by
$\Psi(P,x,x_0,t_r,t_{0,r})$.
Temporarily assume that
\begin{equation}
\mu_j(x,t_r)\neq \mu_j'(x,t_r) \text{ for } j\neq  j'
\text{ and } (x,t_r)\in\widetilde{\Om}_\mu\subseteq\Om_\mu,
\end{equation}
where $\widetilde{\Om}_\mu$ is open and connected.
In order to prove that $\psi=\Psi$ one
uses (\ref{718}), (\ref{732}), the time-dependent analog of
(\ref{620}), and
\begin{equation}
F_m(\phi_x+\phi^2) + (G_m -2^{-1}F_{m,x})\phi +H_m = y,
\end{equation}
 to compute
\begin{align}
I_r &= \widetilde{F}_r(\phi_x + \phi^2) + (\widetilde{G}_r -
\fr{1}{2}\widetilde{F}_{r,x} ) \phi + \widetilde{H}_r \no \\
&= \fr{1}{F_m} \Big( y \widetilde{F}_r + ( F_m \widetilde{H}_r
- \widetilde{F}_r H_m ) + \big(
F_m ( \widetilde{G}_r - \fr{1}{2}\widetilde{F}_{r,x} ) -
\widetilde{F}_r (G_m - \fr{1}{2}F_{m,x}) \big)
\phi \Big) \no \\
&=
\fr{1}{3} \fr{D_{m,t_r}}{D_m} +
\fr{1}{F_m} \Big( y \widetilde{F}_r
+ \big(F_m(\widetilde{G}_r-\fr{1}{2}\widetilde{F}_{r,x})-
 F_r(G_m-\fr{1}{2}F_{m,x} )\big) \no \\
&\hspace{24mm}  \times
\big(  F_m y^2 + A_m y + \fr{2}{3}F_m S_m  \big)
\ep(m) D_m^{-1}  \Big) \no \\
&=
\fr{2}{3}\fr{F_m ( \widetilde{G}_r -
\fr{1}{2}\widetilde{F}_{r,x} ) -  \widetilde{F}_r
 ( G_m - \fr{1}{2}F_{m,x})}
{\ep(m) D_m}\big( 3 y^2 + S_m \big) -
\fr{1}{3}\sum_{k=1}^{m-1} \fr{\mu_{j,t_r}}{z-\mu_k}
+ \fr{ y \widetilde{F}_r }{F_m} \no \\
&=- \fr{\mu_{j,t_r}}{z-\mu_j} +
\fr{ y \widetilde{F}_r}{F_m} + O(1)
= - \fr{\mu_{j,t_r}}{z-\mu_j} + O(1)
\end{align}
as $P\to{\hat \mu}_j (x,t_r).$ More concisely,
\begin{equation}
I_r(P,x_0,s) =\fr{\pa}{\pa s}(z-\mu_j(x_0,s))
\text{ for $P$ near } \hat{\mu}_j(x_0,t_r).
\end{equation}
Hence
\begin{equation}
\begin{split}
&\exp\Big(\int_{t_{0,r}}^{t_r} ds\big(\fr{\pa}{\pa s}
\ln(z-\mu_j(x_0,s))+O(1)\big)\Big)\\
&= \begin{cases}
(z-\mu_j(x_0,t_r))O(1) &
\text{ for $P$ near } \hat{\mu}_j(x_0,t_r)
 \neq  \hat{\mu}_j(x_0,t_{0,r}), \\
 O(1) & \text{ for $P$ near } \hat{\mu}_j(x_0,t_r)
 = \hat{\mu}_j(x_0,t_{0,r}), \\
(z-\mu_j(x_0,t_{0,r}))^{-1}O(1) &
\text{ for $P$ near } \hat{\mu}_j(x_0,t_{0,r})
 \neq  \hat{\mu}_j(x_0,t_r),
\end{cases}
\end{split} \lb{e:singt}
\end{equation}
where $O(1)\neq  0$ in \eqref{e:singt}.
Consequently, all zeros and poles
of $\psi$ and $\Psi$ on $\ca{K}_{m-1}\backslash\{\py\}$
are simple and
coincide.  It remains to identify the essential
singularity of $\psi$ and
$\Psi$ at $\py$.  By (\ref{e:IrOm}) we see that
the singularities in
the exponential terms of $\psi$ and $\Psi$ coincide.
 The uniqueness result in Lemma~\ref{lem34} for
Baker-Akhiezer functions completes the proof that
$\psi=\Psi$ on $\widetilde{\Om}_\mu$.  The extension of the
result from $(x,t_r)\in\widetilde{\Om}_\mu$ to
$(x,t_r)\in\Om_\mu$ follows
from the continuity of $\ul{\al}_{P_0}$ and the
hypothesis that $\ca{D}_{\hat{\ul{\mu}}(x,t_r)}$
is nonspecial for $(x,t_r)\in\Om_\mu$.
\end{proof}
The straightening out of the Bsq flows by the Abel map is
contained in our next result.
\begin{theorem} \lb{t:lint}
Assume that the curve $\ca{K}_{m-1}$ is nonsingular, and let
$(x,t_r),$ $(x_0,t_{0,r})\in\cz^2$.  Then
\begin{equation}
\abel{\ca{D}_{\muv{x,t_r}}} = \abel{\ca{D}_{\muv{x_0,t_{0,r}}}}
+ \ul{U}_2^{(2)}(x-x_0)
+ \ul{\widetilde{U}}_{r+1}^{(2)}(t_r-t_{0,r}),
\lb{e:abmu}
\end{equation}
and
\begin{multline}
\ul{A}_{P_0}(\hat{\nu}_0(x,t_r)) + \abel{\ca{D}_{\nuv{x,t_r}}}
 \\ =
\ul{A}_{P_0}(\hat{\nu}_0(x_0,t_{0,r})) +
\abel{\ca{D}_{\nuv{x_0,t_{0,r}}}} + \ul{U}_2^{(2)}(x-x_0)
+ \ul{\widetilde{U}}_{r+1}^{(2)}(t_r-t_{0,r}).
\lb{e:abnut}
\end{multline}
\end{theorem}
\begin{proof}
As in the context of Theorem~\ref{t:lin}, it suffices to prove
(\ref{e:abmu}).  Temporarily assume that
$\ca{D}_{\hat{\ul{\mu}}(x,t_r)}$ is nonspecial for
$(x,t_r)\in\Om_\mu\subseteq\cz^2$,
where $\Om_\mu$ is open and connected.
Introduce the meromorphic differential
\begin{equation}
\Om (x,x_0,t_r,t_{0,r})= \fr{\pa}{\pa z}
\ln (\psi(\dott,x,x_0,t_r,t_{0,r})) dz.
\end{equation}
{}From the representation (\ref{e:psit}) one infers
\begin{equation}
\Om (x,x_0,t_r,t_{0,r})=-(x-x_0)\dsk -
(t_r-t_{0,r})\widetilde\Om_{\py,r+1}^{(2)}
-\sum_{j=1}^{m-1} \om^{(3)}_{\hat{\mu}_j(x_0,t_{0,r}),
\hat{\mu}_j(x,t_r)}+\om,
\end{equation}
where $\om$ denotes a holomorphic differential on
$\ca{K}_{m-1}$,
that is, $\om=\sum_{j=1}^{m-1} e_j \om_j$
for some $e_j\in\cz,\,j=1,\dots,m-1$. Since
$\psi (\dott,x,x_0,t_r,t_{0,r})$
is single-valued on $\ca{K}_{m-1}$, all $a$ and $b$-periods of
$\Om$ are integer multiples of $2\pi i$ and hence
\begin{equation}
2\pi i m_k= \int_{a_k} \Om (x,x_0,t_r,t_{0,r})
=\int_{a_k} \om =e_k,
\quad j=1,\dots,m-1
\end{equation}
for some $m_k\in\zz$.
Similarly, for some $n_k\in\zz$,
\begin{align}
2\pi in_k &=\int_{b_k} \Om (x,x_0,t_r,t_{0,r})
 =-(x-x_0)\int_{b_k} \dsk - (t_r-t_{0,r}) \int_{b_k}
\widetilde\Om_{\py,r+1}^{(2)} \no \\[1mm] &
\hspace{5mm} -\sum_{j=1}^{m-1} \int_{b_k}
\om^{(3)}_{\hat{\mu}_j(x_0,t_{0,r}),\hat{\mu}_j(x,t_r)}
+ 2\pi i \sum_{j=1}^{m-1} m_j \int_{b_k} \om_j \no \\[1mm] &
\hspace{5mm} =-(x-x_0)\int_{b_k} \dsk - (t_r-t_{0,r}) \int_{b_k}
\widetilde\Om_{\py,r+1}^{(2)}
 -2\pi i\sum_{j=1}^{m-1}
\int_{\hat{\mu}_j(x,t_r)}^{\hat{\mu}_j(x_0,t_{0,r})}
\om_k \no \\[1mm] & \hspace{5mm}
+ 2\pi i \sum_{j=1}^{m-1} m_j \int_{b_k} \om_j
 =-2\pi i (x-x_0)\,U_{2,k}^{(2)} - 2\pi i(t_r-t_{0,r})\,
\widetilde{U}_{r+1,k}^{(2)}
\no \\[1mm] & \hspace{5mm}
+ 2\pi i\al_{P_0,k}(\ca{D}_{\muv{x,t_r}}) -
2\pi i \al_{P_0,k}(\ca{D}_{\muv{x_0,t_{0,r}}})
+ 2\pi i \sum_{j=1}^{m-1} m_j\tau_{j,k},
\end{align}
where we used \eqref{aa25e}.
By symmetry of $\tau$ (see Theorem \ref{taa4}) this is
equivalent to
\begin{equation}
\abel{\ca{D}_{\muv{x,t_r}}} =
\abel{\ca{D}_{\muv{x_0,t_{0,r}}}}
+\ul{U}^{(2)}_2 (x-x_0) +
\ul{\widetilde{U}}_{r+1}^{(2)}(t_r-t_{0,r}),
\end{equation}
for $(x,t_r)\in\Om_\mu$.  This result extends
from $(x,t_r)\in\Om_\mu$
to $(x,t_r)\in\cz^2$ using the continuity of
$\ul{\al}_{P_0}$ and
the fact that positive nonspecial divisors are dense in
 the space of positive
divisors (cf.\ \cite{fk}, p.~95).
\end{proof}
Our principal result, the theta function representation of the
class of time-dependent algebro-geometric quasi-periodic Bsq
solutions now quickly follows from the material prepared
thus far.
\begin{theorem} \lb{t:tq0q1}
Assume that the curve $\ca{K}_{m-1}$ is nonsingular
and let $(x,t_r)\in\Om_\mu$, where
$\Om_\mu\subseteq\cz^2$ is open and connected.
Suppose also that $\ca{D}_{\ul{\hat{\mu}}(x,t_r)}$,
or equivalently,
$\ca{D}_{\ul{\hat{\nu}}(x,t_r)}$ is nonspecial.
Then
\begin{align}
q_0(x,t_r) &= 3\,\pa_{\ul{U}_3^{(2)}}\pa_x
\ln(\vez{\py}{\muv{x,t_r}}) + (3/2)w, \\
q_1(x,t_r) &= 3\,\pa_x^2\ln(\vez{\py}{\muv{x,t_r}}) + 3u,
\end{align}
where $u$ and $w$ are defined by \emph{(\ref{itu1})}
and \emph{(\ref{itw1})}, respectively, and
$\pa_{\ul{U}_3^{(2)}}$ denotes the
directional derivative introduced in \eqref{dir}.
\end{theorem}
\begin{proof}
The proof carries over
\textit{ad verbatim} from the stationary case,
Theorem \ref{t:q0q1}.
\end{proof}
{\bf Acknowledgments.} K.~U.~would like to thank G. Teschl
for numerous helpful discussions. Moreover, he is
indebted to the Department of  Mathematics at the
University of
Missouri, Columbia for the extraordinary hospitality
extended
to him during a stay in the Spring of 1998.

%%%%%%%%%%%%%%%%%%%%%%%%%%%%%%%%%%%%%%%%%%%%%%%%%%%%%%%%%%
\appendix
%%%%%%%%%%%%%%%%%%%%%%%%%%%%%%%%%%%%%%%%%%%%%%%%%%%%%%%%%%%%%
\section{Algebraic Curves and their Theta Functions in
a Nutshell}\lb{app-a}
%%%%%%%%%%%%%%%%%%%%%%%%%%%%%%%%%%%%%%%%%%%%%%%%%%%%%%%%%%%%

This appendix treats some of the basic aspects of
complex algebraic curves
and their theta functions as used at numerous places in
this paper.  The material below is standard (see,
 e.g., \cite{bries}, \cite{fk}, \cite{Gunning:1972}, \cite{kraz},
and \cite{mir}), and we
include it for two major reasons:  On the one hand
it allows us to introduce a large part of the notation used
in Sections~\ref{S:2.3} and \ref{S:3.2} (which otherwise would
take up considerable space and disrupt the flow of arguments in
these sections) and on the other hand, it permits a fairly
self-contained presentation of the Bsq hierarchy and its
algebro-geometric solutions in this paper.
\begin{defn}\lb{da1}
An affine plane (complex) algebraic curve
$\ca{K}$ is the locus of zeros in $\cz^2$ of a
(nonconstant) polynomial $\ca{F}(z,y)$ in
two variables.  The polynomial $\ca{F}$ is
called nonsingular at a root $(z_0,y_0)$ if
\begin{equation}
\nabla\ca{F}(z_0,y_0)=(\ca{F}_z(z_0,y_0),
\ca{F}_y(z_0,y_0))\neq  0. \lb{aaa1}
\end{equation}
The affine plane curve $\ca{K}$ of roots of
$\ca{F}$ is called nonsingular at $P_0=(z_0,y_0)$ if
$\ca{F}$ is nonsingular at $P_0$. The curve
 $\ca{K}$  is called nonsingular, or smooth, if it is
nonsingular at each of its points.
\end{defn}
The Implicit Function Theorem allows one
to conclude that a smooth affine curve $\ca{K}$ is
locally a graph and to introduce complex charts
on $\ca{K}$ as follows.  If $\ca{F}(P_0)=0$ with
$\ca{F}_y(P_0)\neq  0$, there is a holomorphic
function $g_{P_0}(z)$ such that in a neighborhood
$U_{P_0}$ of $P_0$, the curve $\ca{K}$ is
characterized by the graph $y=g_{P_0}(z)$.  Hence the
projection
\begin{equation}
\ti{\pi}_z \colon U_{P_0}\to\ti{\pi}_z(U_{P_0})
\subset\cz, \quad
(z,y)\mapsto z, \lb{aa1}
\end{equation}
yields a complex chart on $\ca{K}$.
If, on the other hand, $\ca{F}(P_0)=0$
with $\ca{F}_z(P_0)\neq  0$,
then the projection
\begin{equation}
\ti{\pi}_y \colon U_{P_0}\to\ti{\pi}_y(U_{P_0})
\subset\cz, \quad
(z,y)\mapsto y, \lb{aa2}
\end{equation}
defines a chart on $\ca{K}$.  In this way, as
 long as $\ca{K}$ is nonsingular, one arrives at a
complex atlas on $\ca{K}$.  The space $\ca{K}
\subset\cz^2$ is second countable and Hausdorff.  In
order to obtain a Riemann surface one needs
 connectedness of $\ca{K}$ which is implied by adding the
assumption of irreducibility of the polynomial
$\ca{F}$.  Thus,  $\ca{K}$ equipped with charts
\eqref{aa1} and \eqref{aa2} is a Riemann surface
if $\ca{F}$ is nonsingular and irreducible.  Affine
plane curves $\ca{K}$ are unbounded as subsets
of $\cz^2$, and hence noncompact.  The
compactification of
$\ca{K}$ is conveniently described in terms of
the projective plane $\cz\pz^2$, the
set of all one-dimensional (complex) subspaces
of $\cz^3$.

In order to simplify notations, we temporarily abbreviate
$x_0=x$, $x_1=y$, and $x_2=z$. Moreover, we
denote the linear span of $(x_2,x_1,x_0)\in\cz^3
\backslash\{0\}$ by $[x_2:x_1:x_0]$.  In
particular,  $[x_2:x_1:x_0]\in\cz\pz^2$ with
$L_\infty=\{[x_2:x_1:x_0]\in\cz\pz^2 \mid x_0=0\}$
 representing the line at infinity.
Since the homogeneous coordinates $[x_2:x_1:x_0]$
satisfy
\begin{equation}
[x_2:x_1:x_0]=[cx_2:cx_1:cx_0], \quad
 c\in\cz\backslash\{0\}, \lb{aa4}
\end{equation}
the space $\cz\pz^2$ can be viewed as the
 quotient space of $\cz^3\backslash\{0\}$ by the
multiplicative action of $\cz\backslash\{0\}$, that is,
$\cz\pz^2=(\cz^3\backslash\{0\})/(\cz\backslash\{0\})$,
 and hence
$\cz\pz^2$ inherits a Hausdorff topology which is the
 quotient topology  induced by the natural
map
\begin{equation}
\iota\colon \cz^3\backslash\{0\}\to\cz\pz^2,\quad
(x_2,x_1,x_0)\mapsto[x_2:x_1:x_0].\lb{aa4a}
\end{equation}

Next, define the open sets
\begin{equation}
U^m=\{[x_2:x_1:x_0]\in\cz\pz^2 \mid x_m\neq  0\},
\quad m=0, 1, 2. \lb{aa5}
\end{equation}
Then
\begin{equation}
f^0 \colon U^0\to \cz^2, \quad [x_2:x_1:x_0]\mapsto
\bigg(\fr{x_2}{x_0}, \fr{x_1}{x_0}\bigg) \lb{aa6}
\end{equation}
with inverse
\begin{equation}
(f^0)^{-1} \colon \cz^2\to U^0, \quad (x_2,x_1)
\mapsto [x_2:x_1:1], \lb{aa7}
\end{equation}
and analogously for functions $f^1$ and $f^2$
(relative to sets $U^1$ and $U^2$, respectively), are
homeomorphisms.  In particular, $U^0$, $U^1$,
and $U^2$ together cover $\cz\pz^2$.
Moreover,
$\cz\pz^2$ is compact since it is covered by
the closed unit (poly)disks in $U^0$, $U^1$, and
$U^2$.

Let $\ca{P}$ be a (nonconstant) homogeneous
 polynomial of degree $d$ in $(x_2,x_1,x_0)$, that is,
\begin{equation}
\ca{P}(cx_2,cx_1,cx_0)=c^d\ca{P}(x_2,x_1,x_0), \lb{aa8}
\end{equation}
and introduce
\begin{equation}
\barr{\ca{K}}=\{[x_2:x_1:x_0]\in\cz\pz^2
\mid \ca{P}(x_2,x_1,x_0)=0\}. \lb{aa9}
\end{equation}
The set $\barr{\ca{K}}$ is well-defined (even
though $\ca{P}(u,v,w)$  is not for
$[u:v:w]\in\cz\pz^2$) and closed in $\cz\pz^2$.
 The intersections,
\begin{equation}
\ca{K}^m=\barr{\ca{K}}\cap U^m, \quad m=0, 1, 2 \lb{aa10}
\end{equation}
are affine plane curves when transported to $\cz^2$, that is,
\begin{equation}
\ca{K}^0\cong \{(x_2,x_1)\in\cz^2 \mid
\ca{P}(x_2,x_1,1)=0\} \lb{aa11}
\end{equation}
represents the affine curve $\ca{F}(z,y)=0$,
where $\ca{F}(x_2,x_1)=\ca{P}(x_2,x_1,1)$, and
analogously for $\ca{K}^1$ and $\ca{K}^2$.
 We recall that $\ca{F}(x_2,x_1)$ is irreducible if and
only if $\ca{P}(x_2,x_1,x_0)$ is irreducible.

Given the affine curve defined by $\ca{F}(x_2,x_1)=0$,
 the associated homogeneous polynomial
$\ca{P}(x_2,x_1,x_0)$ can be obtained from
\begin{equation}
\ca{P}(x_2,x_1,x_0)=x_0^d \ca{F}\bigg(\fr{x_2}{x_0},
\fr{x_1}{x_0}\bigg), \lb{aa11a}
\end{equation}
where $d$ denotes the degree of $\ca{F}$ (and $\ca{P}$).

The element $[x_2:x_1:0]\in\cz\pz^2$ represents
 the point at infinity along the direction
$x_2:x_1$ in $\cz^2$ (identifying $[x_2:x_1:0]
\in\cz\pz^2$ and $[x_2:x_1]\in\cz\pz^1$).  The set
of all such elements then represents the line
at infinity,
$L_\infty$, and yields the compactification
 $\cz\pz^2$ of $\cz^2$. In other words,
$\cz\pz^2\cong\cz^2\cup L_\infty$, $\cz\pz^1
\cong\cz\cup \{\infty\}$, and
$L_\infty\cong\cz\pz^1$. The projective plane
curve
$\barr{\ca{K}}$ then intersects $L_\infty$ in
 a finite number of points (the points at infinity).
\begin{defn} \lb{da2}
A projective plane (complex) algebraic curve
 $\barr{\ca{K}}$ is the locus of zeros in $\cz\pz^2$ of
a homogeneous polynomial $\ca{P}$ in three variables.

A homogeneous (nonconstant) polynomial
$\ca{P}(x_2,x_1,x_0)$ is called nonsingular
if there are no common solutions
$(x_{2,0},x_{1,0},x_{0,0})\in\cz^3\backslash\{0\}$ of
\begin{gather}
\ca{P}(x_{2,0},x_{1,0},x_{0,0})=0,  \\
\nabla\ca{P}(x_{2,0},x_{1,0},x_{0,0})=(\ca{P}_{x_2},
\ca{P}_{x_1},
\ca{P}_{x_0})(x_{2,0},x_{1,0},x_{0,0})=0. \lb{aa12}
\end{gather}
The set $\barr{\ca{K}}$ is called a smooth projective
plane curve
(of degree $d\in\nz$\/) if $\ca{P}$ is nonsingular
(and of degree
 $d\in\nz$\/).
\end{defn}
One verifies that the homogeneous polynomial
 $\ca{P}(x_2,x_1,x_0)$ is nonsingular if and only if
each $\ca{K}^m$ is a smooth affine plane curve in $\cz^2$.
 Moreover, any
nonsingular homogeneous polynomial $\ca{P}(x_2,x_1,x_0)$
is irreducible and consequently each
$\ca{K}^m$ is a Riemann surface for $m=0$, $1$, and $2$.
 The coordinate
charts on each $\ca{K}^m$ are simply the projections, that
 is, $x_2/x_0$ and $x_1/x_0$ for $\ca{K}^0$,
$x_2/x_1$ and $x_0/x_1$ for
$\ca{K}^1$, and finally, $x_1/x_2$ and  $x_0/x_2$ for
 $\ca{K}^2$.  These separate complex structures
on $\ca{K}^m$ are compatible on $\barr{\ca{K}}$ and
 hence induce a complex structure on $\barr{\ca{K}}$.

The zero locus in $\cz\pz^2$ of a nonsingular
homogeneous polynomial
$\ca{P}(x_2,x_1,x_0)$ defines a smooth projective
plane curve $\barr{\ca{K}}$ which is a compact
Riemann surface.  Topologically, this Riemann
surface is a sphere with $g$ handles where
\begin{equation}
g=  (d-1)(d-2)/2, \lb{aa13}
\end{equation}
with $d$ the degree of $\ca{P}(x_2,x_1,x_0)$.
 In particular, $\barr{\ca{K}}$ has topological genus
$g$ and we indicate this by writing $\barr{\ca{K}}_g$
in our main text, or simply $\ca{K}_g$
if no confusion can arise. In general, the projective
curve $\ca{K}_g$ can be singular even though
the associated affine curve
$\ca{K}^0_g$ is nonsingular.  In this case one
 has to account for the singularities at infinity and
properly amend the genus formula \eqref{aa13}
according to results of Clebsch, Noether, and
Pl\"ucker.

If $\ca{K}_g$ is a nonsingular projective curve,
 associated with the
homogeneous polynomial
$\ca{P}(z,y,x)$ of degree $d$, the set of finite
 branch points of $\ca{K}_g$ is given by
\begin{equation}
\{[z:y:1]\in\cz\pz^2 \mid \ca{P}(z,y,1)=\ca{P}_y(z,y,1)=0
  \}.\lb{aa14}
\end{equation}
Similarly, branch points at infinity are defined by
\begin{equation}
\{[z:y:0]\in\cz\pz^2 \mid
\ca{P}(z,y,0)=\ca{P}_y(z,y,0)=0  \}.\lb{aa15}
\end{equation}
The set of branch points $\ca{B}$ of $\ca{K}_g$
then being the union of points in \eqref{aa14} and
\eqref{aa15}.  Given $\ca{B}=\{P_1,\dots,P_r\}$
one can cut the complex plane along smooth
nonintersecting curves $\ca{C}_q$ (e.g., straight
lines if $P_1,\dots,P_r$ are arranged suitably)
connecting $P_q$ and $P_{q+1}$ for $q=1,\dots,r-1$,
 and defines holomorphic functions $f_1, \dots,
f_d$ on the cut plane $\Pi=
\cz\backslash\cup_{q=1}^{r-1}\ca{C}_q$ such that
\begin{equation}
\ca{P}(z,y,1)=0
\text{  for $y\in\Pi$ if and only if
 $y=f_j(z)$ for some $j\in\{1,\dots,d\}$.}
\lb{aa16}
\end{equation}
This yields a topological construction of
$\ca{K}_g$ by appropriately gluing together $d$ copies
of the cut plane $\Pi$, the result being a
sphere with $g$ handles ($g$ depending on the order of
the branch points in $\ca{B}$).  If $\ca{K}_g$ is singular,
this procedure requires appropriate
modifications.

Next, choose a homology basis $\{a_j,b_j\}_{j=1}^g$
on $\ca{K}_g$ for some $g\in\nz$ in such a way
that the intersection matrix of the cycles satisfies
\begin{equation}
a_j\circ b_k=\delta_{j,k}, \quad j,k=1,\dots,g \lb{aa16a}
\end{equation}
(with $a_j$ and $b_k$ intersecting to form a
 right-handed coordinate system).

Turning  briefly to meromorphic differentials
($1$-forms) on $\ca{K}_g$, we state the following
result.
\begin{theorem}[Riemann's period relations] \lb{taa3}
Let $g\in\nz$ and suppose $\om$ and $\nu$ to be closed
differentials
(\/$1$-forms) on $\ca{K}_g$.  Then \\
(i)
\begin{equation}
\iint\limits_{\ca{K}_g}\om\wedge\nu=
\sum_{j=1}^g\left(\big(\int_{a_j}\om
\big)\big(\int_{b_j}\nu
\big)-\big(\int_{b_j}\om\big)
\big(\int_{a_j}\nu\big) \right). \lb{aa17}
\end{equation}
If, in addition $\om$ and $\nu$ are holomorphic
$1$-forms on $\ca{K}_g$, then
\begin{equation}
\sum_{j=1}^g\left(\big(\int_{a_j}\om\big)
\big(\int_{b_j}\nu\big)
-\big(\int_{b_j}\om\big)\big(\int_{a_j}\nu\big)
\right)=0. \lb{aa17a}
\end{equation}
(ii) If $\om$ is a nonzero holomorphic $1$-form
on $\ca{K}_g$, then
\begin{equation}
\Im\left(\sum_{j=1}^g\big(\int_{a_j}\om
\big)\big(\int_{b_j}\om\big)\right)>0. \lb{aa18}
\end{equation}
\end{theorem}
The proof of Theorem \ref{taa3} is usually
 based on Stokes' theorem and a canonical dissection of
$\ca{K}_g$ along its cycles yielding the
simply connected interior $\hatt{\ca{K}}_g$ of the
fundamental polygon $\pa\hatt{\ca{K}}_g$ given by
\begin{equation}
\pa\hatt{\ca{K}}_g=a_1b_1a_1^{-1}b_1^{-1}a_2b_2a_2^{-1}
b_2^{-1}\dots a_g^{-1}b_g^{-1}. \lb{aa19}
\end{equation}
Given the cycles $\{a_j,b_j\}_{j=1}^g$, we denote by
$\{\om_j\}_{j=1}^g$ a normalized basis of
the space of holomorphic differentials (also called
 Abelian differentials of the first kind,
denoted \textit{dfk}) on $\ca{K}_g$, that is,
\begin{equation}
\int_{a_j}\om_k=\delta_{j,k}, \quad  j,k=1,\dots, g.
\lb{aa19a}
\end{equation}
The $b$-periods of $\om_k$ are then defined by
\begin{equation}
\tau_{j,k}=\int_{b_j}\om_k, \quad j,k=1,\dots, g. \lb{aa20}
\end{equation}
Theorem \ref{taa3} then implies the following result.
\begin{theorem} \lb{taa4}
The matrix $\tau$ is symmetric, that is,
\begin{equation}
\tau_{j,k}=\tau_{k,j}, \quad j,k=1,\dots, g, \lb{aa21}
\end{equation}
with a positive definite imaginary part,
\begin{equation}
\Im(\tau)= (\tau-\tau^*) /(2i)  >0. \lb{aa22}
\end{equation}
\end{theorem}
Abelian differentials of the second kind (abbreviated
{\textit{dsk}), say $\om^{(2)}$, are
characterized by the property that all their residues vanish.
  They are normalized by the vanishing
of all their $a$-periods (achieved by adding a suitable
linear combination of \textit{dfk}'s)
\begin{equation}
\int_{a_j}\om^{(2)}=0, \quad j=1,\dots, g,\lb{aa23}
\end{equation}
which determines them uniquely.  (We will always
assume that the poles of \textit{dsk}'s on
$\ca{K}_g$ lie in $\hatt{\ca{K}}_g$, that is,
do not lie on $\pa\hatt{\ca{K}}_g$.  This can always
be achieved by an appropriate choice of the
cycles $a_j$ and $b_j$.)  We may add in this context
that the sum of the residues of any meromorphic
 differential $\nu$ on $\ca{K}_g$ vanishes, the
residue at a pole $Q_0\in\ca{K}_g$ of $\nu$
being defined by
\begin{equation}
\resN_{Q_0}(\nu)=\fr{1}{2\pi i}\int_{\ga_{Q_0}} \nu,
\lb{aa24}
\end{equation}
where $\ga_{Q_0}$ is a smooth, simple, closed contour,
 oriented counter-clockwise, encircling $Q_0$,
but no other pole of $\nu$.
\begin{theorem}\lb{ta5}
Let $g\in\nz$.  Assume $\om_{Q_1,n}^{(2)}$ to be
a \textit{dsk} on
$\ca{K}_g$, whose only pole is $Q_1\in\hatt{\ca{K}}_g$
with principal
part $\zeta_{Q_1}^{-n}d\zeta_{Q_1}$ for some $n\in\nz_0$ and
$\om^{(1)}$ a \textit{dfk} on  $\ca{K}_g$ of the type
$\om^{(1)}=\sum_{m=0}^\infty c_m(Q_1)\zeta_{Q_1}^m
 d\zeta_{Q_1}$ near $Q_1$.
Then
\begin{equation}
\sum_{j=1}^g\left(\big(\int_{a_j}\om^{(1)} \big)
\big(\int_{b_j}\om^{(2)}_{Q_1,n}\big) -
\big(\int_{b_j}\om^{(1)} \big)
\big(\int_{a_j}\om^{(2)}_{Q_1,n}\big)\right)=
\fr{2\pi i}{(n-1)}c_{n-2}(Q_1), \quad n\geq 2. \lb{aa25}
\end{equation}
In particular, if $\om_{Q_1,n}^{(2)}$ is normalized
 and $\om^{(1)}=\om_j=\sum_{m=0}^\infty
c_{j,m}(Q_1)\zeta_{Q_1}^m\, d\zeta_{Q_1}$, then
\begin{equation}
\int_{b_j}\om_{Q_1,n}^{(2)}=\fr{2\pi i}{(n-1)}
c_{j,n-2}(Q_1), \quad n\geq 2,
\quad j=1,\dots, g.\lb{aa25a}
\end{equation}
\end{theorem}
Any meromorphic differential $\om^{(3)}$ on
$\ca{K}_g$ not of the first or
second kind is said to be of the third
kind, written \textit{dtk}.
It is common to  normalize a \textit{dtk}
$\om^{(3)}$, by the vanishing of its
$a$-periods, that is, by
\begin{equation}
\int_{a_j} \om^{(3)} =0, \quad j=1,\dots,g.
\lb{aa25b}
\end{equation}
A normal \textit{dtk}, denoted  $\om_{Q_1, Q_2}^{(3)}$,
 associated
with two distinct points $Q_1, Q_2 \in \hatt{\ca{K}}_g$
 by definition has
simple poles at $Q_\ell$ with residues $(-1)^{\ell+1}$
for $\ell=1$ and $2,$ vanishing $a$-periods, and is
holomorphic anywhere else.
\begin{theorem}\lb{taa6}
Let $g\in\nz$.  Suppose $\om^{(3)}$ to be a
\textit{dtk} on $\ca{K}_g$
whose only singularities are simple poles at
$Q_n\in\hatt{\ca{K}}_g$ with residues $c_n$
for $n=1,\dots,N$.  Denote by $\om^{(1)}$ a
\textit{dfk} on $\ca{K}_g$.  Then
\begin{equation}
\sum_{j=1}^g\left(\big(\int_{a_j}\om^{(1)}
\big)\big(\int_{b_j}\om^{(3)}\big)
-\big(\int_{b_j}\om^{(1)}\big)
\big(\int_{a_j}\om^{(3)}\big) \right)
=2\pi i\sum_{n=1}^N c_n \int_{Q_0}^{Q_n}\om^{(1)},
\lb{aa25c}
\end{equation}
where $Q_0\in\hatt{\ca{K}}_g$ is any fixed base point.
In particular, if $\om^{(3)}$ is normalized and
$\om^{(1)}=\om_j$, then
\begin{equation}
\int_{b_j}\om^{(3)}=2 \pi i \sum_{n=1}^N c_n
\int_{Q_0}^{Q_n}\om_j,
\quad j=1,\dots, g.
\lb{aa25d}
\end{equation}
Moreover, if $\om^{(3)}_{Q_1,Q_2}$ is a normal
\textit{dtk} on $\ca{K}_g$ holomorphic on
$\ca{K}_g\backslash\{Q_1,Q_2\}$, then
\begin{equation}
\int_{b_j}\om^{(3)}_{Q_1,Q_2}=2 \pi i
 \int_{Q_2}^{Q_1}\om_j,
\quad j=1,\dots, g.
\lb{aa25e}
\end{equation}
\end{theorem}
We shall always assume (without loss
of generality) that all poles of \textit{dsk}'s and
\textit{dtk}'s on $\ca{K}_g$ lie on
$\hatt{\ca{K}}_g$ (i.e., not on $\pa\hatt{\ca{K}}_g$) and
that integration paths on the right
 hand side of \eqref{aa25c}--\eqref{aa25e} do not
touch any cycles $a_j$ or $b_k$.

Next, we turn to divisors on $\ca{K}_g$
and the Jacobi variety
$J(\ca{K}_g)$ of $\ca{K}_g$. Let
$\ca{H}(\ca{K}_g)$ ($\ca{M}(\ca{K}_g)$) and
$\ca{H}^1(\ca{K}_g)$ ($\ca{M}^1(\ca{K}_g)$) denote the set of
holomorphic (meromorphic) functions (i.e., $0$-forms) and
holomorphic (meromorphic) $1$-forms on
$\ca{K}_g$ for some $g\in\nz_0$.
\begin{defn}\lb{daa7}
Let $g\in\nz_0$.  Suppose $f\in\ca{M}(\ca{K}_g)$,
$\om=h(\zeta_{Q_0})d\zeta_{Q_0}\in\ca{M}^1(\ca{K}_g)$,
 and $(U_{Q_0},\zeta_{Q_0})$ a
chart near $Q_0\in\ca{K}_g$. \\
(i) If $(f\circ\zeta_{Q_0}^{-1})(\zeta)=\sum_{n=m_0}^\infty
c_n(Q_0)\zeta^n$ for some $m_0\in\zz$ (which turns out
to be independent of the chosen chart), the
order $\nu_f(Q_0)$ of $f$ at $Q_0$ is defined by
\begin{equation}
\nu_f(Q_0)=m_0. \lb{aa26}
\end{equation}
One defines $\nu_f(P)=\infty$ for all $P\in\ca{K}_g$
if $f$ is identically zero on
$\ca{K}_g$. \\
(ii) If $h_{Q_0}(\zeta_{Q_0})=\sum_{n=m_0}^\infty
d_n(Q_0)\zeta^n_{Q_0}$ for some $m_0\in\zz$
(which again is independent of the chart chosen),
the order $\nu_\om(Q_0)$ of $\om$ at $Q_0$
is defined by
\begin{equation}
\nu_\om(Q_0)=m_0.  \lb{aa27}
\end{equation}
\end{defn}
\begin{defn}\lb{daa8}
Let $g\in\nz_0$. \\
(i) A divisor $\ca{D}$ on $\ca{K}_g$ is a map
$\ca{D}\colon\ca{K}_g\to\zz$, where $\ca{D}(P)\neq  0$
for only
finitely many $P\in\ca{K}_g$.  On the set of
all divisors $\Div(\ca{K}_g)$ on $\ca{K}_g$ one
introduces the partial ordering
\begin{equation}
\ca{D}\geq\ca{E} \text{ if  } \ca{D}(P)\geq
\ca{E}(P), \quad P\in\ca{K}_g. \lb{aa28}
\end{equation}
(ii) The degree $\deg(\ca{D})$ of $\ca{D}
\in\Div(\ca{K}_g)$ is defined by
\begin{equation}
\deg(\ca{D})=\sum_{P\in\ca{K}_g}\ca{D}(P). \lb{aa29}
\end{equation}
(iii) $\ca{D}\in\Div(\ca{K}_g)$ is called
nonnegative (or effective) if
\begin{equation}
\ca{D}\geq 0,  \lb{aa30}
\end{equation}
where $0$ denotes the zero divisor $0(P)=0$
for all $P\in\ca{K}_g$.  \\
(iv) Let $\ca{D},\ca{E}\in\Div(\ca{K}_g)$.
Then $\ca{D}$ is called a multiple of $\ca{E}$ if
\begin{equation}
\ca{D}\geq\ca{E}. \lb{aa31}
\end{equation}
$\ca{D}$ and $\ca{E}$ are called relatively prime if
\begin{equation}
\ca{D}(P)\ca{E}(P)=0, \quad P\in\ca{K}_g.  \lb{aa32}
\end{equation}
(v) If $f\in\ca{M}(\ca{K}_g)\backslash\{0\}$
and $\om\in\ca{M}^1(\ca{K}_g)\backslash\{0\}$, then the
divisor $(f)$ of $f$ is defined by
\begin{equation}
(f)\colon\ca{K}_g\to \zz, \quad P\mapsto \nu_f(P) \lb{aa33}
\end{equation}
(thus $f$ is holomorphic, $f\in\ca{H}(\ca{K}_g),$ if
and only if
$(f)\geq0$),
 and the divisor of $\om$ is defined by
\begin{equation}
(\om)\colon\ca{K}_g\to\zz, \quad P\mapsto \nu_\om(P)
\lb{aa34}
\end{equation}
(thus $\om$ is a \textit{dfk},
$\omega\in\ca{H}^1(\ca{K}_g),$ if
and only if
$(\om)\geq0$). The divisor $(f)$ is called a
principal divisor, and $(\om)$ a canonical divisor. \\
(vi)  The divisors $\ca{D},\ca{E}\in\Div(\ca{K}_g)$
are called equivalent, written
$\ca{D} \sim\ca{E}$, if
\begin{equation}
\ca{D}-\ca{E}=(f) \lb{aa35}
\end{equation}
for some $f\in\ca{M}(\ca{K}_g)\backslash\{0\}$. The
divisor class
$[\ca{D}]$ of $\ca{D}$ is
defined by
\begin{equation}
[\ca{D}]=\{\ca{E}\in\Div(\ca{K}_g) \mid \ca{E}\sim\ca{D}\}.
\lb{aa36}
\end{equation}
\end{defn}
Clearly, $\Div(\ca{K}_g)$ forms an Abelian group with respect to
addition of divisors.  The
principal divisors form a subgroup $\DivP(\ca{K}_g)$
of $\Div(\ca{K}_g)$.  The quotient
group $\Div(\ca{K}_g)/\DivP(\ca{K}_g)$ consists of
the cosets of divisors, the divisor
classes defined in \eqref{aa36}.  Also the set of
divisors of degree zero, $\Div_0(\ca{K}_g)$, forms
a subgroup of
$\Div(\ca{K}_g)$.  Since $\DivP(\ca{K}_g)\subset
\Div_0(\ca{K}_g)$, one can
introduce the quotient group $\text{Pic}(\ca{K}_g)=
\Div_0(\ca{K}_g)/\DivP(\ca{K}_g)$
called the Picard group of $\ca{K}_g$.
\begin{theorem} \lb{taa9}
Let $g\in\nz_0$.  Suppose $f\in\ca{M}(\ca{K}_g)$ and
$\om\in\ca{M}^1(\ca{K}_g)$.  Then
\begin{equation}
\deg((f))=0 \text{ and }
\deg((w))=2(g-1). \lb{aa37a}
\end{equation}
\end{theorem}
\begin{defn} \lb{da10}
Let  $g\in\nz_0$, and define
\begin{equation}
\ca{L}(\ca{D})=\{f\in\ca{M}(\ca{K}_g) \mid
 (f)\geq \ca{D}\},  \quad
\ca{L}^1(\ca{D})=\{\om\in\ca{M}^1(\ca{K}_g)
\mid (\om)\geq \ca{D}\}.  \lb{aa39}
\end{equation}
\end{defn}
Both $\ca{L}(\ca{D})$ and $\ca{L}^1(\ca{D})$ are
 linear spaces over $\cz$.  We denote their (complex)
dimensions by
\begin{equation}
r(\ca{D})=\dim\ca{L}(\ca{D}), \quad
i(\ca{D})=\dim\ca{L}^1(\ca{D}). \lb{aa41}
\end{equation}
$i(\ca{D})$ is also called the index of specialty of
$\ca{D}$.
\begin{lem} \lb{laa11}
Let $g\in\nz_0$ and $\ca{D}\in\Div(\ca{K}_g)$.
 Then  $\deg(\ca{D})$, $r(\ca{D})$, and
$i(\ca{D})$ only depend on the divisor
class $[\ca{D}]$ of $\ca{D}$ (and not on the particular
representative $\ca{D}$).  Moreover, for
 $\om\in\ca{M}^1(\ca{K}_g)\backslash\{0\}$ one infers
\begin{equation}
i(\ca{D})=r(\ca{D}-(\om)),
\quad \ca{D}\in\Div(\ca{K}_g). \lb{aa42}
\end{equation}
\end{lem}
\begin{theorem}[Riemann-Roch]\lb{taa12}
Let $g\in\nz_0$ and $\ca{D}\in\Div(\ca{K}_g)$.
 Then $r(-\ca{D})$ and $i(\ca{D})$ are
finite and
\begin{equation}
r(-\ca{D})=\deg(\ca{D})+i(\ca{D})-g+1.  \lb{aa43}
\end{equation}
In particular, Riemann's inequality
\begin{equation}
r(-\ca{D})\geq\deg(\ca{D})-g+1  \lb{aa44}
\end{equation}
holds.
\end{theorem}
Next we turn to the Jacobi variety and the Abel map.
\begin{defn} \lb{taa13}
Let $g\in\nz$ and define the period lattice $L_g$
in $\cz^g$ by
\begin{equation}
L_g=\{\ul{z}\in\cz^g \mid \ul{z}=\ul{N}+\tau\ul{M}, \;
\ul{N},\ul{M}\in\zz^g\}. \lb{aa45}
\end{equation}
Then the Jacobi variety $J(\ca{K}_g)$ of $\ca{K}_g$ is
defined by
\begin{equation}
J(\ca{K}_g)=\cz^g/L_g,  \lb{aa45a}
\end{equation}
and the Abel maps are defined by
\begin{align}
\ul{A}_{P_0} \colon \ca{K}_g \to J(\ca{K}_g),\quad
P\mapsto \ul A_{P_0} (P) &=(A_{P_0,1} (P),\dots,A_{P_0,g} (P))
\no \\ &=(\int_{P_{0}}^P \om_1,\dots,\int_{P_{0}}^P \om_g)
\, (\text{mod }L_g), \lb{aa46}
\end{align}
and
\begin{equation}
\ul \al_{P_0}  \colon
\Div(\ca{K}_g) \to J(\ca{K}_g),\quad
\ca{D} \mapsto \ul \al_{P_0} (\ca{D})
=\sum_{P \in \ca{K}_g} \ca{D} (P) \ul A_{P_0} (P),
\lb{aa47}
\end{equation}
where $P_0\in\ca{K}_g$ is a fixed base point and (for
convenience only) the same path is chosen from
$P_0$ to $P$ for all $j=1,\dots,g$ in \eqref{aa46} and
 \eqref{aa47}\footnote{This convention
allows one to avoid the multiplicative version
 of the Riemann-Roch Theorem at various places in
this paper.}.
\end{defn}
Clearly, $\ul A_{P_0}$  is well-defined since
changing the path from $P_0$ to $P$  amounts to
adding a closed cycle whose contribution in
the integral \eqref{aa46} consists in adding a vector in
$L_g$.  Moreover, $\ul \al_{P_0}$ is a group
 homomorphism and $J(\ca{K}_g)$ is a complex torus of
(complex) dimension $g$ that depends on the
choice of the homology basis $\{a_j,b_j\}_{j=1}^g$.
However, different homology bases yield
isomorphic Jacobians, see \cite{fk}, p.\ 137,
and \cite{Gunning:1972}, Section 8(b).
\begin{theorem}[Abel's theorem] \lb{taa14}
Let $g\in\nz$.  Then  $\ca{D}\in \Div(\ca{K}_g)$ is principal
if and only if
\begin{equation}
\deg (\ca{D}) =0 \text{ and } \ul \al_{P_0} (\ca{D})=\ul{0}.
\lb{aa48}
\end{equation}
\end{theorem}
Next, we turn to Riemann theta functions and a
constructive approach to the Jacobi inversion
problem.  We assume $g\in\nz$ for the remainder of
this appendix.

Given the curve
$\ca{K}_g$, the homology basis $\{a_j,b_j\}_{j=1}^g$,
 and the matrix $\tau$ of $b$-periods of the
\textit{dfk}'s $\{\om_j\}_{j=1}^g$, the Riemann theta
function associated
with $\ca{K}_g$ and the homology basis is defined as
\begin{equation}
\tta (\ul z) =\sum_{\ul n \in\zz^g}
 \exp \big(2\pi i (\ul n,\ul z) + \pi i (\ul n,
\tau \ul n)\big), \quad \ul z \in\cz^g,
\lb{aa49}
\end{equation}
where $(\ul u, \ul v)=\sum_{j=1}^g \overline{u}_j v_j$
denotes the scalar product in $\cz^g$. Because of
\eqref{aa22}, $\tta$ is well-defined and
represents an entire function on $\cz^g$.
Elementary properties of $\tta$ are, for instance,
\begin{align}
& \tta(z_1, \dots, z_{j-1}, -z_j, z_{j+1},
\dots, z_n) =\tta(\ul z),\quad \ul z=(z_1,
\dots,z_g)\in\cz^g, \lb{aa50}\\
& \tta (\ul z +\ul m +\tau \ul n)
=\tta (\ul z) \exp \big(-2\pi i (\ul n,\ul
z) -\pi i (\ul n, \tau
\ul n) \big), \quad \ul m, \ul n \in\zz^n,
 \; \ul z\in\cz^g. \lb{aa51}
\end{align}
\begin{lem} \lb{laa16}
Let $\ul \xi\in\cz^g$ and define
\begin{equation}
F\colon\hatt{\ca{K}}_g\to\cz, \quad
P\mapsto\tta(\ul{\hatt A}_{P_0}(P)-\ul\xi),\lb{aa51a}
\end{equation}
where
\begin{align}
\ul{\hatt A}_{P_0}\colon\hatt{\ca{K}}_g\to\cz^g,
\quad
P\mapsto\ul{\hatt A}_{P_0}(P) &=\big(\hatt A_{P_0,1}(P),
\dots,\hatt A_{P_0,g}(P)\big)
\no \\ &=
\bigg(\int_{P_0}^P\om_1,\dots,\int_{P_0}^P\om_g\bigg).
\lb{aa52}
\end{align}
Suppose $F$ is not identically zero on $\hatt{\ca{K}}_g$,
that is,
  $F\not\equiv0$.  Then $F$ has
precisely $g$ zeros on $\hatt{\ca{K}}_g$ counting
multiplicities.
\end{lem}
Lemma \ref{laa16} is traditionally proven by
integrating $d\ln(F)$
along $\pa\hatt{\ca{K}}_g$.
\begin{theorem} \lb{taa17}
Let $\ul\xi\in\cz^g$ and define $F$ as in \eqref{aa51a}.
Assume that $F$ is not identically zero on
$\hatt{\ca{K}}_g$, and let $Q_1,\dots,Q_g\in\ca{K}_g$ be
the zeros of $F$ (multiplicities included)
given by Lemma \ref{laa16}.  Define the corresponding
 positive divisor $\ca{D}_{\ul Q}$ of degree
$g$ on
$\ca{K}_g$ by
\begin{multline}
{\ca{D}}_{\ul{Q}} \colon {\ca{K}}_g \to \nz_0,  \\
 P \mapsto  {\ca{D}}_{\ul{Q}}(P)=\begin{cases} m & \text{if
$P$ occurs $m$ times in  $\{Q_1,\dots, Q_g\}$},\\
0& \text{if $P\not\in\{Q_1,\dots, Q_g\}$},
\end{cases} \\
\ul{Q}=(Q_1,\dots, Q_g), \lb{aa53}
\end{multline}
and recall the Abel map $\ul \al_{P_0}$ in \eqref{aa47}.
Then
there exists a vector
$\ul{\Xi}_{P_0}\in\cz^g$, the vector of Riemann constants,
such
that
\begin{equation}
\ul\alpha_{P_0}(\ca{D}_{\ul{Q}})
=(\ul{\xi}-\ul{\Xi}_{P_0})(\text{mod }L_g).
\lb{aa54}
\end{equation}
The vector $\ul{\Xi}_{P_0}=(\Xi_{P_{0,1}}, \dots,
 \Xi_{P_{0,g}})$ is given by
\begin{equation}
\Xi_{P_{0,j}}=\fr{1}{2}(1+\tau_{j,j})-
\sum_{\substack{\ell=1 \\ \ell\neq  j}}^g\int_{a_\ell}
\om_\ell(P)\int_{P_0}^P\om_j,
\quad j=1,\dots,g. \lb{aa55}
\end{equation}
\end{theorem}
For the proof of Theorem \ref{taa17} one integrates
 $\hatt A_{P_{0,j}}(P)d\ln (F(P))$ along
$\pa\hatt{\ca{K}}_g$.  Clearly, $\ul{\Xi}_{P_0}$
depends on the base point $P_0$ and on the choice of
the homology basis $\{a_j,b_j\}_{j=1}^g$.
\begin{rem} \lb{raa18}
Theorem \ref{taa14} yields a partial solution of
Jacobi's inversion problem which can be stated as
follows:  Given $\ul{\xi}\in\cz^g$, find a divisor
 $\ca{D}_{\ul{Q}}\in\Div(\ca{K}_g)$ such that
\begin{equation}
\ul{\al}_{P_0}(\ca{D}_{\ul{Q}})=\ul{\xi}(\text{mod }L_g).
\lb{aa57}
\end{equation}
Indeed, if $\widetilde F(P)=\tta(\ul{\Xi}_{P_0}
-{\ul{\hatt A}}_{P_0}(P)+\ul{\xi})\not\equiv0$  on
$\hatt{\ca{K}}_g$, the zeros
$Q_1,\dots,Q_g\in\hatt{\ca{K}}_g$
of $\widetilde F$ (guaranteed by Lemma
\ref{laa16}) satisfy Jacobi's inversion
 problem by \eqref{aa54}.  Thus it remains to specify
conditions such that $\widetilde F\not\equiv0$  on
$\hatt{\ca{K}}_g$.
\end{rem}

\begin{rem} \lb{raa19}
While $\tta(\ul{z})$ is well-defined (in fact,
entire) for $\ul{z}\in\cz^g$, it is not well-defined on
$J(\ca{K}_g)=\cz^g/L_g$ because of  \eqref{aa51}.
 Nevertheless, $\tta$ is a ``multiplicative
function'' on $J(\ca{K}_g)$ since the multipliers
in \eqref{aa51} cannot vanish.  In particular, if
$\ul{z}_1=\ul{z}_2(\text{mod }L_g)$, then $\tta(\ul{z}_1)=0$ if
 and only if  $\tta(\ul{z}_2)=0$.  Hence it is
meaningful to state that $\tta$ vanishes at points of
$J(\ca{K}_g)$. Since the Abel map
$\ul{A}_{P_0}$ maps $\ca{K}_g$ into  $J(\ca{K}_g)$,
the function $\tta(\ul{A}_{P_0}(P)-\ul{\xi})$ for
$\ul\xi\in\cz^g$, becomes a multiplicative function
on $\ca{K}_g$. Again it makes sense to say that
$\tta(\ul{A}_{P_0}(\dott)-\ul{\xi})$ vanishes at
points of $\ca{K}_g$.
\end{rem}
In the following we use the obvious notation
\begin{align}
X+Y&=\{(\ul{x}+\ul{y})\in J(\ca{K}_g)\mid \ul{x}
\in X, \ul{y}\in Y \}, \no \\
-X&= \{-\ul{x}\in J(\ca{K}_g)\mid \ul{x}\in X \},
 \lb{aa58} \\
X+\ul{z}&=\{(\ul{x}+\ul{z})\in J(\ca{K}_g)\mid
\ul{x}\in X\}, \no
\end{align}
for $X, Y\subset J(\ca{K}_g)$ and  $\ul{z}\in J(\ca{K}_g)$.
 Furthermore, we may identify
the $n$th symmetric power of
$\ca{K}_g$, denoted $\si^n\ca{K}_g$, with the set of
nonnegative
divisors of degree $n\in\nz$ on $\ca{K}_g$. Moreover,
we introduce the convenient notation
($N\in\nz$)
\begin{equation}
\ca{D}_{P_0\ul Q}=\ca{D}_{P_0}+\ca{D}_{\ul Q}, \quad
\ca{D}_{\ul Q}=\ca{D}_{Q_1}+\cdots+\ca{D}_{Q_N}, \quad
\ul Q=(Q_1,\dots,Q_N)\in\si^N\ca{K}_g, \lb{aa61}
\end{equation}
where for any $Q\in\ca{K}_g$,
\begin{equation}
{\ca{D}}_{Q} \colon {\ca{K}}_g \to \nz_0, \quad
P \mapsto  \ca{D}_Q(P)= \begin{cases} 1 & \text{for $P=Q$},\\
0 & \text{for $P\in\ca{K}_g\backslash\{Q\}$}.
\end{cases} \lb{aa61A}
\end{equation}
\begin{defn} \lb{daa20}
(i) Define
\begin{equation}
\ul{W}_0=\{\ul{0}\}\subset J(\ca{K}_g),
\quad \ul{W}_n=\ul{\al}_{P_0}(\si^n\ca{K}_g),
\quad n\in\nz. \lb{aa59}
\end{equation}
(ii) A positive divisor $\ca{D}\in\Div(\ca{K}_g)$
is called special if $i(\ca{D})\geq1$,
otherwise $\ca{D}$ is called nonspecial. \\
(iii) $Q\in\ca{K}_g$ is called a Weierstrass point
 of $\ca{K}_g$ if $i(g\ca{D}_{Q})\geq1$,
where $g\ca{D}_{Q}=\sum_{j=1}^g \ca{D}_Q$.
\end{defn}
\begin{rem} \lb{raa21}
(i)  Since $i(\ca{D}_P)=0$ for all $P\in\ca{K}_1$,
the curve $\ca{K}_1$ has no Weierstrass points.
For
$g\geq2$, and $\ca{K}_g$ hyperelliptic, the
 Weierstrass points of $\ca{K}_g$ are given
precisely by
the $2g+2$ branch points of $\ca{K}_g$. \\
(ii) The special divisors of the type
$\ca{D}_{\ul Q}$ with
$\ul Q=(Q_1,\dots,Q_N)\in\si^N\ca{K}_g$ and
$\deg(\ul Q)=N\geq g$
are precisely the critical points of the Abel map
$\ul{\al}_{P_0}
\colon\si^N\ca{K}_g\to J(\ca{K}_g)$,
that is, the set of points $\ca{D}$ at which the rank of the
differential  $d\ul{\al}_{P_0}$ is less than $g$. \\
(iii) While $\si^m\ca{K}_g\not\subset\si^n\ca{K}_g$ for
$m<n$, one has
$\ul{W}_m\subset\ul{W}_n$ for $m<n$.  Thus
$\ul{W}_n=J(\ca{K}_g)$
 for $n\geq g$ by Theorem
\ref{taa12}.
\end{rem}
\begin{theorem} \lb{taa22}
The set $\ul{W}_{g-1}+\ul{\Xi}_{P_0}\subset J(\ca{K}_g)$ is
the complete set of zeros of
$\tta$ on $J(\ca{K}_g)$, that is,
\begin{equation}
\tta(X)=0 \text{ if and only if } X\in\ul{W}_{g-1}+
\ul{\Xi}_{P_0}
\lb{aa62}
\end{equation}
\emph{(}i.e., if and only if $X=\big(\ul{\al}_{P_0}
(\ca{D})+\ul{\Xi}_{P_0}\big) (\text{mod } L_g) $ for some
$\ca{D}\in\si^{g-1}\ca{K}_g$\emph{)}.  The set
$\ul{W}_{g-1}+\ul{\Xi}_{P_0}$ has complex dimension $g-1$.
\end{theorem}
\begin{theorem}[Riemann's vanishing theorem] \lb{taa23}
Let $\ul{\xi}\in\cz^g$.  \\
(i) If $\tta(\ul{\xi})\neq  0$, then there exists a
 unique $\ca{D}\in\si^g\ca{K}_g$ such that
\begin{equation}
\ul{\xi}=\big(\ul{\al}_{P_0}(\ca{D})+\ul{\Xi}_{P_0}
\big)\, (\text{mod }L_g) \lb{aa63}
\end{equation}
and
\begin{equation}
 i(\ca{D})=0.\lb{aa64}
\end{equation}
(ii) If $\tta(\ul{\xi})=0$ and $g=1$, then
\begin{equation}
\ul{\xi}=\ul{\Xi}_{P_0}(\text{mod }L_1)=2^{-1}(1+\tau)
(\text{mod }L_1),
\quad L_1=\zz+\tau\zz, \quad
-i\tau>0.\lb{aa65}
\end{equation}
(iii) Assume  $\tta(\ul{\xi})=0$ and $g\geq 2$.
Let $s\in\nz$ with $s\leq g-1$ be the smallest
integer such that $\tta(\ul{W}_s-\ul{W}_s-\ul{\xi})\neq  0$
  \emph{(}i.e., there exist
$\ca{E}, \ca{F}\in\si^s\ca{K}_g$ with $\ca{E}\neq \ca{F}$
 such that
$\tta(\ul{\al}_{P_0}(\ca{E})-\ul{\al}_{P_0}(\ca{F})
 -\ul{\xi})\neq  0$\emph{)}.  Then there exists a
$\ca{D}\in\si^{g-1}\ca{K}_g$ such that
\begin{equation}
\ul{\xi}=\big(\ul{\al}_{P_0}(\ca{D})+\ul{\Xi}_{P_0}
\big)\, (\text{mod }L_g) \lb{aa66}
\end{equation}
and
\begin{equation}
 i(\ca{D})=s.\lb{aa67}
\end{equation}
All partial derivatives of $\tta$ with respect to
 $A_{P_0,j}$ for $j=1,\dots, g$ of order
strictly less than $s$ vanish at $\ul{\xi}$, whereas
at least one partial derivative of $\tta$ of
order $s$ is nonzero at $\ul{\xi}$.  Moreover, $s\leq
(g+1)/2$ and the integer $s$ is the same for
$\ul{\xi}$ and $-\ul{\xi}$.
\end{theorem}
Note that there is no explicit reference to the base
 point $P_0$ in the formulation of Theorem
\ref{taa23} since the set $\ul{W}_s-\ul{W}_s\subset
J(\ca{K}_g)$  is independent of
the base point while $\ul{W}_s$ alone is not.
\begin{theorem}[Jacobi's inversion theorem] \lb{taa24}
The map $\ul{\al}_{P_0}$ is surjective.  More precisely,
 given
$\ti{\ul{\xi}}=(\ul{\xi}+\ul{\Xi}_{P_0})\in\cz^g$,
the divisors $\ca{D}$ in \eqref{aa63} and
\eqref{aa66} (resp.\ $\ca{D}=\ca{D}_{P_0}$ if $g=1$)
 solve the Jacobi inversion problem for
$\ul{\xi}\in\cz^g$.
\end{theorem}
We summarize some of this analysis in the following remark.
\begin{rem} \lb{raa25}
Consider the function
\begin{equation}
G(P)=\tta\big(\ul{\Xi}_{P_0}-\ul{\hatt A}_{P_0}(P)+
\sum_{j=1}^g\ul{\hatt A}_{P_0}(Q_j)\big), \quad
 P,Q_j\in \ca{K}_g, \quad j=1,\dots, g \lb{aa68}
\end{equation}
on $\ca{K}_g$.  Then
\begin{align}
G(Q_k)=\tta(\ul{\Xi}_{P_0}+\sum_{\substack{j=1 \\ j
 \neq  k}}^g\ul{\hatt A}_{P_0}(Q_j))
=\tta\big(\ul{\Xi}_{P_0}+\ul{\al}_{P_0}
(\ca{D}_{(Q_1,\dots,Q_{k-1},Q_{k+1},\dots,Q_g)})\big)=0,
\lb{aa69} \\
\hspace*{8cm} k=1,\dots, g \no
\end{align}
by Theorem \ref{taa22}.  Moreover, by Lemma
\ref{laa16} and Theorem \ref{taa23}, the
points $Q_1,\dots,Q_g$ are the only zeros of
$G$ on $\ca{K}_g$ if and only if $\ca{D}_{\ul{Q}}$ is
nonspecial, that is, if and only if
\begin{equation}
i(\ca{D}_{\ul{Q}})=0, \quad \ul{Q}=(Q_1,\dots,Q_g)
\in\si^g\ca{K}_g. \lb{aa70}
\end{equation}
Conversely,  $G\equiv0$  on $\ca{K}_g$ if and only
if $\ca{D}_{\ul{Q}}$ is special,
that is, if and only if $i(\ca{D}_{\ul{Q}})\geq 1$.
\end{rem}
We also mention the elementary change in the Abel
map and in Riemann's vector if one changes the
base point,
\begin{align}
 \ul{A}_{P_1}&=\big(\ul{A}_{P_0}-\ul{A}_{P_0}(P_1)
 \big) \, (\text{mod }L_g), \lb{aa70a} \\
\ul{\Xi}_{P_1}&=\big(\ul{\Xi}_{P_0}+(g-1)\ul{A}_{P_0}(P_1)
\big) \, (\text{mod }L_g),\quad P_0, P_1\in\ca{K}_g.\lb{aa71}
\end{align}
\begin{rem} \lb{raa27}
Let $\ul{\xi}\in J(\ca{K}_g)$ be given, assume that
$\tta(\ul{\Xi}_{P_0}-\ul{A}_{P_0}(\dott)+\ul{\xi})\not
\equiv0$ on $\ca{K}_g$ and
suppose that
$\ul{A}_{P_0}^{-1}(\ul{\xi})=(Q_1,\dots,Q_g)\in\si^g\ca{K}_g$
 is the unique solution of Jacobi's
inversion problem.  Let $f\in\ca{M}(\ca{K}_g)\backslash\{0\}$
and suppose $f(Q_j)\neq \infty$ for
$j=1,\dots, g$.  Then $\ul{\xi}$ uniquely determines the values
 $f(Q_1),\dots,f(Q_g)$.  Moreover,
any symmetric function of these values is a single-valued
meromorphic function of
$\ul{\xi}\in J(\ca{K}_g)$, that is, an Abelian function
on $J(\ca{K}_g)$.  Any such meromorphic
function on $J(\ca{K}_g)$ can be expressed in terms of the
Riemann theta function on $\ca{K}_g$. For
instance, for the elementary symmetric functions of the
second kind (Newton polynomials) one
obtains from the residue theorem in analogy to the proof
 of Lemma \ref{laa16} that
\begin{equation}
\sum_{j=1}^g f(Q_j)^n
=\sum_{j=1}^g \int_{a_j} f(P)^n\om_j(P)
-\sum_{\substack{P_r\in\ca{K}_g \\ f(P_r)=
\infty}}\resN_{P=P_r}\left(
f(P)^n d\ln(\tta(\ul{\Xi}_{P_0}-\ul{A}_{P_0}+\ul{\xi}))\right),
\lb{aa77}
\end{equation}
where an appropriate homology basis $\{a_j,b_j\}_{j=1}^g$ with
$\pa\hatt{\ca{K}}_g=a_1b_1a_1^{-1}b_1^{-1}\dots
a_g^{-1}b_g^{-1}$ avoiding $\{Q_1,\dots,Q_g\}$
and the poles $\{P_r\}$ of $f$ has been chosen.
(We also note that Lemma \ref{laa16} just
corresponds to the case $n=0$ in \eqref{aa77}.)
\end{rem}
Finally, we formulate the following auxiliary result
(cf., e.g., Lemma 3.4 in \cite{GR96}).
\begin{lem} \lb{lem34}
Let $\psi(\dott,x),\; x\in\ca{U}$, $\ca{U}\subseteq\rz$
open, be
meromorphic on ${\ca{K}}_g\backslash
\{\py\}$ with an essential singularity
at $\py$ \emph{(}and ${\widetilde\Om}_{\py,r+1}^{(2)}$
defined as
in \eqref{clap}\emph{)}
such that ${\widetilde \psi}(\dott,x)$ defined by
\begin{equation}
{\widetilde \psi} (\dott,x) =\psi (\dott,x)
\exp\bigg(-i(x-x_0)\int_{P_0}^P
{\widetilde\Om}_{\py,r+1}^{(2)}\bigg)
\lb{342a}
\end{equation}
is multi-valued meromorphic on $\ca{K}_n$ and its
divisor satisfies
\begin{equation}
({\widetilde \psi} (\dott,x))\geq -{\ca{D}}_{\hat{\ul \mu}(x)}.
\lb{342b}
\end{equation}
Define a divisor ${\ca{D}}_0 (x)$ by
\begin{equation}
({\widetilde \psi} (\dott,x))={\ca{D}}_0 (x)
-\ca{D}_{\hat{\ul \mu} (x)}.
\lb{342c}
\end{equation}
Then
\begin{equation}
{\ca{D}}_0 (x) \in\si^g {\ca{K}}_g, \; \ca{D}_0 (x) \geq 0,
\; \deg (\ca{D}_0 (x))=g.
\lb{342d}
\end{equation}
Moreover, if $\ca{D}_0 (x)$ is nonspecial for all
$x\in\ca{U}$, that is, if $i (\ca{D}_0 (x) ) =0$,
then $\psi (\dott,x)$ is unique up to a constant
multiple (which may depend on $x\in\ca{U}$).
\end{lem}

%%%%%%%%%%%%%%%%%%%%%%%%%%%%%%%%%%%%%%%%%%%%%%%%%%%%%%%%%%%%%
\section{Trigonal Curves of Boussinesq-Type} \lb{app-b}
%%%%%%%%%%%%%%%%%%%%%%%%%%%%%%%%%%%%%%%%%%%%%%%%%%%%%%%%%%%%%%

We give a brief summary of some of the fundamental properties
and notations needed from the theory of trigonal curves of
Boussinesq-type (i.e., those with a triple point at infinity).

First we investigate what happens at the point
(or possibly points) at
infinity on our Bsq-type curves.
Fix $g\in\nz$. The Bsq-type curve $\ca{K}_g$
of arithmetic genus $g=m-1$ is defined by
\begin{align}
\ca{F}_{m-1}(z,y) &=
y^3 + y\,S_m(z) - T_m (z) = 0, \no \\
S_m(z) &= \sum_{p=0}^{2\,n-1+\ep} s_{m,p} z^p, \quad
T_m(z) = z^{m}+ \sum_{q=0}^{m-1} t_{m,q} z^q, \lb{bb1}\\
  &  \hspace{10mm} \mm . \no
\end{align}

Following the treatment in \cite{mum} one substitutes the
variable $u=z^{-1}$ into \eqref{bb1} to obtain
\begin{align}
u^{3n+\ep}y^3 + \big(s_{m,0}u^{2n-1+\ep} + \dots +
 s_{m,2n-1+\ep}\big)u^{n+1}y  -
\big(t_{m,0}u^{3n+\ep} + \dots + t_{m,m-1}u + 1\big) = 0.
\lb{eq5}
\end{align}
Let $v=u^{n+1} y$ in \eqref{eq5} to obtain
\begin{align}
v^3 + (s_{m,0}u^{2n-1+\ep} + \dots
+ s_{m,2n-1+\ep})u^{3-\ep}v   -
(t_{m,0}u^{3n+\ep} + \dots + t_{m,3n-1+\ep}u + 1)u^{3-\ep}
= 0. \lb{eq6}
\end{align}
Let $u\to 0$ (corresponding to $z\to\infty$) in \eqref{eq6} to
 obtain $v^3=0$.
This corresponds to one point of multiplicity three at
infinity (in both cases $\ep=1$ and $\ep=2$), given by
$(u,v)=(0,0)$.  We therefore use the coordinate
$\zeta= z^{-1/3}$ at the branch point at infinity, denoted
by $\py$.

The curve \eqref{bb1} is compactified by adding the
point $\py$ at infinity.  In homogeneous coordinates, the
point at infinity we add is $[1:0:0]\in\cz\pz^2$ if $g=0$
or $g=1$, otherwise the point at infinity we add is
$[0:1:0]\in\cz\pz^2$.  The point $\py$ is singular in all
cases except when $g=1$, or when $g=2$ and $s_{m,0}=-1/3$.

Although not directly associated with the Bsq hierarchy,
we note
that the case $\ep=0$ in \eqref{bb1} is analogous to AKNS,
Toda,
and Thirring-type hyperelliptic curves, which are not
branched at
infinity. In fact, a similar argument to that above, with the
coordinate $v= u^{n} y  $  in \eqref{eq5},
yields the equation $v^3=1$ as $u\to 0$.
This corresponds to three distinct points $P_{\infty,j},\,
j=1,2,3$ at infinity (each with multiplicity one), given by the
three points
$(u,v)=(0,\om_j)$ for $j=1,2,3$, where the $\om_1$,
$\om_2$, and
$\om_3$
are the third roots of unity.  As each point at infinity has
multiplicity one, none are branch points, and consequently
each admits the local coordinate $u=1/z$ for $|z|$
sufficiently large.

In \cite{bc2}, p.~561, Burchnall and Chaundy define
the $g$-number of an algebraic curve as the maximum number
of double points possible in the finite plane.
For Bsq-type curves the $g$-number is $g=m-1$.
For a curve that is smooth in the finite plane, the $g$-number
coincides with the arithmetic genus of the curve, but in
the presence of
double points, the $g$-number remains the same, while the genus
is diminished (according to results of Clebsch, Noether, and
Pl\"ucker, see, e.g., \cite{bries} and \cite{mir}).  We
now prove
that the $g$-number of $\ca{K}_g$,
and hence the arithmetic genus of $\mathcal{K}_g$ if
$\mathcal{K}_g$ is smooth in the finite plane,
is $m-1$ using a special case of the Riemann-Hurwitz theorem.
%%%%%%%%%%%%%%%%%%%%%%%%%%%%%%%%%%%%%%%%%%%%%%%%%%%%%%%%%%%%%
\begin{theorem} \lb{th2}
Let $\tilde \pi_z \colon \ca{K}_g \to \cz\pz^1$ be the
projection map
with respect to
the $z$ coordinate.  Then
\begin{equation}
\sum_{P\in \ca{K}_g}\big[\nu_P(\tilde \pi_z)-1\big] =
2g + 4, \lb{eq7}
\end{equation}
where $\nu_P(\tilde \pi_z)$ denotes the multiplicity of
$\tilde \pi_z$ at
$P\in\ca{K}_g$,
and $g$ is the arithmetic genus of the curve $\ca{K}_g$.
\end{theorem}
%%%%%%%%%%%%%%%%%%%%%%%%%%%%%%%%%%%%%%%%%%%%%%%%%%%%%%%%%%%%%
If equation \eqref{bb1} has only double points, this implies
that
the discriminant $\Delta (z)$ of the curve \eqref{bb1},
defined
by
\begin{equation}
\Delta (z) = 27 T_m(z)^2 + 4\,S_m(z)^3 \lb{eq9}
\end{equation}
(modulo constants), is non-zero.
$\Delta (z)$ is easily seen to be a polynomial of degree $2m$.
Hence in the finite complex plane, the Riemann surface defined
by the compactification of \eqref{bb1} can have at most $2m$
double points, corresponding to the possible $2m$ zeros of
$\Delta (z)$.  If all finite branch points are distinct
double points
(taking into account the triple point at infinity) one obtains
$\sum_{P\in \ca{K}_g}\big[\nu_P(\tilde \pi_z)-1\big]=2m+2$,
and so by
\eqref{eq7}, one infers $g=m-1$.

Let $\ca{B}$ denote the set of branch points and let
$|\ca{B}|$ denote the number of branch points counted according
to multiplicity.
In the case of Bsq-type curves, $\deg (\tilde \pi_z)=3$, and
$\nu_P(\tilde \pi_z)=1$ for all $P\in\ca{K}_g\backslash\ca{B}$.
Moreover, $\nu_P(\tilde \pi_z)\in\{2,3\}$ for all $P\in\ca{B}$.
Hence
$|\ca{B}|\leq\sum_{P\in \ca{K}_g}\big[\nu_P(\tilde
\pi_z)-1\big]\leq
2|\ca{B}|$, and \eqref{eq7} reduces to
\begin{equation}
g+2 \leq |\ca{B}| \leq 2g+4 \lb{eq8}.
\end{equation}
Thus one arrives at an upper and lower bound on the
number of branch points on $\ca{K}_g$.

When $m=1$, corresponding to $g=0$, there are
no non-zero holomorphic differentials on $\ca{K}_g$.
When $m=2$, corresponding to $g=1$, the only holomorphic
differential on $\ca{K}_g$ is $dz/(3y(P)^2+S_m(z))$.
Recall also that $m\neq 0(\text{mod }3)$, so we need
not consider
holomorphic differentials for the case $m=3$.
One verifies that $dz/(3y(P)^2+S_m(z))$ and
$y(P)dz/(3y(P)^2+S_m(z))$
are holomorphic differentials $\ca{K}_g$ with zeros at $\py$
of order $2(m-2)$ and $(m-4),$ respectively, for $m\geq 4$.
It follows that the differentials ($m=3n+\varepsilon,$
$\varepsilon\in\{1,2\}$)
\begin{equation}
\eta_\ell(P)=\fr{1}{3y(P)^2+S_m(z)}
\begin{cases}
z^{\ell -1}dz & \text{ for } 1\leq\ell\leq g-n, \\
y(P)z^{\ell+n-g-1}dz & \text{ for } g-n+1\leq\ell\leq g,
\end{cases} \lb{e:basis}
\end{equation}
form a basis in the space of holomorphic differentials
$\ca{H}^1(\ca{K}_g)$.
Introducing the invertible matrix $\Upsilon\in GL(g,\cz)$,
\begin{align}
\begin{split}
\Upsilon & =(\Upsilon_{j,k})_{j,k=1,\dots,g},
\quad \Upsilon_{j,k}
= \int_{a_k} \eta_j, \\
\ul{e}(k) & = (e_1(k), \dots,
e_g(k)), \quad e_j (k) =
\big(\Upsilon^{-1}\big)_{j,k} \, , \lb{B.7}
\end{split}
\end{align}
the normalized differentials $\om_j$ for $j=1,\dots,g$,
\begin{equation}
\om_j = \sum_{\ell=1}^g e_j (\ell) \eta_\ell,
\quad \int_{a_k} \om_j =
\delta_{j,k}, \quad j,k=1,\dots,g
\lb{b26}
\end{equation}
form a canonical basis for $\ca{H}^1(\ca{K}_g)$.
Near $\py$ one infers
\begin{equation}
\ul{\om}  = (\om_1,\dots,\om_g)
\ztz \big(\ul{\al}_0^{(\varepsilon)} +
\ul{\al}_1^{(\varepsilon)} \zeta +
\ul{\al}_3^{(\varepsilon)} \zeta^3
 + O(\zeta^4)\big)d\zeta , \lb{abc}
\end{equation}
where
\begin{align}
\ul{\al}_0^{(\varepsilon)} &= -
\begin{cases}
\ul{e}(g), & \varepsilon =1, \\
\ul{e}(g-n), & \varepsilon =2,
\end{cases} \lb{def} \\[1mm]
\ul{\al}_1^{(\varepsilon)} &=
\begin{cases}
- \ul{e}(g-n), &  \varepsilon =1,\\
\big( d_0^{(2)}\ul{e}(g-n) - \ul{e}(g) \big),
& \varepsilon =2,
\end{cases} \\[1mm]
\ul{\al}_3^{(\varepsilon)} &=
\begin{cases}
\big( d_1^{(1)}\ul{e}(g) + c_1^{(1)}\ul{e}(g-n) - \ul{e}(g-1)
\big),  & \varepsilon =1, \\
\big( (2c_1^{(2)}-(d_0^{(2)})^3)\ul{e}(g-n) - \ul{e}(g-n-1)
+ (d_0^{(2)})^2\ul{e}(g) \big),
& \varepsilon =2,
\end{cases} \\
& \text{etc.,} \no
\end{align}
and
\begin{equation}
y(P) \ztz \big( c_0^{(\varepsilon)}
+ d_0^{(\varepsilon)} \zeta +
c_1^{(\varepsilon)} \zeta^3 + d_1^{(\varepsilon)} \zeta^4 +
O(\zeta^6)
\big)\zeta^{-3n-2} \text{ as } P\to\py, \lb{b27a}
\end{equation}
with
\begin{equation}
(c_0^{(\varepsilon)},d_0^{(\varepsilon)}) =
\begin{cases} (0,1),
& \varepsilon =1, \\ (1,d_0^{(2)}), & \varepsilon =2,
\end{cases}
\qquad d_0^{(2)}\in\cz.
\end{equation}
In particular, using \eqref{aa25a}, \eqref{abc}, and
\eqref{def},  one
obtains
\begin{equation}
\fr{1}{2\pi i}\int_{b_j}\dsk = \al_{0,j}^{(\varepsilon)}
\quad\text{and}\quad
\fr{1}{2\pi i}\int_{b_j}\dsk[3] =
\fr{1}{2}\al_{1,j}^{(\varepsilon)}. \lb{ben1}
\end{equation}
Finally, we turn our attention to special divisors.

{}From the theory of elementary symmetric polynomials one
infers the following lemma.
%%%%%%%%%%%%%%%%%%%%%%%%%%%%%%%%%%%%%%%%%%%%%%%%%%%%%%%%%%%%%%
\begin{lem} \lb{le1}
Pick $z\in\cz$, and denote by $y_1(z),\,y_2(z)$, and
$y_3(z)$, the
three solutions of (\ref{bb1}).  These solutions are
distinct if
and only if the discriminant $\Delta (z)\neq  0$. Moreover,
introduce $Q_j=(z,y_j)\in\ca{K}_g$ for $j=1,2,3$. Then
\begin{enumerate}
\item[(i)] $\sum_{j=1}^3 y_j(z) = 0$.
\item[(ii)] $\sum_{j<k}^3 y_j(z)y_k(z) = S_m(z)$.
\item[(iii)] $\prod_{j=1}^3 y_j(z) = T_m(z)$.
\item[(iv)] $\sum_{j=1}^3 y_j(z)^2 = -2S_m(z)$.
\item[(v)] $\sum_{j=1}^3 y_j(z)^3=3T_m(z)$.
\item[(vi)] $\sum_{j\neq k}^3 y_j(z)^2y_k(z) = -3T_m(z)$.
\item[(vii)] $\sum_{j<k}^3 y_j(z)^2y_k(z)^2 = S_m(z)^2$.
\item[(viii)] $\prod_{j=1}^3 \big(3y_j(z)^2+S_m(z)\big)
= \Delta
(z)$.
\end{enumerate}
\end{lem}
%%%%%%%%%%%%%%%%%%%%%%%%%%%%%%%%%%%%%%%%%%%%%%%%%%%%%%%%%%%%%
\begin{lem} \lb{sar}
Let $m_1,\dots,m_r\in\nz$ with $\sum_{j=1}^r m_j = g$ and
$Q_j=(z,y_j),$ $j=1,2,3$ as in Lemma \ref{le1}.
Suppose $P_1,\dots,P_r\in\ca{K}_g$. If
\begin{equation}
\{Q_1,Q_2,Q_3\}\subseteq\{P_1,\dots,P_r\},
\end{equation}
then the divisor
$\ca{D}_{m_1P_1+\dots+m_rP_r}\in\sigma^g\ca{K}_g$
is special.  In particular, if one of the points
$P_j\in\{P_1,\dots,P_r\}$ is a triple point, then
the divisor $\ca{D}_{m_1P_1+\dots+m_rP_r}\in\sigma^g\ca{K}_g$
is special.
\end{lem}
%%%%%%%%%%%%%%%%%%%%%%%%%%%%%%%%%%%%%%%%%%%%%%%%%%%%%%%%%%%%%%
\begin{proof}
Using the identities in Lemma \ref{le1}, one readily computes
\begin{equation}
\sum_{j=1}^3 \fr{1}{3y_j(z)^2+S_m(z)} = 0, \quad
\sum_{j=1}^3 \fr{y_j(z)}{3y_j(z)^2+S_m(z)} = 0. \lb{1}
\end{equation}
Thus, choosing for simplicity the base point $P_0=P_\infty,$
a comparison of \eqref{aa46}, \eqref{e:basis}, and
\eqref{1} yields
\begin{equation}
\sum_{j=1}^3 {\ul A}_{P_{\infty}}(Q_j) = 0 \, (\text{mod }L_g).
\end{equation}
Thus $\ca{D}_{m_1P_1+\dots+m_rP_r}\in\sigma^g\ca{K}_g$
is special by Theorem~\ref{taa22}.
\end{proof}

%%%%%%%%%%%%%%%%%%%%%%%%%%%%%%%%%%%%%%%%%%%%%%%%%%%%%%%%%%%%%

\end{document}